\documentclass[fleqn,10pt]{article}
\usepackage{latexsym, graphicx, epsfig, amsmath,  amssymb,amsfonts}
\usepackage{natbib,amsthm,version}
\usepackage{amsbsy,bm,multirow}
\usepackage[titletoc,page]{appendix}
\usepackage[mathscr]{eucal}
\usepackage{color}

\title{
A Rotational Pressure-Correction Scheme for
Incompressible Two-Phase Flows with Open Boundaries
} 
\author{
  S. Dong\thanks{Author of correspondence. Email: sdong@purdue.edu}, \; X. Wang \\
  Center for Computational and Applied Mathematics \\
  Department of Mathematics \\
  Purdue University, USA 
 } 

\date{} 
\begin{document}
\maketitle



\begin{abstract}

Two-phase outflows refer to situations where
the interface formed between two immiscible incompressible fluids
passes through open portions of the domain
boundary. 
We present in this paper
several new forms of open boundary conditions
for two-phase outflow simulations within 
the phase field framework. In addition, we also
present a rotational 
pressure-correction based algorithm
for numerically treating these  open 
boundary conditions. Our algorithm
gives rise to linear algebraic systems for the velocity
and the pressure 
that involve
only constant and time-independent coefficient matrices
after discretization,
despite the variable density and variable viscosity
of the two-phase mixture.
By comparing simulation results with the theory and the experimental
data, we show that the method developed herein
produces physically accurate results. 
We also present numerical experiments to demonstrate
the long-term stability of the method in situations where
large density contrast, large viscosity contrast,
and backflows are present at the
two-phase outflow/open boundaries.

\end{abstract}


\vspace{0.05cm}
Keywords: {\em 
two-phase outflow; open boundary condition; pressure correction; phase field;
outflow boundary condition; two-phase open boundary condition
}

\section{Introduction}
\label{sec:intro}

The current work focuses on the motion of a mixture of
two immiscible incompressible fluids in a domain
that is open on part of its boundary.
The domain boundary is open in the sense that 
the fluids can freely leave or even enter the domain 
through such boundaries.
In particular, we assume that the interface formed
between the two fluids will pass through the
open portions of the domain boundary.
Therefore, the problem will involve
truly two-phase outflow or open boundaries.


Two-phase outflows are encountered in many situations:
oil plumes in the deep sea, wakes of surface ships,
and ocean waves generated by the wind shear are
some of the examples.
These problems usually involve physically unbounded
flow domains. Numerical simulation of such problems
will therefore need to truncate the domain to
a finite size, and some open/outflow boundary
condition (OBC) will be required on the artificial boundary.
The presence of fluid interfaces  at
the open boundary calls for appropriate
two-phase open boundary conditions
in such problems.

Several challenges are associated with the design of
two-phase open boundary conditions.
Some of the challenges are common to those encountered
with single-phase outflows, for example,
the instability associated with
strong vortices or backflows at
the open boundary 
(a.k.a. backflow instability) \cite{DongKC2014}.
Others  are new and unique to
two-phase outflows. 
For example, owing to the presence of fluid interfaces,
two-phase outflow problems
involve viscosity contrasts,
density contrasts, and surface tension
at the open boundaries.
Large viscosity ratios and large density ratios 
at the open boundary can cause severe stability
difficulties \cite{Dong2014obc}.


While outflow/open boundary conditions for single-phase flows
have been under intensive investigations for decades,
very scarce work exists for two-phase outflows or open boundaries.
In \cite{LouGS2013} the zero-flux (Neumann), 
convective, and extrapolation boundary conditions
originated from single-phase flows have been studied
for the two-phase lattice-Boltzmann equation.
The zero-flux condition is also employed for
the outflow boundary in \cite{AlbadawiDRMD2013}
within the context of a coupled level-set/volume-of-fluids
method, and in \cite{Son2001} in the context of
a level set method where the outflow boundary involves
only a single type of fluid. 
The outflow condition for two immiscible fluids in
a porous medium is discussed in \cite{LenzingerS2010}.
The works of \cite{Munkejord2006,DesmaraisK2014} 
have both considered the outflow condition 
for two-phase compressible flows in one dimension.


In a recent work \cite{Dong2014obc} we have proposed a set of effective
two-phase outflow (and also inflow) boundary conditions
within the phase field framework. 
A salient characteristic of
these boundary conditions is that
they ensure the energy stability of the two-phase system.
By looking into the two-phase energy balance relation,
we have shown, at the continuum level, that
with these boundary conditions  the total
energy of the two-phase system will not increase 
over time,
even in situations where there are strong vortices or backflows,
large viscosity contrast, and large density
contrast at the outflow/open boundaries.
In \cite{Dong2014obc} 
we have further developed an algorithm for numerically treating
these open boundary conditions based on
a velocity-correction type splitting strategy.


Inspired by the two-phase energy balance relation from
\cite{Dong2014obc}, we propose in the current work
several new forms of outflow/open boundary conditions 
for the two-phase momentum equations within
the phase field framework.
We also discuss a generalized form of the two-phase
open boundary conditions, which includes these new forms 
and those forms proposed in \cite{Dong2014obc}
as particular cases.
In addition, we present an
algorithm for numerically treating
these new open boundary conditions.
Different than that of \cite{Dong2014obc},
the current algorithm 
is based on a rotational pressure-correction
type strategy for solving the two-phase momentum
equations. 
This algorithm shares one common property
with those of \cite{Dong2014obc,DongS2012},
namely, after discretization it gives rise to
linear algebraic systems for the pressure and
the velocity that involve
only {\em constant}
and {\em time-independent} coefficient matrices,
despite the variable density and variable
viscosity of the two-phase mixture.
Therefore, these coefficient matrices can be
pre-computed during pre-processing.
This makes the current algorithm computationally 
very efficient and attractive.  


It is  commonly observed that,
with traditional splitting type schemes,
 the variable density in
the Navier-Stokes equation has resulted in
a variable (time-dependent) coefficient
matrix for the pressure linear algebraic system
after discretization 
\cite{BellM1992,PuckettABMR1997,GuermondQ2000,
DingSS2007,LiuW2007,PyoS2007,CalgaroCG2008,LiMGS2013}.
This creates a severe computational and performance issue,
due to the need for the frequent re-computation
of the coefficient matrix and
the challenge in efficiently solving the resultant linear
algebraic system at large density ratios.
%
Guermond \& Salgado \cite{GuermondS2009} have advocated a penalty point
of view toward the projection idea,
leading to a Poisson type equation for the pressure;
see also \cite{GuermondS2011,ShenY2010}.
%
Dong \& Shen \cite{DongS2012} have proposed a different
strategy for coping with the variable density.
By a reformulation of the pressure term in
the variable-density Navier-Stokes equation, 
they have developed a 
scheme which requires the solution of 
a pressure Poisson equation with
constant (time-independent) coefficient matrix;
see also \cite{Dong2012,Dong2014nphase,Dong2014obc,DoddF2014,Dong2015}.
%


The novelties of this paper lie in two aspects:
(i) the several new forms of outflow/open 
boundary conditions for the two-phase momentum equations,
and (ii) the pressure-correction based algorithm
for numerically treating these two-phase
open boundary conditions.
On the other hand,
we would like to point out that
the  method for solving the phase field equation
employed in the current paper is not new. 
It was originally developed in 
\cite{DongS2012}.


The numerical algorithm presented herein has been implemented
using $C^0$ continuous high-order spectral elements 
\cite{SherwinK1995,KarniadakisS2005,ZhengD2011} for 
spatial discretizations 
in the current paper.
It should however be noted that the algorithm is general and
can also be implemented with other spatial
discretization techniques.


\section{Pressure Correction Scheme for Two-Phase Outflows}
\label{sec:method}

In this section we present several
 new open boundary conditions for the two-phase
momentum equations, together with an algorithm
for numerically treating these  open boundary conditions based on a rotational
pressure correction-type strategy.

\subsection{Governing Equations and Open Boundary Conditions}
\label{sec:formulation}

Let $\Omega$ denote a flow domain in
two or three dimensions (2-D or 3-D), and
$\partial\Omega$ denote the boundary of $\Omega$.
Consider the mixture of two immiscible incompressible
fluids contained in $\Omega$.
We use $\rho_1$ and $\rho_2$ to denote
the constant densities of the two individual
fluids, and use $\mu_1$ and $\mu_2$
to denote their constant dynamic viscosities.
With the phase field approach,
the motion of this mixture can be
described by the following system of
equations \cite{LiuS2003,YueFLS2004,DongS2012},
\begin{subequations}
\begin{equation}
\rho\left(
  \frac{\partial\mathbf{u}}{\partial t} + \mathbf{u}\cdot\nabla\mathbf{u}
\right) 
= -\nabla p + \nabla\cdot\left[\mu \mathbf{D}(\mathbf{u})  \right]
- \lambda\nabla\cdot(\nabla\phi\otimes\nabla\phi)
+ \mathbf{f}
\label{equ:nse}
\end{equation}
\begin{equation}
\nabla\cdot \mathbf{u} = 0
\label{equ:continuity}
\end{equation}
\begin{equation}
\frac{\partial\phi}{\partial t} + \mathbf{u}\cdot\nabla\phi
= -\lambda\gamma_1\nabla^2\left[
  \nabla^2\phi - h(\phi)
\right] + g(\mathbf{x},t)
\label{equ:CH}
\end{equation}
\end{subequations}
where $\mathbf{x}$ and $t$ are respectively the spatial
coordinates and time,
$\mathbf{u}(\mathbf{x},t)$ is the velocity,
$p(\mathbf{x},t)$ is pressure,
$\mathbf{D}(\mathbf{u}) = \nabla\mathbf{u}+\nabla\mathbf{u}^T$
(the superscript $T$ denotes transpose),
$\mathbf{f}(\mathbf{x},t)$ denotes some
external body force, and 
$\otimes$ represents the tensor product.
%
$\phi(\mathbf{x},t)$ is the phase field function,
$-1\leqslant \phi\leqslant 1$.
Regions with $\phi=1$ denote the first fluid,
and the regions with $\phi=-1$ denote the second fluid.
The function $h(\phi)$ is given by
$
h(\phi) = \frac{1}{\eta^2}\phi(\phi^2-1),
$
where $\eta$ is the characteristic scale of
the interfacial thickness.
$\lambda$ is referred to as the mixing energy
density coefficient, and is given by \cite{YueFLS2004}
\begin{equation}
\lambda = \frac{3}{2\sqrt{2}}\sigma\eta
\end{equation}
where $\sigma$ is the surface tension 
and  assumed to be constant in the current 
paper.
The constant $\gamma_1>0$ is the mobility coefficient
associated with the interface.
%
$\rho(\phi)$ and $\mu(\phi)$ are respectively the 
density and dynamic viscosity of the mixture, given by
\begin{equation}
\rho(\phi) = \frac{1}{2}(\rho_1+\rho_2) 
   + \frac{1}{2}(\rho_1-\rho_2) \phi, \qquad
\mu(\phi) = \frac{1}{2}(\mu_1+\mu_2)
   + \frac{1}{2}(\mu_1-\mu_2) \phi.
\label{equ:rho_mu}
\end{equation}
%
The function $g(\mathbf{x},t)$ in \eqref{equ:CH}
is a prescribed source term for the purpose of
numerical testing only, and will be set to
$g(\mathbf{x},t)=0$ in actual simulations.
Equation \eqref{equ:CH} with $g=0$ is
the Cahn-Hilliard equation.


We assume that 
the domain boundary consists of three types 
which are non-overlapping with one another:
\begin{equation}
\partial\Omega = \partial\Omega_{i} \cup \partial\Omega_w
  \cup \partial\Omega_o.
\end{equation}
We refer to $\partial\Omega_i$ as the inflow boundary,
$\partial\Omega_w$ as the wall boundary,
and $\partial\Omega_o$ as the outflow or
open boundary.
On the inflow and the wall boundaries,
the velocity $\mathbf{u}$ is assumed to be known.
In addition, the phase field function $\phi$
is also assumed to be known on the inflow boundary.
On the wall boundary we assume that 
the wettability property (i.e. contact angle)
is known. 
On the other hand, 
at the outflow/open boundary $\partial\Omega_o$
none of the flow variables (velocity $\mathbf{u}$,
pressure $p$, phase field function $\phi$)
is known.

Inspired by the two-phase energy balance 
discussed in \cite{Dong2014obc},
we propose in the following several new boundary conditions
for the open boundary $\partial\Omega_o$:
\begin{subequations}
\begin{equation}
-p\mathbf{n} + \mu\mathbf{n}\cdot\mathbf{D}(\mathbf{u})
- \left[\frac{\lambda}{2}\nabla\phi\cdot\nabla\phi + F(\phi) \right]\mathbf{n}
- \frac{1}{4}\rho\left[
  |\mathbf{u}|^2\mathbf{n} + (\mathbf{n}\cdot\mathbf{u})\mathbf{u}
\right]\Theta_0(\mathbf{n},\mathbf{u})
= \mathbf{f}_b(\mathbf{x},t),
\ \text{on} \ \partial\Omega_o;
\label{equ:obc_v_C}
\end{equation}
\begin{equation}
-p\mathbf{n} + \mu\mathbf{n}\cdot\mathbf{D}(\mathbf{u})
- \left[\frac{\lambda}{2}\nabla\phi\cdot\nabla\phi + F(\phi) \right]\mathbf{n}
- \frac{1}{2}\rho\left[
  |\mathbf{u}|^2\mathbf{n} + (\mathbf{n}\cdot\mathbf{u})\mathbf{u}
\right]\Theta_0(\mathbf{n},\mathbf{u})
= \mathbf{f}_b(\mathbf{x},t),
\ \text{on} \ \partial\Omega_o;
\label{equ:obc_v_D}
\end{equation}
\begin{equation}
-p\mathbf{n} + \mu\mathbf{n}\cdot\mathbf{D}(\mathbf{u})
- \left[\frac{\lambda}{2}\nabla\phi\cdot\nabla\phi + F(\phi) \right]\mathbf{n}
- \left[\rho
  |\mathbf{u}|^2\mathbf{n} \right]
  \Theta_0(\mathbf{n},\mathbf{u})
= \mathbf{f}_b(\mathbf{x},t), 
\ \text{on} \ \partial\Omega_o;
\label{equ:obc_v_A}
\end{equation}
\begin{equation}
-p\mathbf{n} + \mu\mathbf{n}\cdot\mathbf{D}(\mathbf{u})
- \left[\frac{\lambda}{2}\nabla\phi\cdot\nabla\phi + F(\phi) \right]\mathbf{n}
- \left[\rho
  (\mathbf{n}\cdot\mathbf{u})\mathbf{u} \right]
  \Theta_0(\mathbf{n},\mathbf{u})
= \mathbf{f}_b(\mathbf{x},t),
\ \text{on} \ \partial\Omega_o.
\label{equ:obc_v_B}
\end{equation}
\end{subequations}
%
%
In the above equations \eqref{equ:obc_v_C}--\eqref{equ:obc_v_B},
$\mathbf{n}$ is the outward-pointing unit vector
normal to $\partial\Omega_o$,
$|\mathbf{u}|$ denotes the magnitude of $\mathbf{u}$,
$\mu$ and $\rho$ are respectively
the mixture dynamic viscosity and density given
by \eqref{equ:rho_mu}.
The function $F(\phi)$ is given by
$
F(\phi) = \frac{\lambda}{4\eta^2}(1-\phi^2)^2,
$
and note that
$
\left[
\frac{\lambda}{2}\nabla\phi\cdot\nabla\phi +  F(\phi)
\right]
$
is the free energy density of the 
two-phase system \cite{LiuS2003,YueFLS2004}.
%
$\mathbf{f}_b$ is a prescribed function on $\partial\Omega_o$
for the purpose of numerical testing only,
and will be set to $\mathbf{f}_b=0$ in actual
simulations.
%
$\Theta_0(\mathbf{n},\mathbf{u})$ is a smoothed
step function whose form is given by \cite{DongKC2014,Dong2014obc}
\begin{equation}
\Theta_0(\mathbf{n},\mathbf{u}) = \frac{1}{2}\left(
  1 - \tanh \frac{\mathbf{n}\cdot\mathbf{u}}{U_0\delta}
\right),
\label{equ:theta0_expr}
\end{equation}
where $U_0$ is the characteristic velocity scale,
and $\delta >0$ is a constant that is sufficiently small.
$\delta$ controls the sharpness of the smoothed
step function, and 
$\Theta_0$ approaches the step function as $\delta \rightarrow 0$.
When $\delta$ is sufficiently small, 
$\Theta_0(\mathbf{n},\mathbf{u})$ essentially
assumes the unit value where $\mathbf{n}\cdot\mathbf{u}<0$
and vanishes otherwise.

The boundary conditions \eqref{equ:obc_v_C}--\eqref{equ:obc_v_B}
belong to the following family of 
 boundary conditions for $\partial\Omega_o$,
\begin{multline}
-p\mathbf{n} + \mu\mathbf{n}\cdot\mathbf{D}(\mathbf{u})
- \left[\frac{\lambda}{2}\nabla\phi\cdot\nabla\phi 
     + F(\phi) \right]\mathbf{n} \\
- \left[ \left(1-\theta+\beta_1\right) \frac{1}{2}\rho
   (\mathbf{n}\cdot\mathbf{u})\mathbf{u}
  + \left(\theta + \beta_2  \right) \frac{1}{2}\rho
   |\mathbf{u}|^2\mathbf{n}
\right]\Theta_0(\mathbf{n},\mathbf{u})
= \mathbf{f}_b(\mathbf{x},t),
\ \text{on} \ \partial\Omega_o,
\label{equ:obc_v_gobc}
\end{multline}
where $0\leqslant \theta \leqslant 1$,
$\beta_1\geqslant 0$ and $\beta_2\geqslant 0$
are constant parameters. One can verify that 
the general form of open boundary condition \eqref{equ:obc_v_gobc},
as $\delta \rightarrow 0$ and assuming $\mathbf{f}_b=0$,
is conducive to the stability of the two-phase 
energy balance equation
given in \cite{Dong2014obc}.
The boundary conditions \eqref{equ:obc_v_C}--\eqref{equ:obc_v_B}
are particular cases of \eqref{equ:obc_v_gobc}.
For example, \eqref{equ:obc_v_C} corresponds to 
$(\theta,\beta_1,\beta_2)=(1/2,0,0)$ and
\eqref{equ:obc_v_B} corresponds to 
$(\theta,\beta_1,\beta_2)=(0,1,0)$ in \eqref{equ:obc_v_gobc}.
This general form also contains the following
two open boundary conditions as particular
cases, which are 
proposed in \cite{Dong2014obc} and can be obtained by 
 respectively setting
$(\theta,\beta_1,\beta_2)=(1,0,0)$ and
$(\theta,\beta_1,\beta_2)=(0,0,0)$ in \eqref{equ:obc_v_gobc},
\begin{subequations}
\begin{equation}
-p\mathbf{n} + \mu\mathbf{n}\cdot\mathbf{D}(\mathbf{u})
- \left[\frac{\lambda}{2}\nabla\phi\cdot\nabla\phi + F(\phi) \right]\mathbf{n}
- \left[ \frac{1}{2}\rho
  |\mathbf{u}|^2\mathbf{n} 
\right]\Theta_0(\mathbf{n},\mathbf{u})
= \mathbf{f}_b(\mathbf{x},t),
\ \text{on} \ \partial\Omega_o;
\label{equ:obc_v_E}
\end{equation}
\begin{equation}
-p\mathbf{n} + \mu\mathbf{n}\cdot\mathbf{D}(\mathbf{u})
- \left[\frac{\lambda}{2}\nabla\phi\cdot\nabla\phi + F(\phi) \right]\mathbf{n}
- \left[\frac{1}{2}\rho
   (\mathbf{n}\cdot\mathbf{u})\mathbf{u}
\right]\Theta_0(\mathbf{n},\mathbf{u})
= \mathbf{f}_b(\mathbf{x},t),
\ \text{on} \ \partial\Omega_o.
\label{equ:obc_v_F}
\end{equation}
\end{subequations}

%

The above boundary conditions on $\partial\Omega_o$ are for
the momentum equations \eqref{equ:nse}--\eqref{equ:continuity}.
In addition to them,
one also needs to supply appropriate boundary 
conditions on $\partial\Omega_o$ 
for the phase field equation \eqref{equ:CH}.
Note that two independent conditions will be 
needed on each boundary, due to the fourth spatial
order of equation \eqref{equ:CH}.
For the phase field function $\phi$, on
the outflow boundary $\partial\Omega_o$
we will employ the boundary conditions
developed in \cite{Dong2014obc}
\begin{subequations}
\begin{equation}
\mathbf{n}\cdot\nabla\left[
  \nabla^2\phi - h(\phi)
\right] = g_{a1}(\mathbf{x},t), 
\quad \text{on} \ \partial\Omega_o
\label{equ:obc_phi_1}
\end{equation}
\begin{equation}
\mathbf{n}\cdot\nabla\phi = -D_0\frac{\partial\phi}{\partial t} 
   + g_{a2}(\mathbf{x},t), 
\quad \text{on} \ \partial\Omega_o,
\label{equ:obc_phi_2}
\end{equation}
\end{subequations}
where $g_{a1}$ and $g_{a2}$ are prescribed source terms
on $\partial\Omega_o$ for the purpose of numerical testing
only, and will be set to $g_{a1}=0$ and $g_{a2}=0$
in actual simulations.
The constant $D_0\geqslant 0$ is
a chosen non-negative constant, and $\frac{1}{D_0}$
plays the role of a convection velocity
at the outflow boundary $\partial\Omega_o$.

The boundary conditions for the other types of boundaries
(wall and inflow)
will be set in accordance with previous works \cite{Dong2012,Dong2014obc}.
We impose a Dirichlet condition for the velocity
on the inflow and wall boundaries,
\begin{equation}
\mathbf{u} = \mathbf{w}(\mathbf{x},t),
\quad \text{on} \ \partial\Omega_i \cup \partial\Omega_w,
\label{equ:dbc_v}
\end{equation} 
where $\mathbf{w}$ is the boundary velocity.
For the phase field function, we
impose the following condition from \cite{Dong2014obc}
on the inflow boundary,
\begin{subequations}
\begin{equation}
\phi = \phi_b(\mathbf{x},t), \quad \text{on} \ \partial\Omega_i,
\label{equ:ibc_phi_1}
\end{equation}
\begin{equation}
\nabla^2\phi - h(\phi) = g_b(\mathbf{x},t),
\quad \text{on} \ \partial\Omega_i,
\label{equ:ibc_phi_2}
\end{equation}
\end{subequations}
where $\phi_b$ denotes the distribution of the 
phase field function on the inflow boundary,
and $g_b$ is a prescribed source term for 
numerical testing only and will be
set to $g_b=0$ in actual simulations.
On the wall boundary we employ 
the contact-angle condition of
\cite{Dong2012}, considering only
the effect of the static contact angle,
\begin{subequations}
\begin{equation}
\mathbf{n}\cdot\nabla\left[
  \nabla^2\phi - h(\phi)
\right] = g_{c1}(\mathbf{x},t), 
\quad \text{on} \ \partial\Omega_w,
\label{equ:wbc_phi_1}
\end{equation}
\begin{equation}
\mathbf{n}\cdot\nabla\phi = \frac{3\sigma}{4\lambda}\cos\theta_s
   (1-\phi^2)
   + g_{c2}(\mathbf{x},t), 
\quad \text{on} \ \partial\Omega_w,
\label{equ:wbc_phi_2}
\end{equation}
\end{subequations}
where $\theta_s$ is the static (equilibrium) contact
angle formed between the fluid interface and
the wall measured on the side of the first fluid,
$g_{c1}$ and $g_{c2}$ are two prescribed
source terms for the purpose of the numerical testing only
and will be set to $g_{c1}=0$ and $g_{c2}=0$
in actual simulations.

Finally, we assume that the following initial conditions for the
velocity and the phase field function are known
\begin{equation}
\mathbf{u}(\mathbf{x},0) = \mathbf{u}_{in}(\mathbf{x}), \quad
\phi(\mathbf{x},0) = \phi_{in}(\mathbf{x}),
\label{equ:ic}
\end{equation}
where the initial velocity $\mathbf{u}_{in}$ and
the initial phase field function $\phi_{in}$
should be compatible with the above boundary conditions
and the governing equations.


\subsection{Two-Phase Momentum Equations: Algorithm and Implementation}
\label{sec:alg_momentum}

The system of equations \eqref{equ:nse}--\eqref{equ:CH},
the boundary conditions \eqref{equ:obc_phi_1}--\eqref{equ:wbc_phi_2},
and one of the conditions among \eqref{equ:obc_v_C}--\eqref{equ:obc_v_B},
together with the initial conditions \eqref{equ:ic} for 
the velocity and the phase field function,
constitute the overall system that need to be
solved in numerical simulations.
We next consider the numerical 
solution of this system.

Because the phase field equation \eqref{equ:CH}
is coupled to the momentum equations 
\eqref{equ:nse}--\eqref{equ:continuity} only
through the convection term, it is possible
and convenient to treat the momentum equations
and the phase field equation individually.
Indeed, by treating 
the convection term in \eqref{equ:CH} explicitly,
one can de-couple the computation for
the phase field function from those for 
the momentum equations.
On can first solve \eqref{equ:CH}
for the phase field function, and then solve
the momentum equations for the pressure and the velocity.

In the following 
we will first concentrate
on the momentum equations \eqref{equ:nse}--\eqref{equ:continuity},
together with the associated boundary
conditions \eqref{equ:obc_v_C}--\eqref{equ:obc_v_B} for $\partial\Omega_o$
and \eqref{equ:dbc_v} for $\partial\Omega_i$ and $\partial\Omega_w$.
We defer the discussion of the solution
to the phase field equation to an Appendix
(see the subsequent Section \ref{sec:overall_method} for detail).
In subsequent discussions of this section 
we assume that 
the variables $\phi$ and $\nabla^2\phi$ have been
computed in appropriate ways and are
already available.


To facilitate the following discussions we introduce an
auxiliary pressure 
\begin{equation}
P = p + \frac{\lambda}{2}\nabla\phi\cdot\nabla\phi
\end{equation}
which will also be loosely called pressure where
no confusion arises.
Then equation \eqref{equ:nse} can be transformed into
\begin{equation}
\frac{\partial\mathbf{u}}{\partial t}
+ \mathbf{u}\cdot\nabla\mathbf{u} = 
-\frac{1}{\rho}\nabla P 
+ \frac{1}{\rho}\nabla\mu\cdot\mathbf{D}(\mathbf{u})
+ \frac{\mu}{\rho}\nabla^2\mathbf{u}
-\frac{\lambda}{\rho}\nabla^2\phi\nabla\phi
+ \frac{1}{\rho}\mathbf{f}.
\label{equ:nse_trans_1} 
\end{equation}
We further re-write the open boundary conditions
\eqref{equ:obc_v_C}--\eqref{equ:obc_v_B} or 
\eqref{equ:obc_v_E}--\eqref{equ:obc_v_F}
into
a unified compact form
\begin{equation}
-P\mathbf{n} + \mu\mathbf{n}\cdot\mathbf{D}(\mathbf{u})
- F(\phi)\mathbf{n} - \mathbf{E}(\rho,\mathbf{n},\mathbf{u})
= \mathbf{f}_b,
\quad \text{on} \ \partial\Omega_o,
\label{equ:obc_v_trans}
\end{equation}
where 
\begin{equation}
\mathbf{E}(\rho,\mathbf{n},\mathbf{u}) = \left\{
\begin{array}{ll}
\frac{1}{4}\rho [
  |\mathbf{u}|^2\mathbf{n} + (\mathbf{n}\cdot\mathbf{u})\mathbf{u}
] \Theta_0(\mathbf{n},u), &
 \text{for boundary condition} \ \eqref{equ:obc_v_C}; \\
\frac{1}{2}\rho[
  |\mathbf{u}|^2\mathbf{n} + (\mathbf{n}\cdot\mathbf{u})\mathbf{u}
] \Theta_0(\mathbf{n},u), &
 \text{for boundary condition} \ \eqref{equ:obc_v_D}; \\
\rho|\mathbf{u}|^2\mathbf{n}\Theta_0(\mathbf{n},u), &
\text{for boundary condition} \ \eqref{equ:obc_v_A}; \\
\rho(\mathbf{n}\cdot\mathbf{u})\mathbf{u}\Theta_0(\mathbf{n},u), &
 \text{for boundary condition} \ \eqref{equ:obc_v_B}; \\
\frac{1}{2}\rho|\mathbf{u}|^2\mathbf{n}\Theta_0(\mathbf{n},u), &
\text{for boundary condition} \ \eqref{equ:obc_v_E}; \\
\frac{1}{2}\rho(\mathbf{n}\cdot\mathbf{u})\mathbf{u}\Theta_0(\mathbf{n},u), &
 \text{for boundary condition} \ \eqref{equ:obc_v_F}. \\
\end{array}
\right.
\label{equ:obc_v_E_expr}
\end{equation}

The following algorithm is for
the equations \eqref{equ:nse_trans_1} and \eqref{equ:continuity},
together with the boundary conditions
\eqref{equ:dbc_v} on $\partial\Omega_i\cup \partial\Omega_w$
and \eqref{equ:obc_v_trans} on $\partial\Omega_o$.
Note that the variables
$\phi$ and $\nabla^2\phi$ are assumed to be known
here, as discussed before.

Let $n$ denote the time step index,
and $(\cdot)^n$ denote the variable $(\cdot)$
at time step $n$.
We use $\tilde{\mathbf{u}}^n$ and $\mathbf{u}^n$
to denote two slightly different approximations of 
the velocity at time step $n$.
Define 
\begin{equation}
\tilde{\mathbf{u}}^0 = \mathbf{u}_{in}, \quad
\mathbf{u}^0 = \mathbf{u}_{in}, \quad
\phi^0 = \phi_{in}.
\end{equation}
%
By enforcing equation \eqref{equ:nse_trans_1}
at time step zero, one can compute
the initial pressure $P^0$ as follows. 
Let 
\begin{equation}
H_{p0}^1(\Omega) = \{ \
  v \in H^1(\Omega) \ : \ v|_{\partial\Omega_o}=0 
\ \},
\label{equ:H1_p0}
\end{equation}
and $q\in H^1_{p0}(\Omega)$ denote the test function.
By taking the inner product between $\nabla q$ and
equation \eqref{equ:nse_trans_1} and integrating by part,
one obtains the weak form about $P^0$,
\begin{multline}
\int_{\Omega} \frac{1}{\rho^0}\nabla P^0\cdot\nabla q 
= \int_{\Omega}\left[
\frac{1}{\rho^0}\mathbf{f}^0
- \tilde{\mathbf{u}}^0\cdot\nabla\tilde{\mathbf{u}}^0
+ \frac{1}{\rho^0}\nabla\mu^0\cdot\mathbf{D}(\tilde{\mathbf{u}}^0)
- \frac{\lambda}{\rho^0}\Psi\nabla\phi^0
+ \nabla\left(\frac{\mu^0}{\rho^0}  \right) \times \tilde{\bm{\omega}}^0
\right] \cdot \nabla q \\
- \int_{\partial\Omega_i\cup\partial\Omega_w\cup\partial\Omega_o}
   \frac{\mu^0}{\rho^0}\mathbf{n}\times\tilde{\bm{\omega}}^0\cdot\nabla q
- \int_{\partial\Omega_i\cup\partial\Omega_w}
   \mathbf{n}\cdot\left.\frac{\partial\mathbf{w}}{\partial t}\right|^0 q,
\quad \forall q\in H_{p0}^1(\Omega),
\label{equ:init_p}
\end{multline}
where 
\begin{equation}
\rho^0 = \rho(\phi^0), \quad
\mu^0 = \mu(\phi^0), \quad
\tilde{\bm{\omega}}^0 = \nabla\times\tilde{\mathbf{u}}^0. 
\label{equ:rho_mu_step_0}
\end{equation}
$\left.\frac{\partial\mathbf{w}}{\partial t}\right|^0$
is the time derivative at time step zero, which can be
numerically computed based on the 
second-order backward differential
formula (BDF2) because the boundary 
velocity $\mathbf{w}(\mathbf{x},t)$ is 
known on $\partial\Omega_i\cup\partial\Omega_w$.
$\Psi$ represents the projection of
$\nabla^2\phi^0$ into the $H^1(\Omega)$ space, and is given
by the following weak form ($\varphi$ denoting the test function),
\begin{equation}
\int_{\Omega} \Psi\varphi 
= -\int_{\Omega} \nabla\phi^0\cdot\nabla\varphi
+ \int_{\partial\Omega_i\cup\partial\Omega_w\cup\partial\Omega_o}
  (\mathbf{n}\cdot\nabla\phi^0)\varphi,
\quad \forall \varphi \in H^1(\Omega).
\label{equ:laplace_phi}
\end{equation}
The weak forms \eqref{equ:init_p}
and \eqref{equ:laplace_phi} can be discretized in space
using $C^0$
spectral elements (or finite elements). 
We solve equation \eqref{equ:init_p}, 
together with the Dirichlet condition
\begin{equation}
P^0 = \mu^0\mathbf{n}\cdot\mathbf{D}(\tilde{\mathbf{u}}^0)\cdot\mathbf{n}
 - F(\phi^0) 
- \mathbf{n}\cdot \mathbf{E}(\rho^0, \mathbf{n}, \tilde{\mathbf{u}}^0)
- \mathbf{f}_b^0\cdot\mathbf{n},
\quad \text{on} \ \partial\Omega_o,
\end{equation} 
to obtain the initial pressure $P^0$,
where $\Psi$ is obtained by 
solving equation \eqref{equ:laplace_phi}.

Given $(\tilde{\mathbf{u}}^n,\mathbf{u}^n,P^n,
\phi^{n+1},\nabla^2\phi^{n+1})$,
where $\phi^{n+1}$ and $\nabla^2\phi^{n+1}$
are assumed known and result from the algorithm for
the phase field equation to be discussed later,
we compute $\tilde{\mathbf{u}}^{n+1}$, $\mathbf{u}^{n+1}$ and $P^{n+1}$, 
together with an auxiliary field variable $\xi^{n+1}$,
successively in a de-coupled fashion 
as follows: \\[0.1in]
\underline{For $\tilde{\mathbf{u}}^{n+1}$:}
\begin{subequations}
\begin{equation}
\begin{split}
\frac{\gamma_0\tilde{\mathbf{u}}^{n+1}-\hat{\mathbf{u}}}{\Delta t}
& + \tilde{\mathbf{u}}^{*,n+1}\cdot\nabla\tilde{\mathbf{u}}^{*,n+1}
+ \frac{1}{\rho_m}\nabla P^n 
- \nu_m\nabla^2\tilde{\mathbf{u}}^{n+1}
= 
\left( \frac{1}{\rho_m} - \frac{1}{\rho^{n+1}} \right) \nabla P^{*,n+1} \\
& 
+ \frac{1}{\rho^{n+1}}\nabla\mu^{n+1}\cdot\mathbf{D}(\tilde{\mathbf{u}}^{*,n+1})
- \left( \frac{\mu^{n+1}}{\rho^{n+1}}-\nu_m \right)
   \nabla\times\nabla\times \tilde{\mathbf{u}}^{*,n+1} \\
&
- \frac{\lambda}{\rho^{n+1}}\nabla^2\phi^{n+1}\nabla\phi^{n+1}
+ \frac{1}{\rho^{n+1}}\mathbf{f}^{n+1}
\end{split}
\label{equ:velocity_1}
\end{equation}
\begin{equation}
\tilde{\mathbf{u}}^{n+1} = \mathbf{w}^{n+1},
\quad \text{on} \ \partial\Omega_i\cup\partial\Omega_w
\label{equ:velocity_2}
\end{equation}
\begin{multline}
\mathbf{n}\cdot\mathbf{D}(\tilde{\mathbf{u}}^{n+1})
=
\left(1 - \frac{\mu^{n+1}}{\mu_0}  \right)
        \mathbf{n}\cdot\mathbf{D}(\tilde{\mathbf{u}}^{*,n+1}) \\
+ \frac{1}{\mu_0}\left[
  P^{*,n+1}\mathbf{n} + F(\phi^{n+1})\mathbf{n}
  + \mathbf{E}(\rho^{n+1},\mathbf{n},\tilde{\mathbf{u}}^{*,n+1})
  + \mathbf{f}_b^{n+1}
\right],
\quad \text{on} \ \partial\Omega_o
\label{equ:velocity_3}
\end{multline}
\begin{equation}
\mathbf{n}\cdot\nabla\tilde{\mathbf{u}}^{n+1}
= \mathbf{n}\cdot\mathbf{D}(\tilde{\mathbf{u}}^{n+1})
- \mathbf{n}\cdot(\nabla\tilde{\mathbf{u}}^{*,n+1})^T,
\quad \text{on} \ \partial\Omega_o.
\label{equ:velocity_4}
\end{equation}
\end{subequations}
\underline{For $\xi^{n+1}$:}
\begin{subequations}
\begin{equation}
\frac{\gamma_0}{\Delta t}\xi^{n+1} - \nu_m\nabla^2\xi^{n+1}
= 
\nabla\cdot \mathbf{G}^{n+1}
+ \nabla\left(\frac{\mu^{n+1}}{\rho^{n+1}}  \right)
  \cdot\nabla\times\nabla\times\tilde{\mathbf{u}}^{n+1}
\label{equ:xi_1}
\end{equation}
\begin{equation}
\mathbf{n}\cdot\nabla\xi^{n+1}
=
\frac{1}{\nu_m}\mathbf{n}\cdot
   \frac{\gamma_0\mathbf{w}^{n+1}-\hat{\mathbf{w}}}{\Delta t}
- \frac{1}{\nu_m}\mathbf{n}\cdot\mathbf{G}^{n+1}
+ \frac{1}{\nu_m}\frac{\mu^{n+1}}{\rho^{n+1}}
   \mathbf{n}\cdot\nabla\times\nabla\times\tilde{\mathbf{u}}^{n+1},
\quad \text{on} \ \partial\Omega_i\cup\partial\Omega_w
\label{equ:xi_2}
\end{equation}
\begin{equation}
\xi^{n+1} = \nabla\cdot\tilde{\mathbf{u}}^{n+1},
\quad \text{on} \ \partial\Omega_o.
\label{equ:xi_3}
\end{equation}
\end{subequations}
\underline{For $P^{n+1}$:}
\begin{subequations}
\begin{equation}
\frac{\gamma_0\mathbf{u}^{n+1}-\gamma_0\tilde{\mathbf{u}}^{n+1}}{\Delta t}
+ \frac{1}{\rho_m} \nabla(P^{n+1} - P^n + \rho_m\nu_m\xi^{n+1}) = 0
\label{equ:pressure_1}
\end{equation}
\begin{equation}
\nabla\cdot\mathbf{u}^{n+1} = 0
\label{equ:pressure_2}
\end{equation}
\begin{equation}
\mathbf{n}\cdot\mathbf{u}^{n+1} = \mathbf{n}\cdot\mathbf{w}^{n+1},
\quad \text{on} \ \partial\Omega_i\cup\partial\Omega_w
\label{equ:pressure_3}
\end{equation}
\begin{equation}
P^{n+1} = \mu^{n+1}\mathbf{n}\cdot
  \mathbf{D}(\tilde{\mathbf{u}}^{n+1})\cdot\mathbf{n}
- F(\phi^{n+1})
- \mathbf{n}\cdot\mathbf{E}(\rho^{n+1},\mathbf{n},\tilde{\mathbf{u}}^{n+1})
- \mu_{\min}\nabla\cdot\tilde{\mathbf{u}}^{n+1},
\quad \text{on} \ \partial\Omega_o.
\label{equ:pressure_4}
\end{equation}
\end{subequations}

The notation employed
in the equations \eqref{equ:velocity_1}--\eqref{equ:pressure_4}
is as follows.
Let $J$ ($J=1$ or $2$) denote the temporal order
of the scheme, and $\chi$ denote a generic
variable.
Then in the above equations,
$\chi^{*,n+1}$ is a $J$-th order explicit
approximation of $\chi^{n+1}$, given by
\begin{equation}
\chi^{*,n+1} = \left\{
\begin{array}{ll}
\chi^n, & J=1 \\
2\chi^n - \chi^{n-1}, & J=2.
\end{array}
\right.
\label{equ:var_star_def}
\end{equation}
The expression 
$
\frac{1}{\Delta t}(\gamma_0\chi^{n+1}-\hat{\chi})
$
denotes an approximation of 
$
\left.\frac{\partial\chi}{\partial t}  \right|^{n+1}
$
by the $J$-th order backward differentiation formula,
where $\Delta t$ is the time step size and
\begin{equation}
\hat{\chi} = \left\{
\begin{array}{ll}
\chi^n, & J=1 \\
2\chi^n - \frac{1}{2}\chi^{n-1}, & J=2,
\end{array}
\right.
\qquad
\gamma_0 = \left\{
\begin{array}{ll}
1, & J=1 \\
\frac{3}{2}, & J=2.
\end{array}
\right.
\label{equ:var_hat_def}
\end{equation}
%
In equations \eqref{equ:xi_1} and \eqref{equ:xi_2}
$\mathbf{G}^{n+1}$ is given by
\begin{multline}
\mathbf{G}^{n+1} = 
\frac{1}{\rho^{n+1}}\mathbf{f}^{n+1}
-\tilde{\mathbf{u}}^{*,n+1}\cdot\nabla\tilde{\mathbf{u}}^{*,n+1}
-\frac{1}{\rho_m}\nabla P^n
+\left(\frac{1}{\rho_m} - \frac{1}{\rho^{n+1}}  \right)\nabla P^{*,n+1} \\
+\frac{1}{\rho^{n+1}}\nabla\mu^{n+1}\cdot\mathbf{D}(\tilde{\mathbf{u}}^{*,n+1})
-\frac{\lambda}{\rho^{n+1}}\nabla^2\phi^{n+1}\nabla\phi^{n+1}.
\label{equ:G_expr}
\end{multline}
The function $\mathbf{E}(\rho,\mathbf{n},\mathbf{u})$ is defined
by equation \eqref{equ:obc_v_E_expr}.
$\rho^{n+1}$ and $\mu^{n+1}$ are given by
equation \eqref{equ:rho_mu} and by using $\phi^{n+1}$.
In equation \eqref{equ:pressure_4} $\mu_{\min} = \min(\mu_1,\mu_2)$.

In the above equations, $\rho_m$ is a chosen
constant that must satisfy the condition
\begin{equation}
0 < \rho_m \leqslant \min(\rho_1,\rho_2).
\label{equ:rho_m_condition}
\end{equation}
This condition is critical to the stability of the scheme.
The scheme is observed to be unstable if this condition is
violated.
We will employ $\rho_m=\min(\rho_1,\rho_2)$ for
the numerical simulations in Section \ref{sec:tests}.
$\nu_m$ is a chosen constant that is sufficiently large,
and a reasonable condition is
\begin{equation}
\nu_m \geqslant \frac{1}{2}\left( 
  \frac{\mu_1}{\rho_1} + \frac{\mu_2}{\rho_2}
\right).
\end{equation}
$\mu_0$ in equation \eqref{equ:velocity_3}
is a chosen constant that is sufficiently large.
In the presence of open boundaries 
and when $\mu_1 \neq\mu_2$,
the scheme is
observed to be unstable if
$\mu_0 \leqslant \min(\mu_1,\mu_2)$.
We will use $\mu_0\geqslant \max(\mu_1,\mu_2)$
in the numerical simulations in Section \ref{sec:tests}.
It is observed that 
increasing $\nu_m$ 
tends to improve the stability.
Increasing $\mu_0$ also tends to improve 
the stability in the presence of open boundaries.
Note that 
the constant $\mu_0$ here should  not  be confused
with the field variable $\mu^{0}=\mu(\phi^0)$
in equation \eqref{equ:rho_mu_step_0},
which represents the distribution of
the dynamic viscosity at time step zero.

We would like to make several comments on the above
scheme:
\begin{itemize}

\item
The computations for the pressure $P^{n+1}$
and the velocity $\tilde{u}^{n+1}$ are de-coupled in
this algorithm, and
the velocity $\mathbf{u}^{n+1}$ can be
evaluated based on equation \eqref{equ:pressure_1}
once $P^{n+1}$ is computed.
One can recognize that the overall construction of the scheme
 resembles 
a rotational pressure-correction type
strategy \cite{GuermondMS2006}.
However, the formulation here contains features 
that distinguish it
from the usual pressure-correction formulations. 
Most notably, the current scheme
involves a discrete equation and associated boundary
conditions, \eqref{equ:xi_1}--\eqref{equ:xi_3},
about an auxiliary variable $\xi^{n+1}$.
In addition, the pressure $P^{n+1}$ from the current
scheme resides in the $H^1(\Omega)$ space.
In contrast, the pressure from 
the usual pressure-correction formulations
resides in the $L^2(\Omega)$ space (see \cite{GuermondMS2006}).

\item
The variable $\xi^{n+1}$ is an approximation of
the quantity $\nabla\cdot\tilde{\mathbf{u}}^{n+1}$.
The equation \eqref{equ:xi_1} about
$\xi^{n+1}$ exists only in the discrete
sense. It is different than the dynamic
equation about $\nabla\cdot\mathbf{u}$
at the continuum level.

\item
The scheme leads to linear algebraic systems involving only
{\em constant} and
{\em time-independent} coefficient matrices
for the pressure, velocity, and
the variable $\xi^{n+1}$ after discretization.
This is due to the reformulations 
of the pressure term and the viscous term, and
the introduction of 
the  constants $\rho_m$
and $\nu_m$ in the scheme.
The reformulation  of the pressure term
for coping with the variable density
is proposed by \cite{DongS2012}.
The idea for the treatment of the viscous term
for dealing with the variable viscosity
can be traced to the early works in the 1970s
(e.g. \cite{GottliebO1977});
see also later works
in e.g. \cite{BadalassiCB2003,DongS2012}.
Because only constant and time-independent 
coefficient matrices are involved, which
can be pre-computed during pre-processing,
the current scheme is computationally
very attractive and efficient.

\item
In the velocity substep we impose a velocity
Neumann-type condition, 
\eqref{equ:velocity_3}--\eqref{equ:velocity_4},
on the open boundary $\partial\Omega_o$.
The discrete condition \eqref{equ:velocity_3}
originates from the open boundary
condition \eqref{equ:obc_v_trans}.
But it contains constructions
involving the constant $\mu_0$, which are
critical to the stability if open boundaries are present.
In the absence of the $\mu_0$ constructions,
the computation is unstable when 
the viscosity ratio of the two fluids becomes 
large and when the fluid interface passes through
the open boundaries.
The idea of the $\mu_0$ construction for treating
the variable viscosity at the open boundary
is first proposed by \cite{Dong2014obc}.
However, there exists a crucial difference 
in terms of stability between the 
current scheme and that of \cite{Dong2014obc}.
The  algorithm of \cite{Dong2014obc}
is based on a velocity-correction type
strategy, and it is observed that a
 smaller $\mu_0$ constant 
tends to improve the stability of that scheme
in the presence of open boundaries \cite{Dong2014obc}.
In contrast, the current scheme is
based on a pressure-correction type strategy,
and we observe that a larger $\mu_0$
constant tends to improve the stability of
the scheme when open boundaries are present.

\item
In the pressure substep we impose
a pressure Dirichlet condition,
\eqref{equ:pressure_4},
on the open boundary $\partial\Omega_o$.
This discrete condition results
essentially from taking the inner product
between $\mathbf{n}$ and 
the open boundary condition \eqref{equ:obc_v_trans}.
However, note that it contains an
extra term $\mu_{\min}\nabla\cdot\tilde{\mathbf{u}}^{n+1}$
in the construction.


\end{itemize}


We employ $C^0$ continuous spectral 
elements \cite{SherwinK1995,KarniadakisS2005,ZhengD2011}
for spatial discretizations in the current paper.
Let us next consider how to implement
the algorithm, represented 
by \eqref{equ:velocity_1}--\eqref{equ:pressure_4},
using $C^0$ spectral elements.
The formulations presented below
with no change also applies to
$C^0$ finite elements.

The main issue with regard to the
implementation arises from
the terms such as
$
\nabla\times\nabla\times\tilde{\mathbf{u}}^{*,n+1},
$
$
\nabla\times\nabla\times\tilde{\mathbf{u}}^{n+1},
$
and 
$
\nabla\cdot\mathbf{G}^{n+1}
$
involved in the algorithm.
These terms cannot be directly computed
in the discrete space of $C^0$ elements.
Note that the term 
$\nabla^2\phi^{n+1}$ itself may
also cause  difficulty to $C^0$ elements.
However, this term
will be computed in a proper fashion using
$C^0$ elements later when discussing how to solve
the phase field equation.
So here we assume that 
$\nabla^2\phi^{n+1}$ is already available
in a suitable form.

We will derive weak forms of the algorithm
for different flow variables. 
In the process the terms causing difficulty to
$C^0$ elements will be treated
in an appropriate way.

Let $\tilde{\bm{\omega}}=\nabla\times\tilde{\mathbf{u}}$
denote the vorticity.
Equation \eqref{equ:velocity_1} can be
re-written as
\begin{equation}
\frac{\gamma_0}{\nu_m\Delta t}\tilde{\mathbf{u}}^{n+1}
- \nabla^2\tilde{\mathbf{u}}^{n+1}
= \frac{1}{\nu_m}\left[
 \mathbf{G}^{n+1} + \frac{\hat{\mathbf{u}}}{\Delta t}
\right]
-\frac{1}{\nu_m}\left(
  \frac{\mu^{n+1}}{\rho^{n+1}} - \nu_m
\right) 
\nabla \times \tilde{\bm{\omega}}^{*,n+1},
\label{equ:velocity_1_reform_1}
\end{equation}
where $\mathbf{G}^{n+1}$ is given
by \eqref{equ:G_expr}.
Let 
\begin{equation}
H_{u0}(\Omega) = \{\ 
v\in H^1(\Omega) \ : \ 
v|_{\partial\Omega_i\cup\partial\Omega_w} = 0
\ \},
\end{equation}
and $\varphi\in H_{u0}^1(\Omega)$
denote the test function.
Taking the $L^2$ inner product
between $\varphi$ and equation
\eqref{equ:velocity_1_reform_1},
and integrating by part,
we get the weak form about $\tilde{\mathbf{u}}^{n+1}$,
\begin{equation}
\begin{split}
\frac{\gamma_0}{\nu_m\Delta t}
  \int_{\Omega} \varphi\tilde{\mathbf{u}}^{n+1}
&+ \int_{\Omega}\nabla\varphi\cdot\nabla\tilde{\mathbf{u}}^{n+1}
= 
\frac{1}{\nu_m}\int_{\Omega}\left[
 \mathbf{G}^{n+1} + \frac{\hat{\mathbf{u}}}{\Delta t}
 + \nabla\left(\frac{\mu^{n+1}}{\rho^{n+1}}  \right)
    \times \tilde{\bm{\omega}}^{*,n+1}
\right]\varphi \\
&
- \frac{1}{\nu_m}\int_{\Omega}\left(
  \frac{\mu^{n+1}}{\rho^{n+1}} - \nu_m
\right)
  \tilde{\bm{\omega}}^{*,n+1}\times\nabla\varphi
- \frac{1}{\nu_m}\int_{\partial\Omega_o}\left(
  \frac{\mu^{n+1}}{\rho^{n+1}} - \nu_m
\right)
  \mathbf{n}\times\tilde{\bm{\omega}}^{*,n+1}\varphi \\
&
+ \int_{\partial\Omega_o}\left\{
  \frac{1}{\mu_0}\left[
    P^{*,n+1}\mathbf{n}+ F(\phi^{n+1})\mathbf{n}
    + \mathbf{E}(\rho^{n+1},\mathbf{n},\tilde{\mathbf{u}}^{*,n+1})
    + \mathbf{f}_b^{n+1}
  \right]
  \right. \\
& \qquad\qquad
  \left.
  +\left(1-\frac{\mu^{n+1}}{\mu_0}  \right)
       \mathbf{n}\cdot\mathbf{D}(\tilde{\mathbf{u}}^{*,n+1})
  - \mathbf{n}\cdot(\nabla\tilde{\mathbf{u}}^{*,n+1})^T
\right\}  \varphi,
\quad \forall \varphi\in H^1_{u0}(\Omega).
\end{split}
\label{equ:vel_weakform}
\end{equation}
When deriving the above weak form we have
used the equations \eqref{equ:velocity_3} and
\eqref{equ:velocity_4}, and the following identity
($K$ denoting a scalar field function)
\begin{equation}
\int_{\Omega}K(\nabla\times\tilde{\bm{\omega}})\varphi
= \int_{\partial\Omega}K (\mathbf{n}\times\tilde{\bm{\omega}})\varphi
- \int_{\Omega}(\nabla K\times\tilde{\bm{\omega}})\varphi
+ \int_{\Omega}K(\tilde{\bm{\omega}}\times\nabla\varphi).
\end{equation}


Let $\vartheta\in H^1_{p0}(\Omega)$ denote the 
test function, where
$H_{p0}^1(\Omega)$ is defined in \eqref{equ:H1_p0}. 
Taking the $L^2$ inner product between
$\vartheta$ and equation \eqref{equ:xi_1},
and integrating by part, we have
\begin{equation}
\begin{split}
\frac{\gamma_0}{\nu_m\Delta t}
 \int_{\Omega}\xi^{n+1}\vartheta
& + \int_{\Omega}\nabla\xi^{n+1}\cdot\nabla\vartheta
= 
-\frac{1}{\nu_m}\int_{\Omega}\left[
  \mathbf{G}^{n+1}
  + \nabla\left(\frac{\mu^{n+1}}{\rho^{n+1}} \right)\times
    \tilde{\bm{\omega}}^{n+1}
\right] \cdot \nabla\vartheta \\
&
+ \frac{1}{\nu_m}\int_{\partial\Omega_i\cup\partial\Omega_w}
   \mathbf{n}\cdot\frac{\gamma_0\mathbf{w}^{n+1}-\hat{\mathbf{w}}}{\Delta t}\vartheta
+ \frac{1}{\nu_m}\int_{\partial\Omega_i\cup\partial\Omega_w}
   \frac{\mu^{n+1}}{\rho^{n+1}}\mathbf{n}\cdot\nabla\times
     \tilde{\bm{\omega}}^{n+1} \vartheta \\
&
- \frac{1}{\nu_m}\int_{\partial\Omega}
    \nabla\left(\frac{\mu^{n+1}}{\rho^{n+1}}  \right)\cdot
    \mathbf{n}\times \tilde{\bm{\omega}}^{n+1} \vartheta, 
\quad \forall \vartheta\in H_{p0}^1(\Omega),
\end{split}
\label{equ:xi_1_reform_1}
\end{equation}
where we have used the fact that $\vartheta\in H_{p0}^1(\Omega)$,
equation \eqref{equ:xi_2}, the divergence 
theorem, and the following identity
($K$ denoting a scalar field function)
\begin{equation}
\nabla K \cdot \nabla\times\tilde{\bm{\omega}}\vartheta
= \nabla\cdot(\tilde{\bm{\omega}}\times\nabla K \vartheta)
+ \nabla K\cdot(\tilde{\bm{\omega}}\times\nabla\vartheta).
\end{equation}
We note the identity 
\begin{equation}
\begin{split}
\int_{\partial\Omega_i\cup\partial\Omega_w}\frac{\mu}{\rho}
  \mathbf{n}\cdot\nabla\times\tilde{\bm{\omega}}\vartheta
&= \int_{\partial\Omega}\frac{\mu}{\rho}
  \mathbf{n}\cdot\nabla\times\tilde{\bm{\omega}}\vartheta \\
&
= \int_{\partial\Omega}
     \nabla\left(\frac{\mu}{\rho}  \right)\cdot
     \mathbf{n}\times\tilde{\bm{\omega}}\vartheta
+ \int_{\partial\Omega}\frac{\mu}{\rho} 
     \mathbf{n}\cdot \tilde{\bm{\omega}}\times\nabla\vartheta,
\quad \forall \vartheta\in H^1_{p0}(\Omega),
\end{split}
\end{equation}
where we have used the fact $\vartheta\in H^1_{p0}(\Omega)$,
and have repeatedly used the divergence theorem.
Then, equation \eqref{equ:xi_1_reform_1}
can be transformed into the final weak form about $\xi^{n+1}$,
\begin{equation}
\begin{split}
\frac{\gamma_0}{\nu_m\Delta t}
 \int_{\Omega}\xi^{n+1}\vartheta
& + \int_{\Omega}\nabla\xi^{n+1}\cdot\nabla\vartheta
= 
-\frac{1}{\nu_m}\int_{\Omega}\left[
  \mathbf{G}^{n+1}
  + \nabla\left(\frac{\mu^{n+1}}{\rho^{n+1}} \right)\times
    \tilde{\bm{\omega}}^{n+1}
\right] \cdot \nabla\vartheta \\
&
+ \frac{1}{\nu_m}\int_{\partial\Omega_i\cup\partial\Omega_w}
   \mathbf{n}\cdot\frac{\gamma_0\mathbf{w}^{n+1}-\hat{\mathbf{w}}}{\Delta t}\vartheta
+ \frac{1}{\nu_m}\int_{\partial\Omega_i\cup\partial\Omega_w\cup\partial\Omega_o}
   \frac{\mu^{n+1}}{\rho^{n+1}}\mathbf{n}\times
     \tilde{\bm{\omega}}^{n+1}\cdot\nabla \vartheta, \\
&
\ \forall \vartheta\in H_{p0}^1(\Omega).
\end{split}
\label{equ:xi_weakform}
\end{equation}


Let $q \in H^1_{p0}(\Omega)$ denote the test function.
Taking the $L^2$ inner product between 
$\nabla q$ and equation \eqref{equ:pressure_1}
and integrating by part, we obtain 
the weak form about $P^{n+1}$,
\begin{multline}
\int_{\Omega}\nabla P^{n+1}\cdot\nabla q 
= 
\int_{\Omega} \left[
  \frac{\gamma_0\rho_m}{\Delta t}\tilde{\mathbf{u}}^{n+1}
  + \nabla\left(P^n - \rho_m\nu_m\xi^{n+1}  \right)
\right] \cdot \nabla q \\
- \frac{\gamma_0\rho_m}{\Delta t}
   \int_{\partial\Omega_i\cup\partial\Omega_w}
     \mathbf{n}\cdot \mathbf{w}^{n+1} q,
\quad \forall q\in H^1_{p0}(\Omega),
\label{equ:p_weakform}
\end{multline}
where we have used the divergence theorem, and the
equations \eqref{equ:pressure_2} and \eqref{equ:pressure_3}.



One can observe that 
the weak forms \eqref{equ:vel_weakform},
\eqref{equ:xi_weakform} and \eqref{equ:p_weakform}
involve no derivatives of order two or higher,
and all the terms can be computed directly
with $C^0$ elements.
These weak forms can be discretized in space
using $C^0$ spectral elements in
the standard way \cite{KarniadakisS2005}.

Given $(\tilde{\mathbf{u}}^n, \mathbf{u}^n,P^n,\phi^{n+1},
\nabla^2\phi^{n+1})$,
our final algorithm for solving
the momentum equations therefore
consists of the following procedure. We refer to
this procedure as {\bf AdvanceMomentum} hereafter.
It produces 
$(\tilde{\mathbf{u}}^{n+1},\mathbf{u}^{n+1},P^{n+1})$
as follows: \\
\underline{{\bf AdvanceMomentum} procedure:}
\begin{itemize}

\item
Solve equation \eqref{equ:vel_weakform},
together with the velocity Dirichlet condition
\eqref{equ:velocity_2} on $\partial\Omega_i\cup\partial\Omega_w$,
for $\tilde{\mathbf{u}}^{n+1}$;

\item
Solve equation \eqref{equ:xi_weakform},
together with the Dirichlet condition \eqref{equ:xi_3}
on $\partial\Omega_o$, for $\xi^{n+1}$;

\item
Solve equation \eqref{equ:p_weakform},
together with the pressure Dirichlet condition
\eqref{equ:pressure_4} on $\partial\Omega_o$, for $P^{n+1}$;

\item
Evaluate $\mathbf{u}^{n+1}$ based on equation
\eqref{equ:pressure_1} in the following form:
\begin{equation}
\mathbf{u}^{n+1} = \tilde{\mathbf{u}}^{n+1}
- \frac{\Delta t}{\gamma_0\rho_m}\nabla\left(
  P^{n+1} - P^n + \rho_m\nu_m \xi^{n+1}
\right).
\label{equ:u_expr_eval}
\end{equation}

\end{itemize}


In the above algorithm,
when imposing the Dirichlet condition \eqref{equ:xi_3}
about $\xi^{n+1}$ on $\partial\Omega_o$ and
when imposing the pressure Dirichlet condition
\eqref{equ:pressure_4} on $\partial\Omega_o$,
it should be noted that with $C^0$ elements
one needs to first project
the Dirichlet data computed from these equations
into the $H^1(\partial\Omega_o)$,
and then impose the projected data as the 
Dirichlet condition.
This is because the expressions for 
the boundary conditions of \eqref{equ:xi_3}
and \eqref{equ:pressure_4} involve derivatives,
which may not be continuous across element boundaries
on $\partial\Omega_o$ for $C^0$ elements.


One can observe that the
{\bf AdvanceMomentum} algorithm has the following 
characteristics: (i) The computations
for the velocity, the pressure, and 
the field variable $\xi^{n+1}$ are all
de-coupled;
(ii) The computations for 
the different components of 
the velocity $\tilde{\mathbf{u}}^{n+1}$
are de-coupled in \eqref{equ:vel_weakform};
(iii) All resultant linear algebraic
systems from the algorithm involve only
constant and time-independent coefficient
matrices, which can be pre-computed.


As discussed in \cite{DongS2012},
the density $\rho^{n+1}$ and the dynamic
viscosity $\mu^{n+1}$ 
computed according to equation \eqref{equ:rho_mu}
based on $\phi^{n+1}$ may 
encounter numerical difficulties 
when the density ratio between the two
fluids becomes very large or conversely
very small.
This is because the numerically-computed
$\phi$ may not exactly lie within
the range $[-1,1]$ and 
may be slightly out of bound at certain
spatial points in the domain,
because of the interaction between mass
conservation and the minimization of
the free energy inherent in the Cahn-Hilliard
dynamics \cite{DongS2012}. 
At large density ratios,
the slightly out-of-range values of $\phi$
may cause the density or
the dynamic viscosity computed
from \eqref{equ:rho_mu} to become negative
at certain points, thus causing numerical
difficulties. 
Following \cite{DongS2012},
when the density ratio becomes large or conversely
small (typically beyond $10^2$ or below $10^{-2}$),
we will use the following modified function
for computing the mixture density and dynamic
viscosity,
\begin{equation}
\hat{\phi} = \left\{
\begin{array}{ll}
\phi, & \text{if} \ |\phi|\leqslant 1, \\
\text{sign}(\phi), & \text{if} \ |\phi|>1;
\end{array}
\right.
\ \
\rho = \frac{1}{2}\left(\rho_1+\rho_2  \right)
+ \frac{1}{2}\left(\rho_1-\rho_2  \right)\hat{\phi};
\ \
\mu = \frac{1}{2}\left(\mu_1+\mu_2  \right)
+ \frac{1}{2}\left(\mu_1-\mu_2  \right)\hat{\phi}.
\label{equ:rho_mu_modified}
\end{equation}


\subsection{Overall Method for Two-Phase Flow Simulations}
\label{sec:overall_method}

%

Let us now consider the numerical solution of the phase field
equation \eqref{equ:CH}, together with 
the boundary conditions \eqref{equ:ibc_phi_1}
and \eqref{equ:ibc_phi_2} for $\partial\Omega_i$,
\eqref{equ:wbc_phi_1} and \eqref{equ:wbc_phi_2}
for $\partial\Omega_w$, and
\eqref{equ:obc_phi_1} and \eqref{equ:obc_phi_2}
for $\partial\Omega_o$.
In a previous work \cite{DongS2012}, we have
developed an algorithm for the phase field equation
\eqref{equ:CH}.
This algorithm 
computes the phase field function $\phi^{n+1}$
and $\nabla^2\phi^{n+1}$ (both in $H^1(\Omega)$ space) by solving
two Helmholtz type equations  in a successive but
un-coupled fashion.
We will employ this algorithm for the
phase field equation in the current work.
For the sake of completeness, we provide a summary
of this algorithm for solving the phase field equation
together with the boundary conditions
in the Appendix of this paper,
and it is referred to as the {\bf AdvancePhase}
procedure (see the Appendix).

Our overall method for simulating incompressible
two-phase flows is a combination of
the algorithm presented in 
Section \ref{sec:alg_momentum} for the 
momentum equations and the algorithm
in the Appendix for the phase field equation.
Specifically, 
given $(\tilde{\mathbf{u}}^n,\mathbf{u}^n,P^n,\phi^n)$,
the overall discrete formulation of the method consists of
equations \eqref{equ:phi_1}--\eqref{equ:phi_8} (in the Appendix),
\eqref{equ:velocity_1}--\eqref{equ:velocity_4},
\eqref{equ:xi_1}--\eqref{equ:xi_3},
and \eqref{equ:pressure_1}--\eqref{equ:pressure_4}.
With $C^0$ spectral-element spatial discretizations,
we go through the developments discussed in 
Section \ref{sec:alg_momentum} and in the Appendix
to obtain the weak forms for the field variables.
The final solution procedure
is composed of
the following steps:
\begin{itemize}

\item
Compute $\phi^{n+1}$ and $\nabla^2\phi^{n+1}$ 
based on the {\bf AdvancePhase} procedure discussed in the Appendix.

\item
Compute $\rho^{n+1}$ and $\mu^{n+1}$ according to
equation \eqref{equ:rho_mu} by using $\phi^{n+1}$
computed above.
When the density ratio becomes large or conversely small
(typically above $10^2$ or below $10^{-2}$), use
equation \eqref{equ:rho_mu_modified} instead.

\item
Compute $(\tilde{\mathbf{u}}^{n+1},\mathbf{u}^{n+1},P^{n+1})$
based on the {\bf AdvanceMomentum} procedure 
discussed in Section \ref{sec:alg_momentum},
using $\phi^{n+1}$, $\nabla^2\phi^{n+1}$,
$\rho^{n+1}$, and $\mu^{n+1}$ computed
above.

\end{itemize}

It can be observed that this  method
has the following characteristics:
(1) The computations for all the flow
variables and auxiliary variables are
completely de-coupled;
(2) All the resultant linear algebraic systems 
after discretization involve only {\em constant}
and {\em time-independent} coefficient matrices,
which can be pre-computed;
(3) Within each time step, the method
involves only the solution of 
individual Helmholtz-type (including Poisson)
equations;
(4) The method is suitable for large density
ratios and large viscosity ratios,
which will be demonstrated using numerical
simulations in Section \ref{sec:tests}.


\section{Representative Numerical Tests}
\label{sec:tests}

\begin{table}
\begin{center}
\begin{tabular*}{1.0\textwidth}{@{\extracolsep{\fill}}
l c | l c}
\hline
variables/parameters & normalization constants &
variables/parameters & normalization constants \\
$\mathbf{x}$, $\eta$ & $L$ & $\sigma$ & $\rho_1U_0^2L$ \\
$\mathbf{u}$, $\mathbf{u}_{in}$, $\mathbf{w}$ & $U_0$ & $D_0$ & $1/U_0$ \\
$t$, $\Delta t$ & $L/U_0$ & $\gamma_1$ & $L/(\rho_1U_0)$ \\
$\mathbf{g}_r$ (gravity) & $U_0^2/L$ & $\lambda$ & $\rho_1U_0^2 L^2$ \\
$p$, $P$, $\mathbf{f}_b$ & $\rho_1U_0^2$ & $\nu_m$ & $U_0L$ \\
$\phi$, $\hat{\phi}$, $\phi_b$, $\phi_{in}$, $\theta_s$ & $1$ & $\mathbf{f}$ & $\rho_1U_0^2/L$ \\
$\rho$, $\rho_1$, $\rho_2$, $\rho_m$ & $\rho_1$ & $\xi^n$, $g$ & $U_0/L$ \\
$\mu$, $\mu_1$, $\mu_2$, $\mu_0$ & $\rho_1U_0L$ & $g_{a1}$ & $1/L^4$   \\
$g_{a2}$, $g_{c2}$ & $1/L$ & $g_b$ & $1/L^2$ \\
$g_{c1}$ & $1/L^3$ \\
\hline
\end{tabular*}
\end{center}
\caption{Normalization constants for the flow variables and parameters.}
\label{tab:normalization}
\end{table}

In this section 
we demonstrate the accuracy  of
the method presented in Section \ref{sec:method}
and its capability 
for coping with two-phase outflows and open boundaries
using several two-phase flow
problems.
These test problems are in two dimensions.
They involve two-phase
open boundaries, 
and large contrasts in densities
and dynamic viscosities of the two fluids. 
Simulation results will be compared
with the experimental data and with the
exact physical solutions from theory
to demonstrate that 
the method developed herein produces physically
accurate results.

We first briefly mention 
the normalization of the governing equations and physical
parameters, which
has been discussed at length in  previous
works \cite{Dong2012,Dong2014obc}. 
Let $L$ denote the characteristic length scale
and $U_0$ denote the characteristic velocity scale.
In Table \ref{tab:normalization} we have listed
the normalization constants for different
physical variables and parameters. For instance,
the non-dimensional mixing energy density coefficient
is given by 
$
\frac{\lambda}{\rho_1 U_0^2L^2}
$
based on this table.
When the flow variables and parameters
are normalized as given by the table,
the forms of the governing equations and the boundary
conditions 
 will remain unchanged
upon normalization.
In the following discussions all the flow variables
and physical parameters are given in non-dimensional
forms unless otherwise noted, with the understanding
that they have all been properly normalized.

\subsection{Convergence Rates}


\begin{figure}
\centerline{
\includegraphics[height=2.3in]{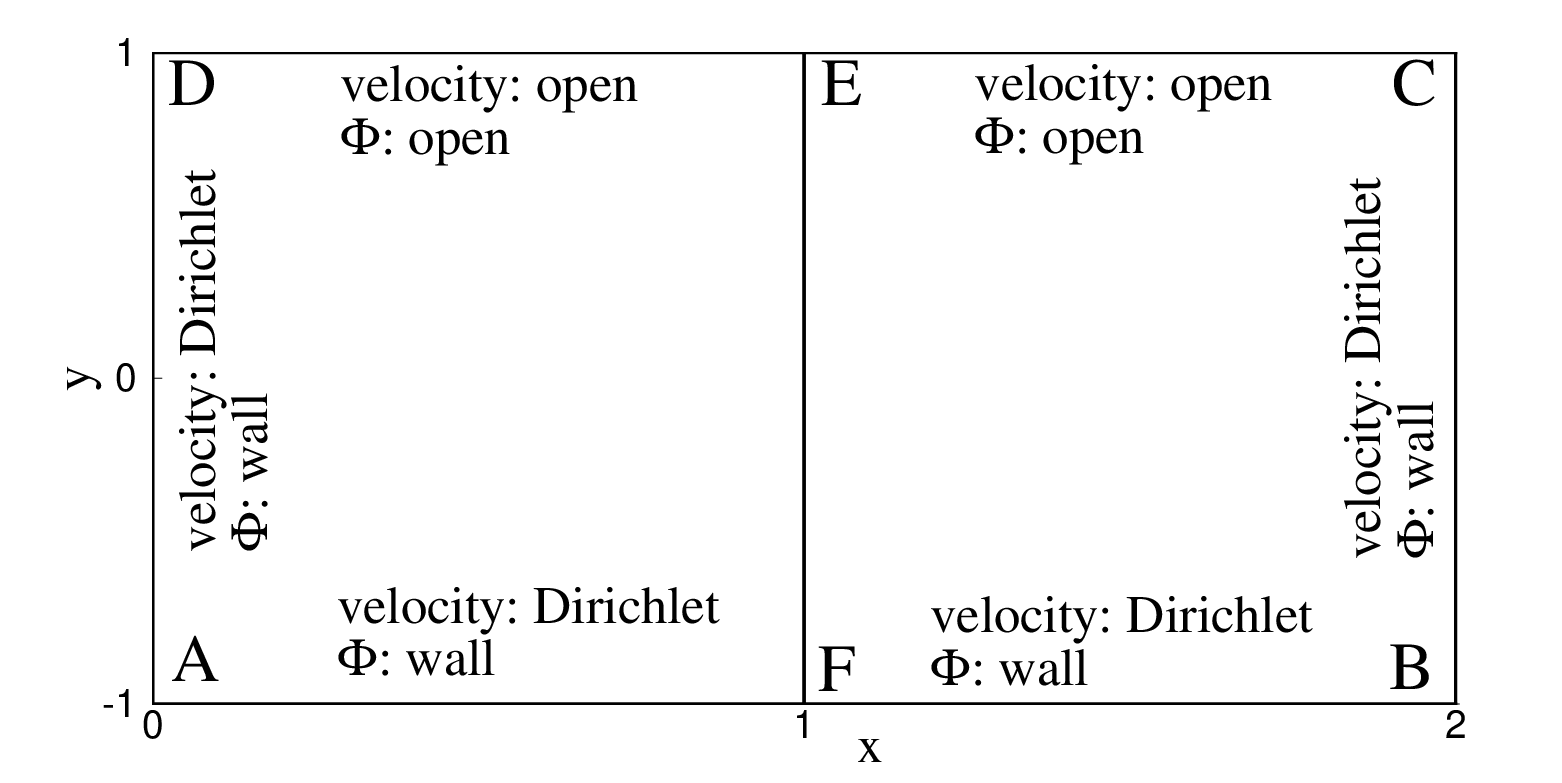}(a)
}
\centerline{
\includegraphics[height=3in]{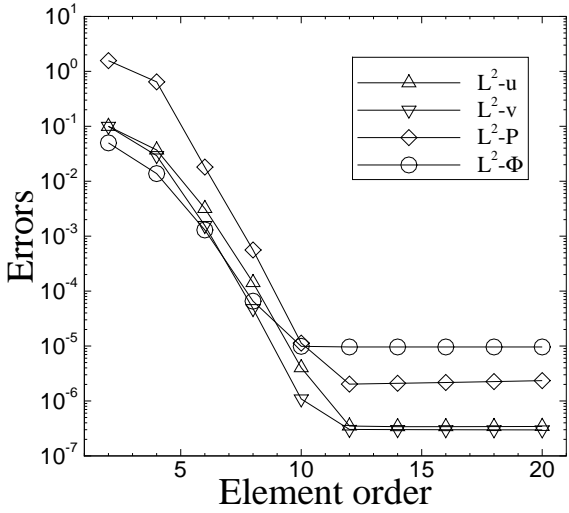}(b)
\includegraphics[height=3in]{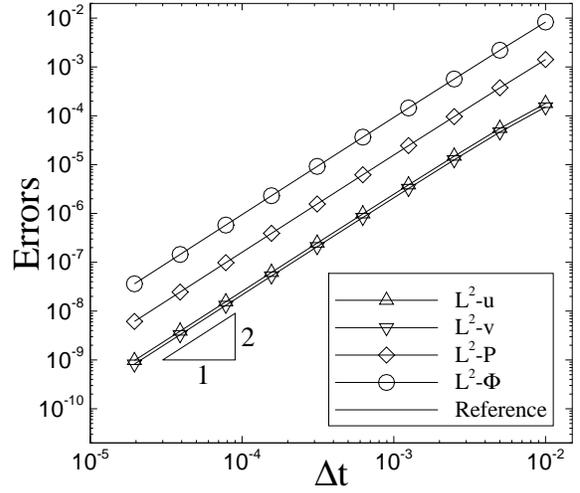}(c)
}
\caption{
Spatial/temporal convergence rates:
(a) Mesh and boundary conditions;
(b) Numerical errors versus element order 
showing spatial exponential convergence
(with fixed $\Delta t=0.001$);
(c) Numerical errors versus $\Delta t$ showing
temporal second-order convergence rate
(element order fixed at $18$).
On the face $\overline{CD}$ the open boundary
condition \eqref{equ:obc_v_D} is used.
}
\label{fig:conv}
\end{figure}

The goal of this section is to study the convergence
behavior of the method from Section \ref{sec:method},
and to demonstrate its spatial and temporal convergence
rates using a contrived analytic solution to
the two-phase governing equations.

The setup of the problem is as follows.
Figure \ref{fig:conv}(a) shows  the rectangular domain
$\overline{ABCD}$ for this problem,
$0\leqslant x\leqslant 2$ and $-1\leqslant y\leqslant 1$.
We consider the following analytic expressions for 
the flow variables 
\begin{equation}
\left\{
\begin{split}
&
u = A\cos \pi y \sin ax \sin Wt \\
&
v = -\frac{Aa}{\pi}\sin \pi y \cos ax \sin Wt \\
&
P = A \sin \pi y \sin ax \cos Wt \\
& 
\phi = B \cos a_1 x \cos b_1 y \sin W_1 t
\end{split}
\right.
\label{equ:anal_soln}
\end{equation}
where $(u,v)$ are the $x$ and $y$ velocity components,
and $A$, $B$, $a$, $W$, $a_1$, $b_1$ and $W_1$ are
prescribed constants to be specified below.
It is evident that the $u$ and $v$ expressions
satisfy the equation \eqref{equ:continuity}.
The external force $\mathbf{f}(\mathbf{x},t)$ in
\eqref{equ:nse_trans_1} and the source term $g(\mathbf{x},t)$
in \eqref{equ:CH} are chosen such that 
the analytic expressions in \eqref{equ:anal_soln}
satisfy the equations 
\eqref{equ:nse_trans_1} and \eqref{equ:CH}.


For the boundary conditions, 
on the sides $\overline{AD}$, $\overline{AB}$
and $\overline{BC}$ we impose the Dirichlet
condition \eqref{equ:dbc_v} for the
velocity
with the boundary velocity $\mathbf{w}$
chosen according to the analytic expressions
of \eqref{equ:anal_soln},
and we impose the contact-angle
conditions \eqref{equ:wbc_phi_1}--\eqref{equ:wbc_phi_2}
for the phase field function,
in which $\theta_s=90^0$ and
$g_{c1}$ and $g_{c2}$ are chosen such that
the $\phi$ expression in \eqref{equ:anal_soln}
satisfies the equations \eqref{equ:wbc_phi_1} and
\eqref{equ:wbc_phi_2}.
On the side $\overline{CD}$ we impose
the open boundary condition \eqref{equ:obc_v_trans},
in which $\mathbf{f}_b$ is chosen such that
the analytic expressions in \eqref{equ:anal_soln}
satisfy \eqref{equ:obc_v_trans},
and we impose the conditions 
\eqref{equ:obc_phi_1}--\eqref{equ:obc_phi_2}
for the phase field function,
in which $D_0=0$ and $g_{a1}$ and $g_{a2}$
are chosen such that the $\phi$
expression in \eqref{equ:anal_soln}
satisfies the equations \eqref{equ:obc_phi_1}
and \eqref{equ:obc_phi_2}.
For the initial conditions \eqref{equ:ic}
we choose $\mathbf{u}_{in}$ and
$\phi_{in}$ according to 
the analytic expressions in \eqref{equ:anal_soln}
by setting $t=0$.

\begin{table}
\begin{center}
\begin{tabular*}{1.0\textwidth}{@{\extracolsep{\fill}} l c | l c}
\hline
parameters & values & parameters & values \\
$A$ & $2.0$ & $\rho_m$ & $\min(\rho_1,\rho_2)$ \\
$B$ & $1.0$ & $\nu_m$ & $\frac{1}{2}\left(\frac{\mu_1}{\rho_1}+\frac{\mu_2}{\rho_2} \right)$ \\
$a$, $a_1$, $b_1$ & $\pi$ & $\mu_0$ & $\max(\mu_1,\mu_2)$ \\
$W$, $W_1$ & $1.0$ & $\delta$ & $\frac{1}{20}$ \\
$\rho_1$ & $1.0$ & $\eta$ & $0.1$  \\
$\rho_2$ & $3.0$ & $\theta_s$ & $90^0$ \\
$\mu_1$ & $0.01$ & $D_0$ & $0.0$ \\
$\mu_2$ & $0.05$ & $J$ (integration order) & $2$ \\
$\sigma$ & $9.428\times 10^{-2}$ \\
$\gamma_1$ & $0.01$ \\
\hline
\end{tabular*}
\end{center}
\caption{Parameter values for convergence tests.}
\label{tab:conv_param}
\end{table}


We partition the domain along the $x$ direction using two 
quadrilateral spectral elements of the same size as
shown in Figure \ref{fig:conv}(a).
The system of governing equations \eqref{equ:nse_trans_1},
\eqref{equ:continuity} and \eqref{equ:CH}
is integrated over time with the algorithm
presented in Section \ref{sec:method}
from $t=0$ to $t=t_f$ ($t_f$ to be specified below).
Then we compute and monitor the errors of the simulation results
at $t=t_f$ against the analytic solution 
given in \eqref{equ:anal_soln}.
The parameters for this problem are listed in
Table \ref{tab:conv_param}.

In the first group of tests we fix the final integration
time at $t_f=0.1$ and the time step size at
$\Delta t=0.001$ ($100$ time steps).
Then we vary the element order systematically
between $2$ and $20$.
Figure \ref{fig:conv}(b) shows the $L^2$ errors of
the velocity, pressure and the phase field
function at $t=t_f$ as a function of
the element order. 
The results correspond to  the open boundary 
condition \eqref{equ:obc_v_D} on the side of 
the domain $\overline{CD}$.
It can be observed that the numerical errors
decrease exponentially as the
element order increases (when below order $10$).
As the element order increases beyond  $12$,
the error curves level off due to the saturation
by the temporal truncation error.

In the second group of tests
we fix the final integration time at $t_f=0.1$
and the element order at a large value $18$,
and then vary the time step size systematically 
between $\Delta t=1.953125\times 10^{-5}$
and $\Delta t=0.01$.
In Figure \ref{fig:conv}(c) we plot the 
$L^2$ errors of the flow variables as
a function of $\Delta t$ in logarithmic
scales. A slope of $2$ has been observed in
the error curves when the time step size
becomes small.

The results of these tests demonstrate that
the method developed in Section \ref{sec:method}
has a spatial exponential convergence rate
and a temporal second-order convergence rate.

\subsection{Capillary Wave}

\begin{figure}
\centerline{
\includegraphics[height=3in]{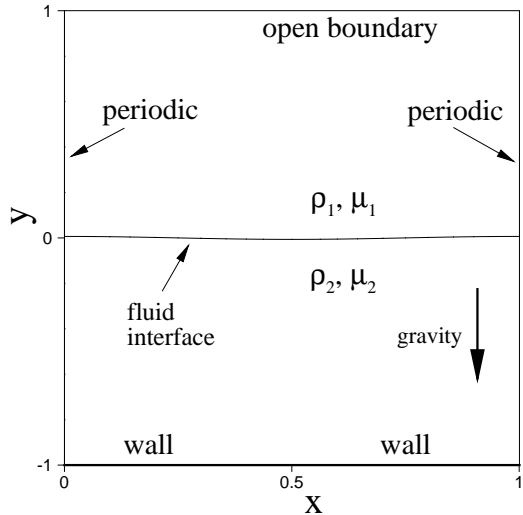}
}
\caption{
Configuration for the capillary wave problem.
}
\label{fig:cwave}
\end{figure}

The goal of this section is to demonstrate the physical
accuracy of our method 
using a two-phase capillary wave problem, whose exact physical
solution is known from the literature \cite{Prosperetti1981}.
The problem involves two fluid phases, density contrast,
viscosity contrast, gravity  and the surface
tension effects. We have  considered this problem 
in a previous work \cite{DongS2012}. It should be noted that
the algorithm tested here is different from that 
of \cite{DongS2012}.

Here is the setting of the problem.
We consider two immiscible incompressible fluids in
an infinite domain. The lighter fluid occupies
the top half of the domain, and the heavier fluid occupies
the bottom half. The gravity is in the vertical direction
and points downward. Without loss of generality
we assume that the first fluid is lighter than 
the second one ($\rho_1\leqslant \rho_2$).
At $t=0$, the interface formed between
the two fluids is perturbed by a small-amplitude
sinusoidal wave from its equilibrium horizontal
position, and starts to oscillate.
The goal of this problem is to study the behavior
of the interface over time.

Prosperetti \cite{Prosperetti1981} reported an 
exact standing-wave (but time-dependent)
solution to this problem under the following condition:
The two fluids may have different densities and
dynamic viscosities, but their kinematic viscosities
must match. The relation of the capillary-wave
amplitude over time has been provided.
We will simulate this problem under the same
condition, and compare with the exact physical solution
from \cite{Prosperetti1981}.

Specifically, we consider a computational
domain as depicted in Figure \ref{fig:cwave} (non-dimensionalized),
$0\leqslant x\leqslant 1$ and $-1\leqslant y\leqslant 1$.
The un-perturbed equilibrium position of the fluid interface coincides 
with the $x$-axis.  We assume that 
the initial perturbation profile of the interface
is given by
\begin{equation}
y = H_0 \cos k x,
\quad k = \frac{2\pi}{\lambda_w} = 2\pi,
\end{equation}
where $\lambda_w=1$ is the wave length
of the perturbation profile,
and $H_0=0.01$ is the initial amplitude of the capillary 
wave. Note that the capillary wave-length $\lambda_w$
is chosen to be the same as  the domain dimension
in the $x$ direction, and that the initial capillary amplitude 
$H_0$ is small compared to the domain dimension in
the $y$ direction.


We employ the algorithm developed in Section \ref{sec:method}
to solve the governing equations \eqref{equ:nse_trans_1}
and \eqref{equ:continuity}--\eqref{equ:CH},
where the external body force in \eqref{equ:nse_trans_1}
is set to $\mathbf{f}=\rho \mathbf{g}_r$ and
$\mathbf{g}_r$ is the gravitational acceleration.
For the boundary conditions, in the
horizontal direction we assume that it is
periodic at $x=0$ and $x=1$.
At the bottom of the domain ($y=-1$), we
assume a solid wall in the simulations,
and impose the Dirichlet condition \eqref{equ:dbc_v}
with $\mathbf{w}=0$ for the
velocity, and impose the boundary conditions 
\eqref{equ:wbc_phi_1}--\eqref{equ:wbc_phi_2}
with $g_{c1}=g_{c2}=0$ and $\theta_s=90^0$.
On the top side ($y=1$) we assume that
the domain is open, and impose the open
boundary condition \eqref{equ:obc_v_trans}
with $\mathbf{f}_b=0$
for the momentum equation, 
and impose the open boundary conditions
\eqref{equ:obc_phi_1}--\eqref{equ:obc_phi_2}
with $g_{a1}=g_{a2}=0$ and $D_0=0$ for
the phase field function.
%
%
We employ the following initial velocity and
phase field function in the simulations
\begin{equation}
\mathbf{u}_{in}(\mathbf{x}) = 0, \quad
\phi_{in}(\mathbf{x}) = \tanh\left(\frac{y-H_0\cos kx}{\sqrt{2}\eta}\right).
\end{equation}

\begin{table}
\begin{center}
\begin{tabular*}{1.0\textwidth}{@{\extracolsep{\fill}} l c | l c}
\hline
parameters & values & parameters & values \\
$|\mathbf{g}_r|$ & $1.0$ & $\rho_m$ & $\min(\rho_1,\rho_2)$ \\
$\sigma$ & $1.0$ & $\nu_m$ 
& $\frac{1}{2}\left(\frac{\mu_1}{\rho_1}+\frac{\mu_2}{\rho_2} \right)$ \\
$H_0$ & $0.01$ & $\mu_0$ & $\mu_1$ \\
$\lambda_w$ & $1.0$ & $\delta$ & $1/100$ \\
$\rho_1$ & $1.0$ & $D_0$ & $0.0$ \\
$\mu_1$ & $0.01$ & $\theta_s$ & $90^0$ \\
$\frac{\mu_2}{\rho_2}$ & $\frac{\mu_1}{\rho_1}$ 
& $J$ (integration order) & $2$  \\
$\rho_2$, $\mu_2$ & (varied) & $\Delta t$ & $2.5\times 10^{-5}$  \\
$\eta$ & $0.002$ & $\lambda$ & $\frac{3}{2\sqrt{2}}\sigma\eta$  \\
$\gamma_1$ & $2.5\eta^2$ & Element order & $14$ \\
Number of elements in mesh & $240$ \\
\hline
\end{tabular*}
\end{center}
\caption{Parameter values for the capillary wave problem.}
\label{tab:capillary_param}
\end{table}

We discretize the domain using $240$ quadrilateral elements,
with $10$ elements in the $x$ direction and $24$ elements
in the $y$ direction.
The elements are uniform along the $x$ direction, and are 
non-uniform along the $y$ direction, clustering around
the region $-0.012\leqslant y\leqslant 0.012$.
We have used an element order $14$ for all the elements.
The non-dimensional time step size is 
fixed at $\Delta t=2.5\times 10^{-5}$
in the simulations.


We choose the physical parameters for this problem
in accordance with those in \cite{DongS2012}.
A summary of the values for the physical and numerical
parameters in this problem is provided in Table
\ref{tab:capillary_param}.
Note that while $\rho_2$ and $\mu_2$ are varied
in different cases in the simulations,
the relation $\frac{\mu_2}{\rho_2}=\frac{\mu_1}{\rho_1}$
is maintained according to the condition of
the exact physical solution by \cite{Prosperetti1981}.


\begin{figure}
\centerline{
\includegraphics[width=3in]{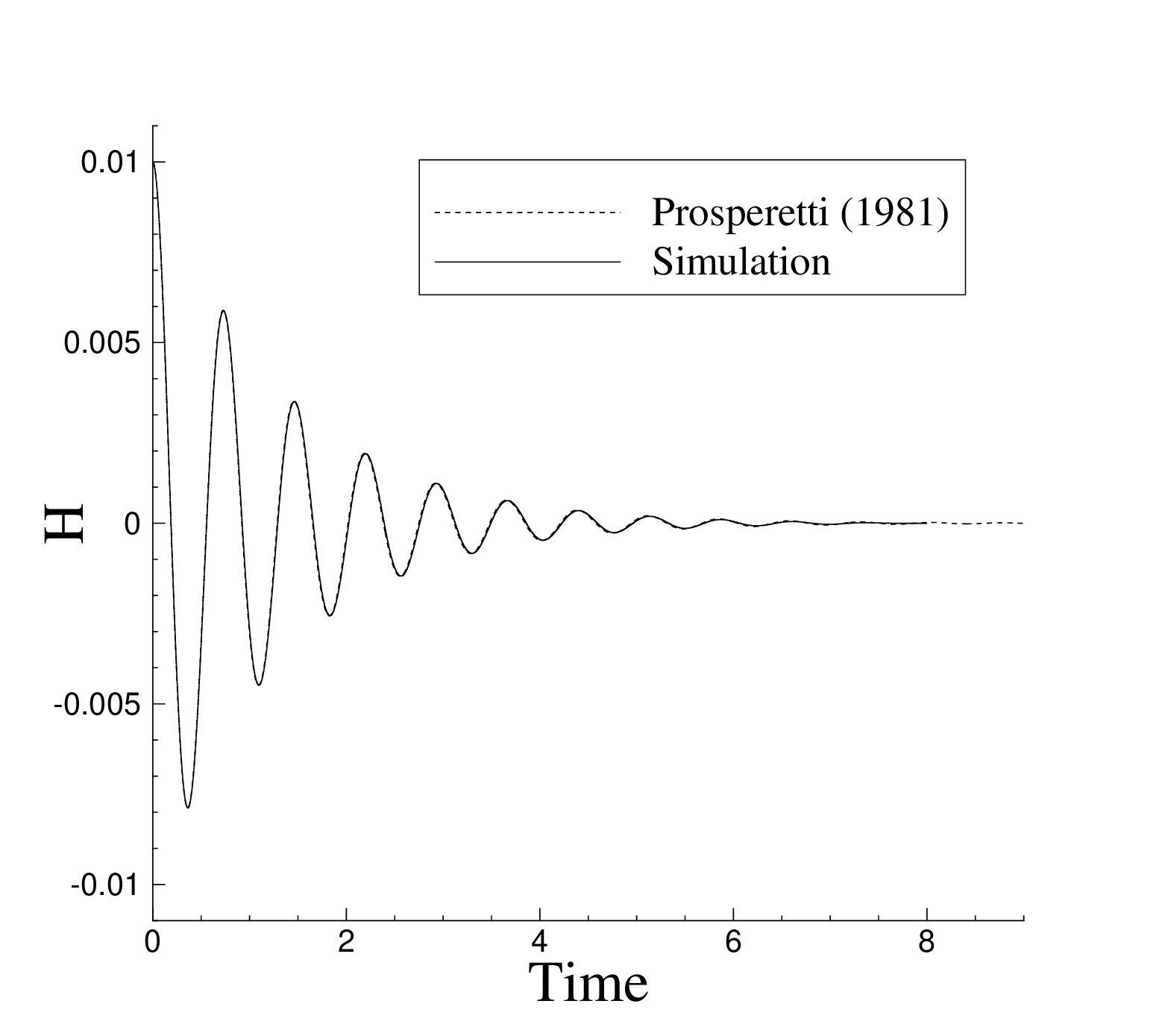}(a)
\includegraphics[width=3in]{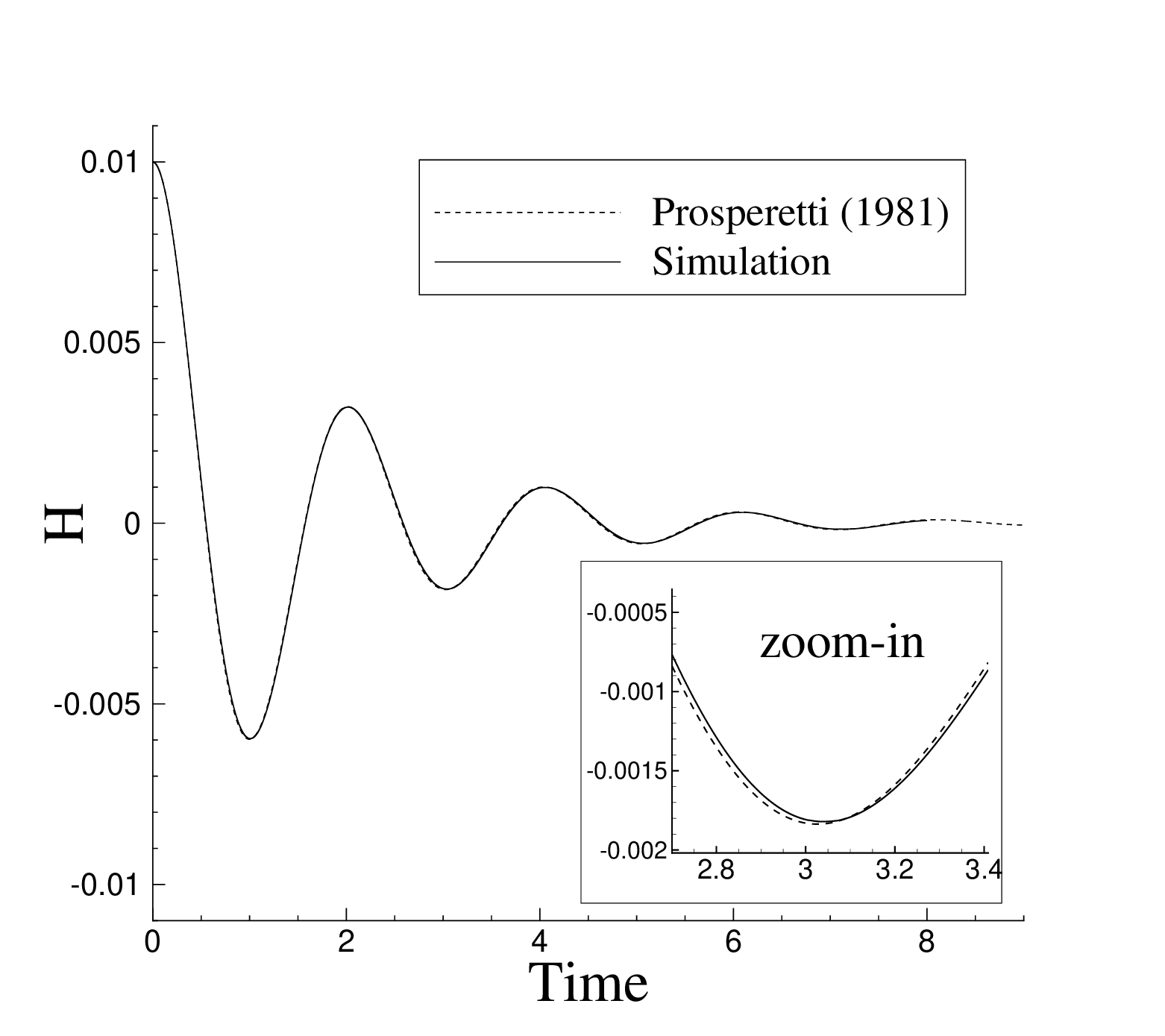}(b)
}
\centerline{
\includegraphics[width=3in]{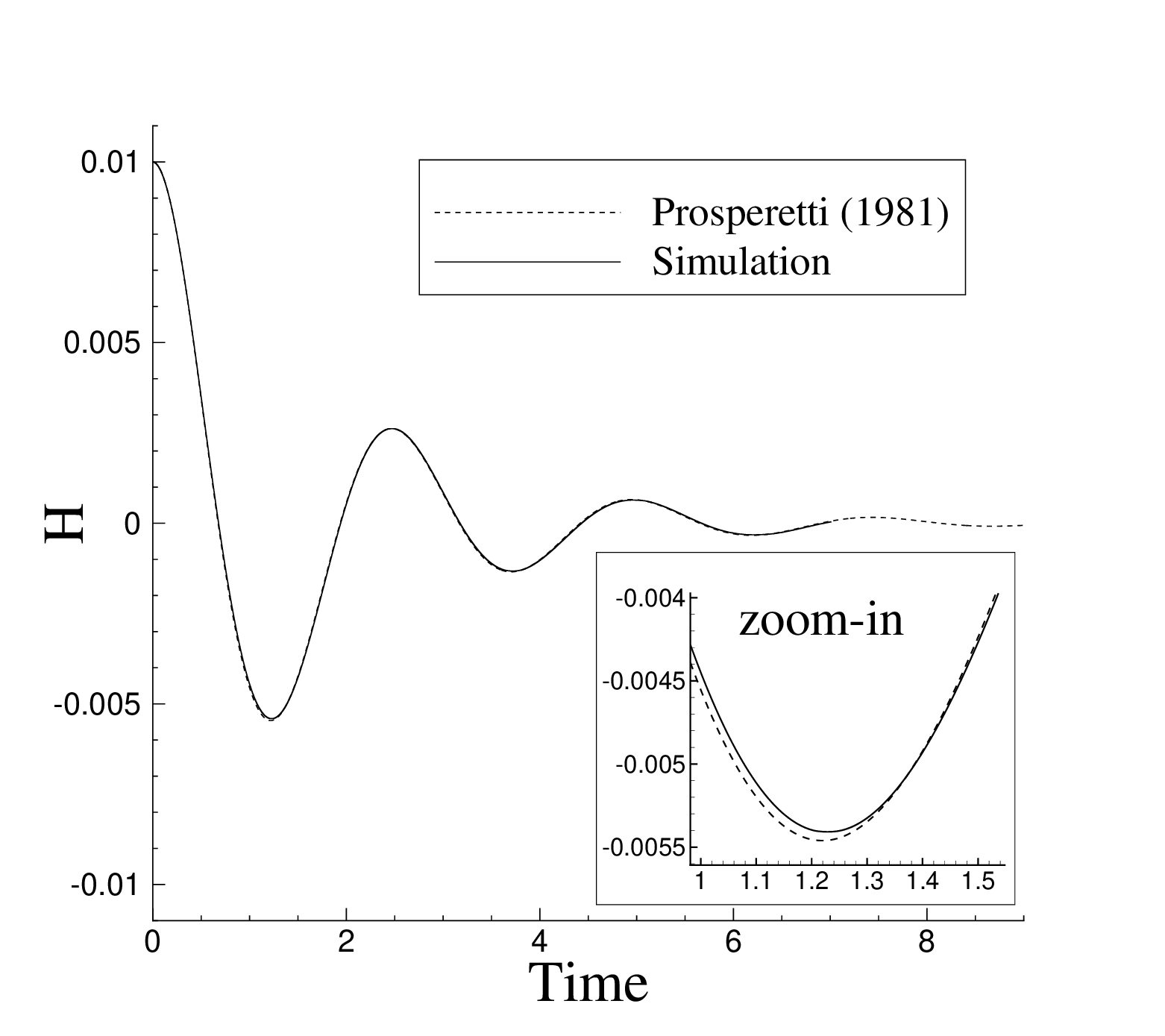}(c)
\includegraphics[width=3in]{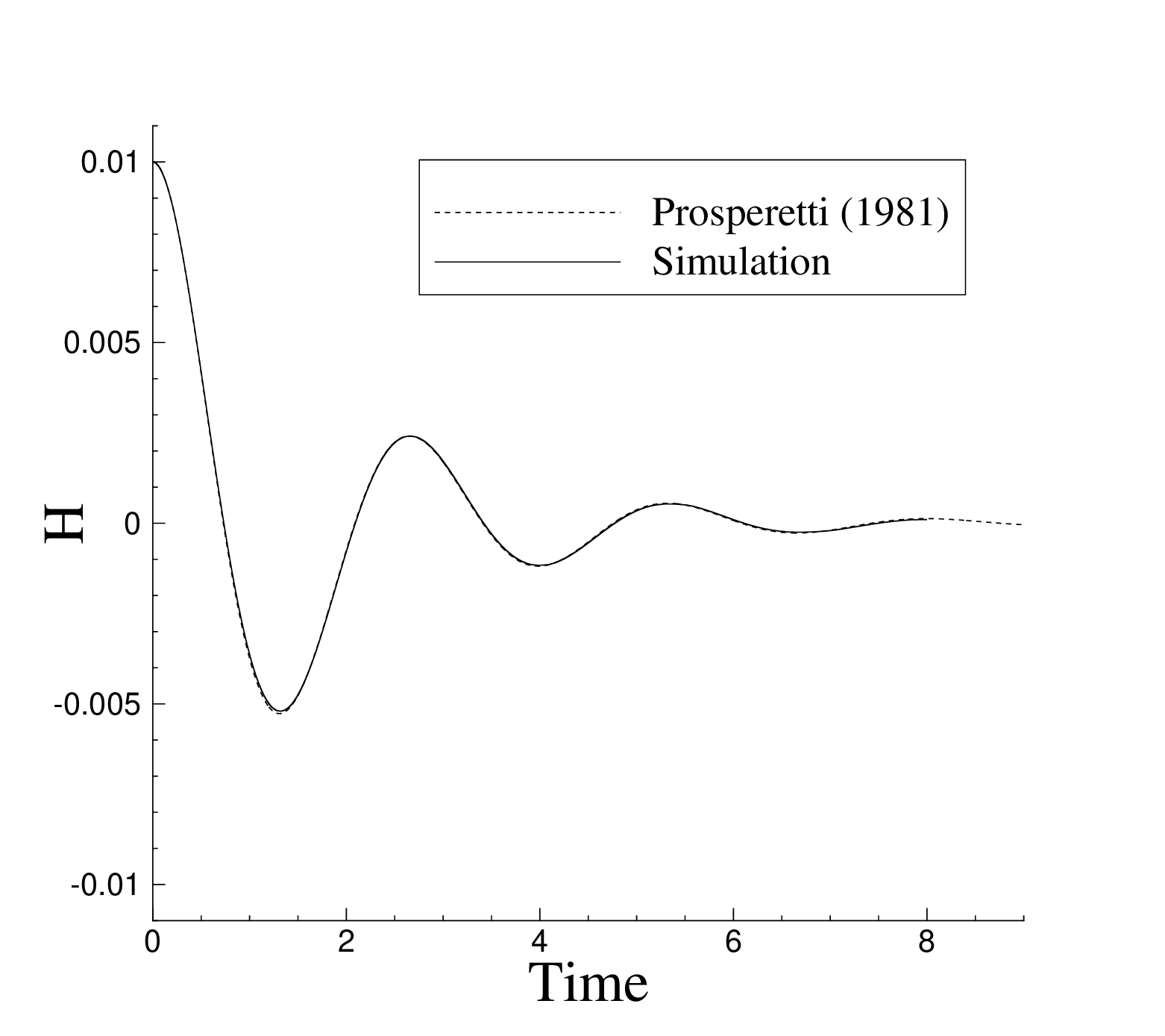}(d)
}
\caption{
Comparison of time histories of the capillary-wave amplitudes
between current simulation and the exact solution
by Prosperetti \cite{Prosperetti1981} for
density ratios
(a) $\frac{\rho_2}{\rho_1}=2$,
(b) $\frac{\rho_2}{\rho_1}=50$,
(c) $\frac{\rho_2}{\rho_1}=200$, and
(d) $\frac{\rho_2}{\rho_1}=1000$.
}
\label{fig:capillary_compare}
\end{figure}

Let us compare the simulation results
with the exact physical solution given by
\cite{Prosperetti1981}.
Figure \ref{fig:capillary_compare}
shows the time histories of the capillary
amplitude $H(t)$ from the simulation
and from the exact solution \cite{Prosperetti1981}
at several density ratios.
Figures \ref{fig:capillary_compare}(a)--(d)
respectively correspond to 
the density ratios 
$\frac{\rho_2}{\rho_1}=2$, 
$50$, $200$, and $1000$.
These results are obtained using the open 
boundary condition \eqref{equ:obc_v_D} at
the upper domain boundary.
It can be observed that
the fluid interface fluctuates about
its equilibrium position with
the amplitude attenuated over time.
The oscillation frequency decreases
with increasing density ratios between
the two fluids.
One can further observe that 
the time-history curves from the simulations
almost exactly overlap with those 
from the physical solution
given by \cite{Prosperetti1981}
for all density ratios. 
The insets of Figure \ref{fig:capillary_compare}(b) 
and Figure \ref{fig:capillary_compare}(c) are
the blow-up views of the curves,
which show that the difference between
the simulation and the exact physical solution
is small.
These results indicate that
our method presented in Section 
\ref{sec:method} has produced physically
accurate results for the capillary
wave problem.

\subsection{Bouncing Water Drop on Superhydrophobic Surface}



\begin{figure}
\centerline{
\includegraphics[width=1.2in]{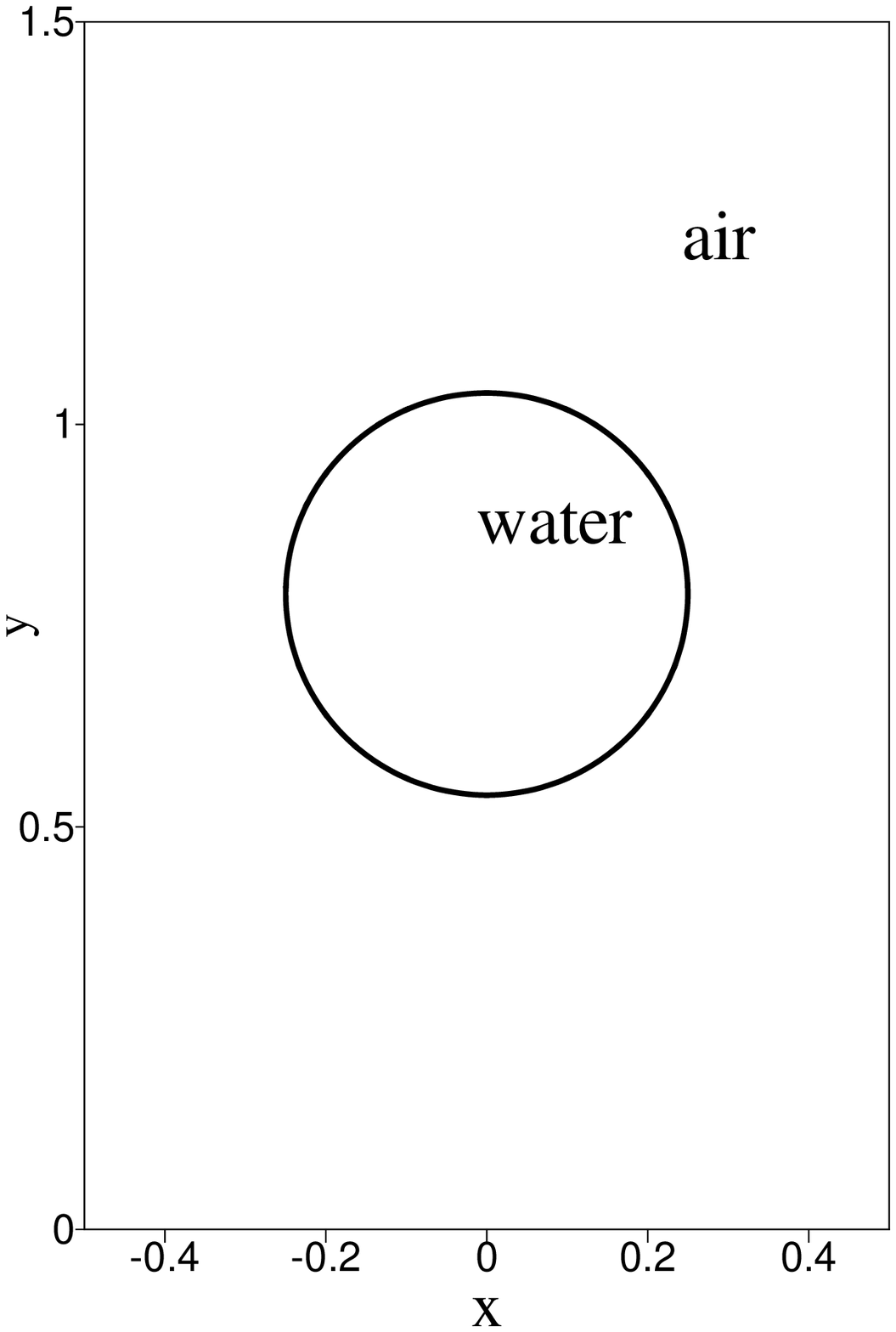}(a)
\includegraphics[width=1.2in]{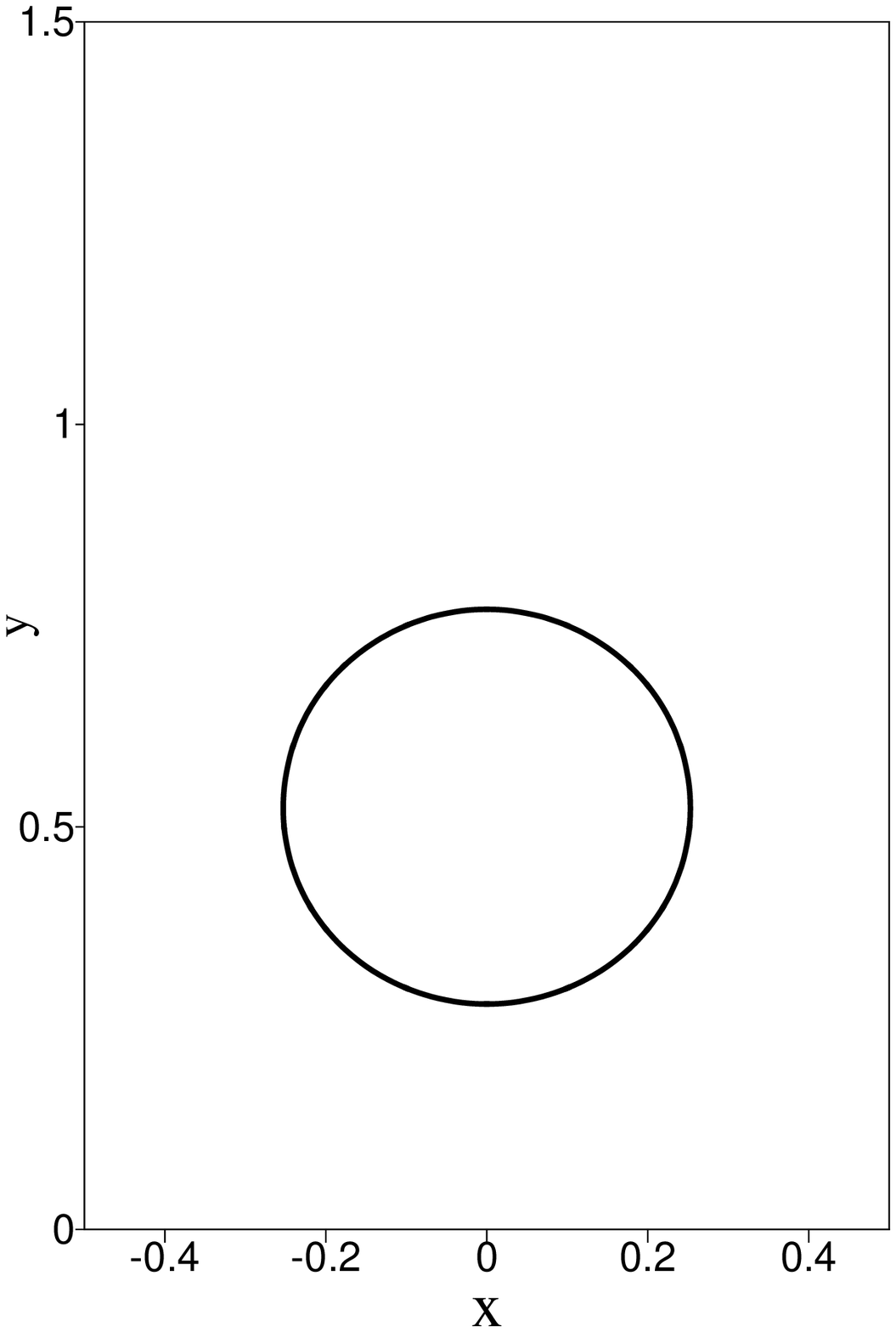}(b)
\includegraphics[width=1.2in]{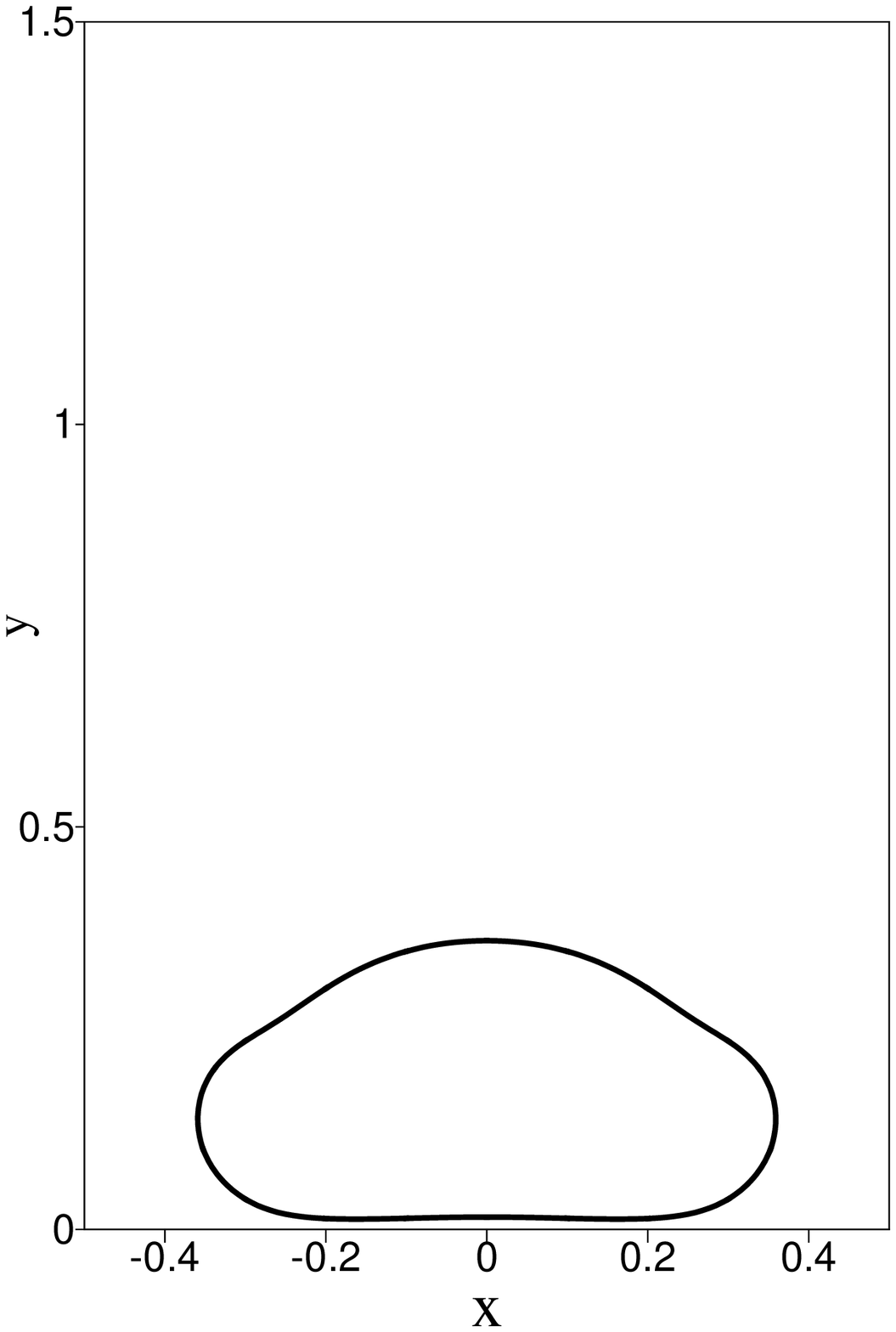}(c)
\includegraphics[width=1.2in]{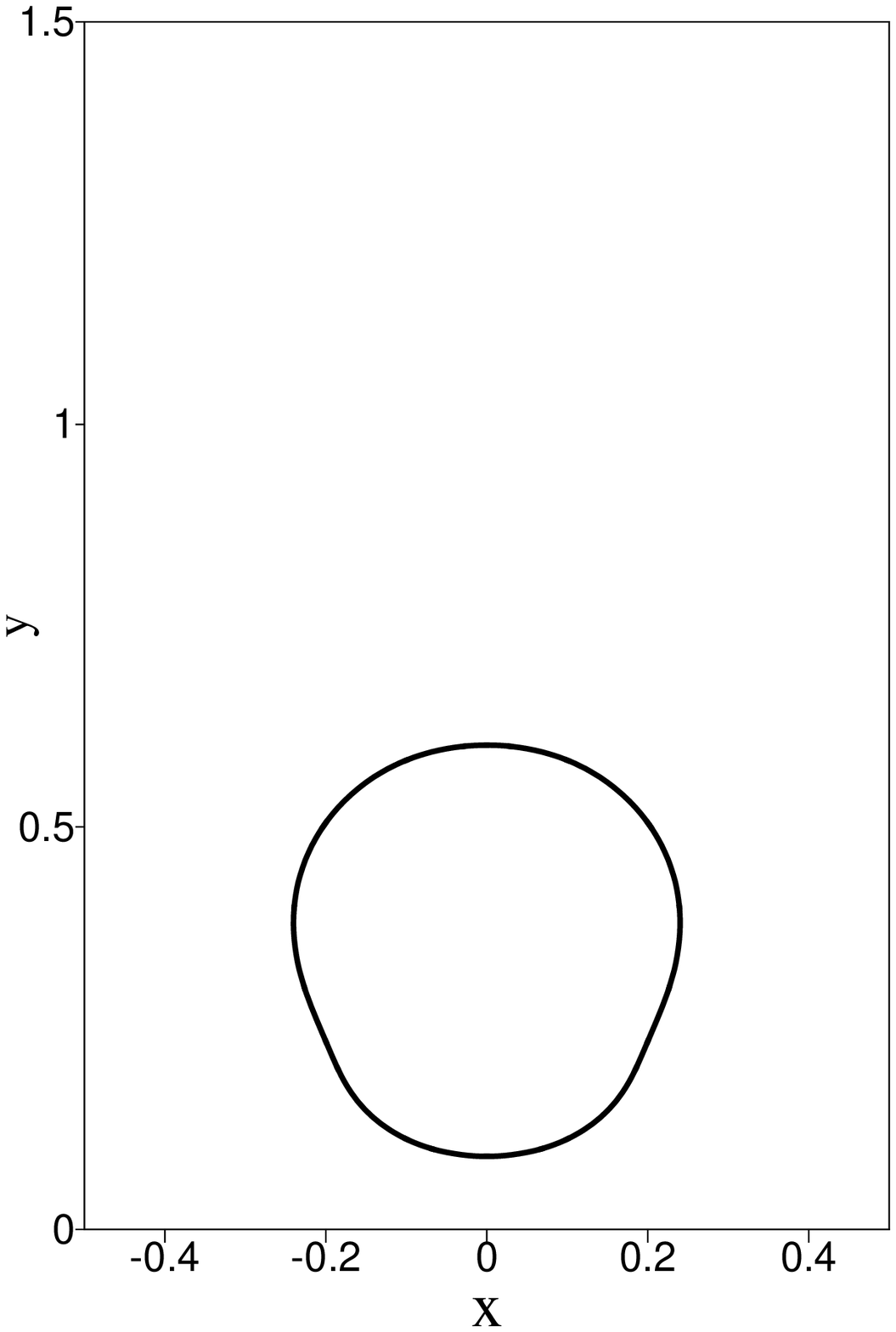}(d)
}
\centerline{
\includegraphics[width=1.2in]{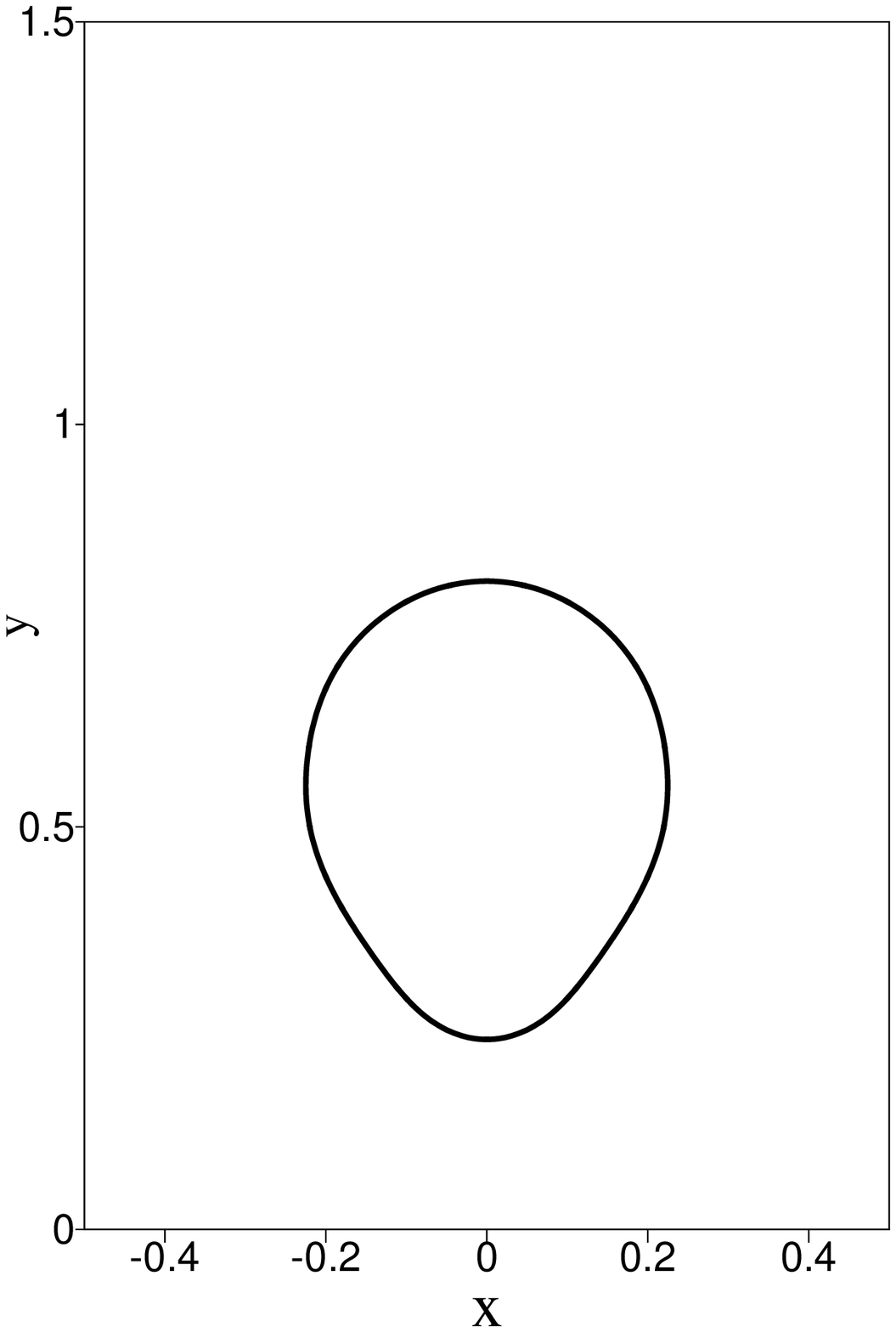}(e)
\includegraphics[width=1.2in]{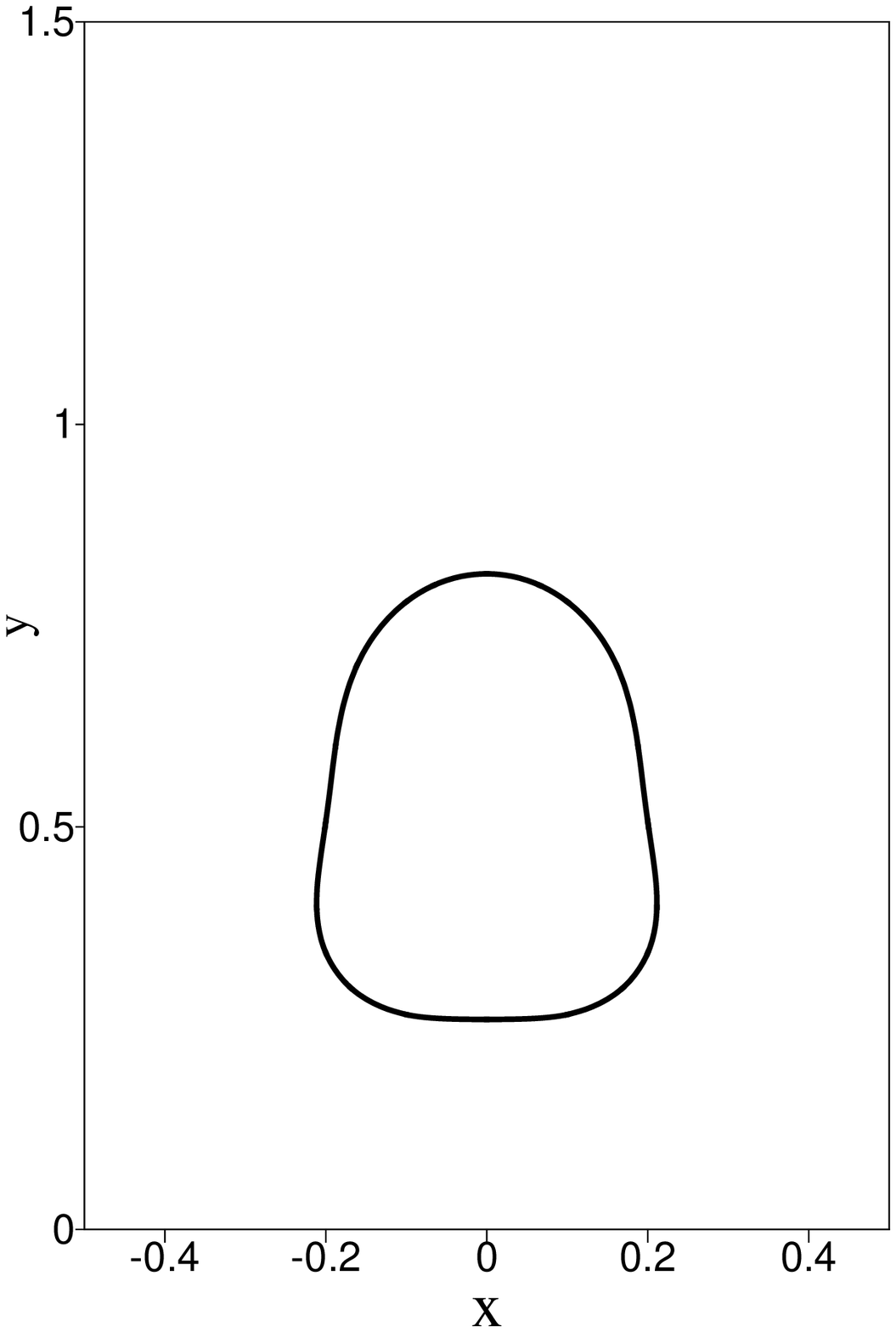}(f)
\includegraphics[width=1.2in]{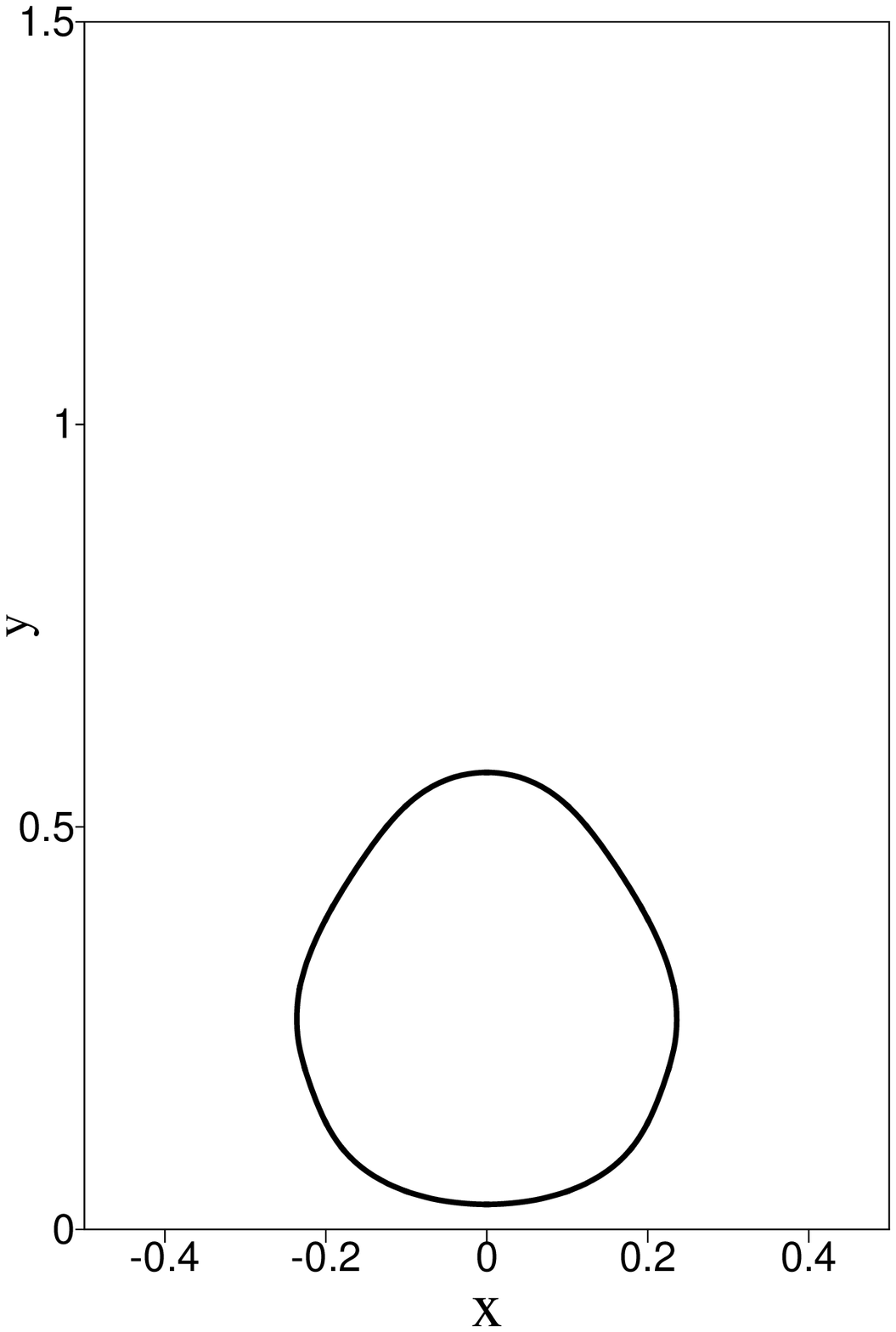}(g)
\includegraphics[width=1.2in]{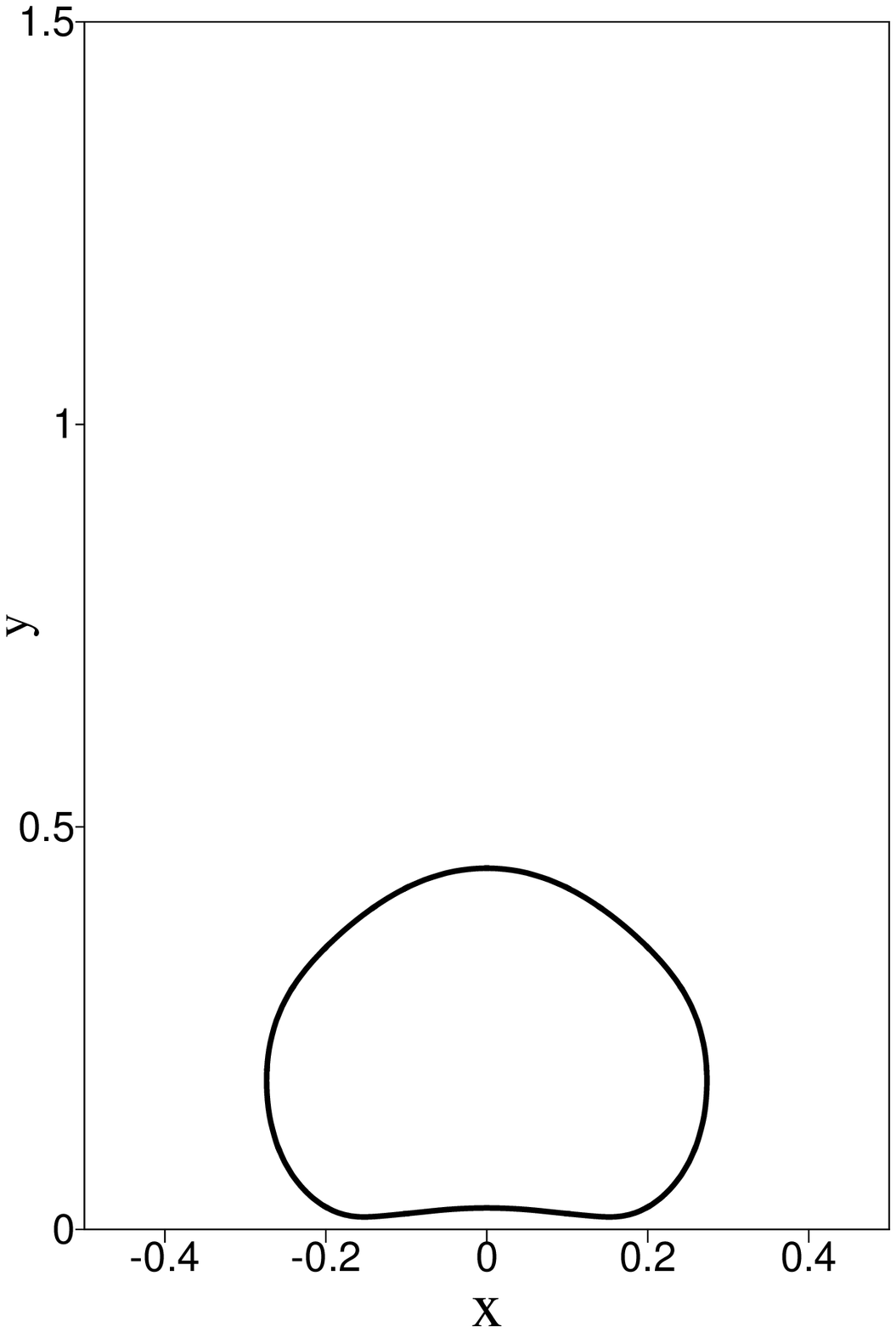}(h)
}
\caption{
Bouncing water drop (initial height $H_0=3.2$mm):
temporal sequence of snapshots of the air-water 
interface at time instants:
(a) $t=0.05$,
(b) $t=0.25$,
(c) $t=0.4$,
(d) $t=0.55$,
(e) $t=0.7$,
(f) $t=0.85$,
(g) $t=1.0$,
(h) $t=1.15$.
}
\label{fig:bouncing_phase_snap}
\end{figure}

The goal of this section is to further
evaluate and demonstrate the 
accuracy of the method developed here
by comparing simulation
results with the experimental measurement.
The test problem considered in this section
involves large density ratio, large viscosity
ratio, and 
superhydrophobic walls 
(i.e. contact angle $\geqslant 150^0$).
A similar problem but under a different 
condition has been considered in a previous
work \cite{Dong2012}.

\begin{table}
\begin{center}
\begin{tabular*}{0.8\textwidth}{@{\extracolsep{\fill}} l  l  l}
\hline
Density $[kg/m^3]$ & air: $1.2041$ & water: $998.207$ \\
Dynamic viscosity $[kg/(m\cdot s)]$ & air: $1.78 \times 10^{-5}$
& water: $1.002\times 10^{-3}$ \\
Surface tension $[kg/s^2]$ & air-water: $7.28 \times 10^{-2}$ \\
Gravity $[m/s^2]$ & $9.8$ \\
\hline
\end{tabular*}
\end{center}
\caption{Physical properties of air and water.}
\label{tab:air_water_param}
\end{table}


We consider a rectangular domain 
(see Figure \ref{fig:bouncing_phase_snap}(a)),
$-\frac{L}{2} \leqslant x \leqslant \frac{L}{2}$
and $0\leqslant y\leqslant \frac{3L}{2}$,
where $L=4$mm.
The domain is periodic in the horizontal direction
at $x=\pm \frac{L}{2}$. 
The top and bottom of the domain are two superhydrophobic
solid walls. 
If the air-water interface
intersects the walls, the contact angle
is assumed to be $170^0$.
The domain is initially filled with air.
A water drop, initially circular with
a radius $R_0=\frac{L}{4}$, is suspended in
the air. The center of the water
drop is initially located at a height $H_0$ above
the bottom wall, that is, 
$(x_0,y_0)=(0,H_0)$,
where $(x_0,y_0)$ is the coordinate of
the center of mass of the water drop.
The gravity is assumed to be in the $-y$ direction.
At $t=0$, the water drop is released,
and falls through the air, impacting 
and bouncing off the bottom wall.
The objective of this problem is to simulate and study
the behavior of the water drop.

The physical properties of 
the air, water and the air-water interface
employed in this problem
are listed in Table \ref{tab:air_water_param}.
The air and the water are respectively assigned
as the first and the second fluids in 
the simulations.
We use $L$ as the characteristic length scale,
and choose the characteristic velocity scale
$U_0 = \sqrt{g_{r0}L}$, where $g_{r0}=1m/s^2$. 
The problem is then non-dimensionalized 
according to Table \ref{tab:normalization}.

\begin{table}
\begin{center}
\begin{tabular*}{1.0\textwidth}{@{\extracolsep{\fill}} l l  | ll}
\hline
parameters & values & parameters & values \\
$\rho_2/\rho_1$ & $829.01$ & $\rho_m$ & $\min(\rho_1,\rho_2)$ \\
$\mu_1/(\rho_1 U_0 L)$ & $5.843\times 10^{-2}$ & $\nu_m$ 
& $\frac{1}{2}\left(\frac{\mu_1}{\rho_1}+\frac{\mu_2}{\rho_2}  \right)$  \\
$\mu_2/\mu_1$ & $56.29$ & $\Delta t U_0/L$ & $2.5\times 10^{-5}$ \\
$\eta/L$ & $0.01$ & $\theta_s$ & $10^0$ \\
$\sigma/(\rho_1 U_0^2 L)$ & $3778.76$ 
& $(\gamma_1\rho_1 U_0)/L$ & $\left(\frac{\eta}{L}\right)^3\frac{1}{\lambda/(\rho_1U_0^2L^2)}$ \\
$\lambda/(\rho_1U_0^2L^2)$ & $\frac{3}{2\sqrt{2}}\frac{\sigma}{\rho_1U_0^2L}\frac{\eta}{L}$ & $g_rL/U_0^2$ (gravity) & $9.8$ \\
Number of elements in mesh & $150$ & Element order & $14$ \\
$H_0$ (water drop initial height) & (varied) &
$J$ (integration order) & $2$ \\
\hline
\end{tabular*}
\end{center}
\caption{Physical and numerical parameter values for the
bouncing water drop problem.}
\label{tab:bouncing_param}
\end{table}


To simulate the problem we discretize the domain
using $150$ equal-sized quadrilateral elements,
with $10$ and $15$ elements in the $x$ and $y$
directions respectively.
We use an element order $14$ for all elements
in the simulations.
The algorithm presented in Section \ref{sec:method}
is employed for marching in time,
with a non-dimensional time 
step size $\Delta t=2.5\times 10^{-5}$. 
%
%
In the horizontal direction we employ
periodic boundary conditions for all flow variables.
At the top and the bottom walls, 
we impose the Dirichlet condition 
\eqref{equ:dbc_v} with $\mathbf{w}=0$
for the velocity, and impose
the contact-angle boundary
conditions \eqref{equ:wbc_phi_1}--\eqref{equ:wbc_phi_2}
with $g_{c1}=g_{c2}=0$ and $\theta_s=10^0$
for the phase field function.
Note that $\theta_s$ in \eqref{equ:wbc_phi_2}
is the angle measured on the side of 
the first fluid, that is, the air for
the current configuration.
%
%
We employ the following initial velocity and
phase field function distributions
\begin{equation}
\mathbf{u}_{in} = 0, \quad
\phi_{in} = \tanh\frac{\left\|\mathbf{x}-\mathbf{X}_0 \right\|-R_0}{\sqrt{2}\eta},
\end{equation}
where $\mathbf{X}_0=(x_0,y_0)$ is the initial
coordinate of the center of  mass of 
the water drop.
%
The values for the physical and numerical parameters
of this problem are summarized in 
Table \ref{tab:bouncing_param}.


Let us first look into the dynamics of 
this air-water two-phase system.
Figure \ref{fig:bouncing_phase_snap}
shows a temporal sequence of snapshots of 
the air-water interface.
The initial height of the water drop
is $H_0=3.2$mm above the bottom wall.
The interface is visualized by the contour levels
$\phi = 0$ at different time instants.
Upon release, the water drop falls
through the air
(Figures \ref{fig:bouncing_phase_snap}(a)--(b)),
and impacts the bottom wall (Figure \ref{fig:bouncing_phase_snap}(c)).
One can observe a notable deformation of
the water drop upon impact of the wall.
Subsequently, the water drop
bounces off the bottom wall (Figure \ref{fig:bouncing_phase_snap}(d))
and rises through the air, reaching 
a maximum height (Figure \ref{fig:bouncing_phase_snap}(e)).
Then the drop falls through the air again
and impacts the bottom wall a second 
time (Figures \ref{fig:bouncing_phase_snap}(f)--(h)).
This process repeats several times,
and the water drop eventually settles down
on the bottom wall.


\begin{figure}
\centerline{
\includegraphics[height=3in]{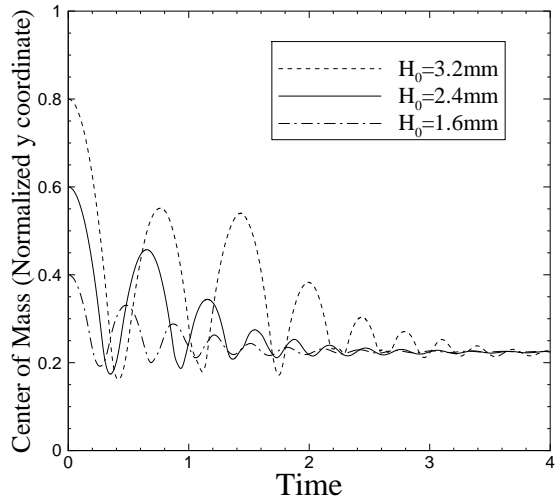}
}
\caption{
Time histories of the water-drop center of mass (y coordinate)
corresponding to several different initial drop heights.
}
\label{fig:bouncing_cofm}
\end{figure}

We have 
monitored the motion of the center of mass of the water drop
for different values of the initial drop height,
ranging from $H_0=1.6$mm to $4$mm.
The drop center of mass is defined by
\begin{equation}
\mathbf{X}_w = (x_w, y_w) = 
\frac{
\int_{\Omega_w} \mathbf{x} d\mathbf{x}
}{
\int_{\Omega_w} d\mathbf{x}
}
\end{equation}
where $\Omega_w(t)$ is the domain occupied by
the water drop at time $t$ and demarcated by
the contour level $\phi=0$.
In Figure \ref{fig:bouncing_cofm}
we show the time histories of the $y$ component (normalized)
of the drop center of mass
for several values of the initial drop height $H_0$.
It can be discerned that the water drop bounces off the 
bottom wall a number of times in all these cases.
One can also discern the oscillation in the drop shape
in later time, before it completely settles down
on the wall.
For larger values of the initial drop height,
 we notice  that
occasionally the water drop can reach a maximum height after
a bounce that
is quite close to that before the bounce; see
for instance the second and third peaks in
the curve for $H_0=3.2$mm of Figure \ref{fig:bouncing_cofm}.
This is likely due to the fact that
a larger initial drop height tends to cause
a more pronounced deformation of the water drop
upon impact  and a more 
pronounced oscillation in the drop shape 
after the bounce-off into the air (see Figures
\ref{fig:bouncing_phase_snap}(c)--(g)).
The elastic energy associated with the drop deformation
can be converted to the kinetic energy
associated with the motion of the drop
center of mass in subsequent impact and lift-off,
thus resulting in a maximum height close to
that before the bounce.


\begin{figure}
\centerline{
\includegraphics[height=3in]{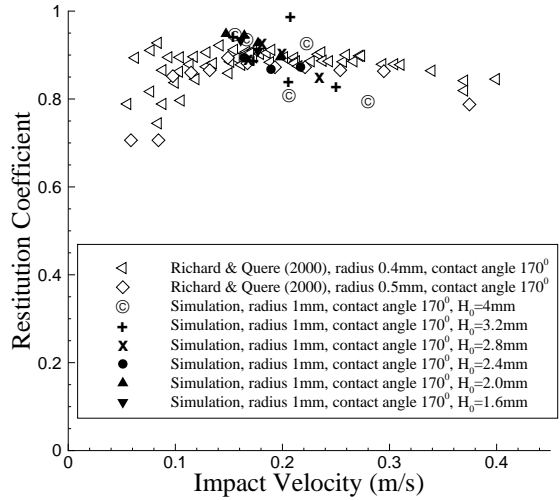}
}
\caption{
Comparison of restitution coefficient as a function
of impact velocity between current simulations
and the experiment \cite{RichardQ2000}.
$H_0$ is the initial height of the water drop.
}
\label{fig:bouncing_rest_coeff}
\end{figure}

We have computed the restitution coefficient 
based on the time histories of the center of mass
of the water drop. We follow \cite{RichardQ2000}
and define the restitution coefficient $C_{res}$ by
\begin{equation}
C_{res} = \sqrt{\frac{H^{\prime}}{H}}
\end{equation} 
where $H$ and $H^{\prime}$
respectively denote 
the maximum heights of the water drop
above the bottom wall
before and after the bounce.
We also follow \cite{RichardQ2000} and
estimate the impact velocity of the water
drop $V_{imp}$ by
\begin{equation}
V_{imp} = \sqrt{2g_r H}
\end{equation}
where $g_r$ is the gravitational acceleration.

In Figure \ref{fig:bouncing_rest_coeff}
we plot the restitution coefficient $C_{res}$ as a function
of the impact velocity $V_{imp}$
from the current simulations.
For comparison, we have also shown
the restitution coefficient data from the 
experiment of \cite{RichardQ2000}.
The restitution coefficients 
corresponding to different initial drop heights $H_0$ 
from the simulations
have been included in this figure.
The drop size in the simulations is a little
larger than that in the experiment of \cite{RichardQ2000}.
The bulk of the restitution coefficients from
current simulations agree  well with the experimentally
determined values. 
On the other hand, some differences can  be observed,
especially for the data points corresponding
to the first few bounces with larger initial
drop-height values.
We observe that for such cases
the restitution coefficients from the simulation
tend to be a little smaller than the bulk of
the experimental data. 
This is likely due to the larger drop deformation upon impact and
stronger drop-shape oscillation after the bounce-off,
associated with a larger initial drop height and a larger
impact velocity.
The elastic energy associated with
the drop deformation may reduce the maximum height
the drop can reach after the bounce, and thus
results in a smaller restitution coefficient.
The outlying data point,
with a large restitution coefficient, 
from the simulation case with an
initial drop height $H_0=3.2$mm
corresponds to the second and the third
peaks in the time-history curve in 
Figure \ref{fig:bouncing_cofm},
which has been discussed in a previous paragraph.

The above comparison indicates  that 
the simulation results obtained using our
method overall are in good agreement with
the experimental measurement.

\subsection{Air Jet in Water with Two-Phase Open Boundaries}


\begin{figure}
\centerline{
\includegraphics[height=3in]{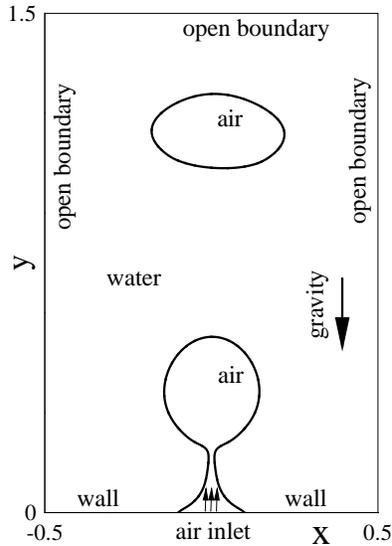}
}
\caption{
Configuration of the air jet in water problem.
}
\label{fig:airjet_config}
\end{figure}

The goal of this section is to demonstrate the effectiveness
of the open boundary conditions and
the numerical algorithm from
Section \ref{sec:method} for two-phase
outflow problems.
The test problem considered in this section
involves open boundaries where the two fluids 
may leave or enter
the domain, large density contrast,
and large viscosity contrast. 
The fluid interface passes through 
the open domain boundary in this problem.

We consider the long-time behavior of an air-water
two-phase flow, 
in which a train of air bubbles continually
forms at a wall inside the water 
and then moves out of the domain
due to buoyancy. 
This flow problem has  been considered
in a previous work \cite{Dong2014obc}. It should be noted that
the open boundary conditions and the numerical
algorithm being tested here are different.

Specifically, we consider the flow
domain shown in Figure \ref{fig:airjet_config},
$-\frac{L}{2}\leqslant x\leqslant \frac{L}{2}$
and $0\leqslant y\leqslant \frac{3L}{2}$,
where $L=3$cm.
The bottom of the domain is a solid wall,
while the other three sides (top, left and right)
are all open, where the fluid can freely leave or
enter the domain.
The domain is initially filled with water,
and the gravity is along the vertical direction
pointing downward.
The bottom wall has an orifice in its center,
with a diameter $d=6$mm.
A stream of air is continuously injected
into the domain through the orifice.
The air velocity has a parabolic profile at
the orifice, with a centerline value
$U_0=17.3cm/s$.
The bottom wall has a neutral wettability,
that is, 
if the air-water interface intersects the wall
 the contact angle would be $90^0$.
Our objective is to simulate and study
the long-time behavior of this system.
The two-phase open boundaries coupled with
the large density ratio between air and water
make this problem very challenging to 
simulate.


The physical parameters concerning the air, water
and the air-water interface have been provided in
Table \ref{tab:air_water_param}. 
We treat the air and the water 
as the first and the second fluids,
respectively.
$L$ and $U_0$ are employed respectively as the characteristic
length and velocity scales.
Normalization of the problem then proceeds according
to Table \ref{tab:normalization}.


The flow domain is discretized using
$600$ quadrilateral spectral elements,
with $20$ and $30$ elements in
the $x$ and $y$ directions respectively.
An element order $12$ has been used for
all elements in the simulations.
%
At the bottom wall, we impose 
the velocity Dirichlet condition \eqref{equ:dbc_v}
with $\mathbf{w}=0$
and the boundary conditions 
\eqref{equ:wbc_phi_1}--\eqref{equ:wbc_phi_2}
with $g_{c1}=g_{c2}=0$ and $\theta_s=90^0$.
At the air inlet we impose 
the velocity Dirichlet condition \eqref{equ:dbc_v}, 
in which $\mathbf{w}$ has zero horizontal component and its
 vertical component takes
 a parabolic profile with a centerline
value $U_0$;
for the phase field function, we impose the 
boundary conditions \eqref{equ:ibc_phi_1}--\eqref{equ:ibc_phi_2},
in which $g_b=0$ and 
\begin{equation}
\phi_b(\mathbf{x},t) = 
-\tanh \frac{x-R}{\sqrt{2}\eta} \left[
    H(x,0) - H(x,R)
  \right]
+ \tanh \frac{x+R}{\sqrt{2}\eta} \left[
    H(x,-R) - H(x,0)
  \right]
\end{equation}
where $R=\frac{d}{2}=3mm$ is the radius of 
the orifice,
and $H(x,a)$ is the heaviside step function taking
unit value if $x\geqslant a$ and vanishing otherwise.
%
On the top, left and right sides of the domain,
we impose the open boundary condition \eqref{equ:obc_v_trans}
with $\mathbf{f}_b=0$
for the momentum equation; 
for the phase field function we impose
the boundary conditions 
\eqref{equ:obc_phi_1}--\eqref{equ:obc_phi_2}
with $g_{a1}=g_{a2}=0$.
%
For the initial conditions, we have used an instantaneous
snapshot of the velocity field and the phase field function
from the simulation of \cite{Dong2014obc}. 
Because long-time simulations have been performed, 
 the initial velocity and phase field distributions
have no effect on the long-time behavior
of the system.

\begin{table}
\begin{center}
\begin{tabular*}{1.0\textwidth}{@{\extracolsep{\fill}} l l  | ll}
\hline
parameters & values & parameters & values \\
$\rho_2/\rho_1$ & $829.01$ & $\rho_m$ & $\min(\rho_1,\rho_2)$ \\
$\mu_1/(\rho_1 U_0 L)$ & $2.845\times 10^{-2}$ & $\nu_m$ 
& $50\max\left(\frac{\mu_1}{\rho_1},\frac{\mu_2}{\rho_2}  \right)$  \\
$\mu_2/\mu_1$ & $56.29$ & $\Delta t U_0/L$ & $1.5\times 10^{-6}$ \\
$\eta/L$ & $0.01$ & $\theta_s$ & $90^0$ \\
$\sigma/(\rho_1 U_0^2 L)$ & $67.178$ 
& $(\gamma_1\rho_1 U_0)/L$ & $0.1\left(\frac{\eta}{L}\right)^3\frac{1}{\lambda/(\rho_1U_0^2L^2)}$ \\
$\lambda/(\rho_1U_0^2L^2)$ & $\frac{3}{2\sqrt{2}}\frac{\sigma}{\rho_1U_0^2L}\frac{\eta}{L}$ & $g_rL/U_0^2$ (gravity) & $9.8$ \\
$D_0 U_0$ & $0.4$ & $\mu_0$ & $20\max (\mu_1, \mu_2)$ \\
Number of elements in mesh & $600$ & Element order & $12$ \\
$-\Delta P/L$ (external pressure gradient) & $\rho_w g_r$ 
& $J$ (integration order) & $2$   \\
$\delta$ & $0.01$ \\
\hline
\end{tabular*}
\end{center}
\caption{Physical and numerical parameter values for the
air jet in water problem.}
\label{tab:airjet_param}
\end{table}

We apply an external pressure gradient 
in the $y$ direction ($-\frac{\Delta P}{L}$) to balance
the weight of water in the simulations, i.e.
\begin{equation}
-\frac{\Delta P}{L} = \rho_w g_r,
\end{equation}
where $\rho_w$ is the water density and
$g_r$ is the magnitude of the gravitational acceleration.


Table \ref{tab:airjet_param} summarizes
the physical and numerical parameter values
in the simulations for this problem.
The $D_0$ in the open
boundary condition \eqref{equ:obc_phi_2}
for the phase field function
is determined based on a preliminary
simulation with $D_0=0$.
Preliminary simulations indicate that
the air bubbles have a 
non-dimensional convection velocity
about $2.0 \sim 3.0$
 at the upper domain
boundary. Because $\frac{1}{D_0}$ plays 
the role of a convection velocity,
we therefore use
an outflow dynamic mobility
$\frac{1}{D_0 U_0} \approx 2.5$ in
the simulations.
A non-dimensional time step size
$1.5 \times 10^{-6}$
has been employed for the current
problem.



\begin{figure}
\centering
\includegraphics[width=3.5in]{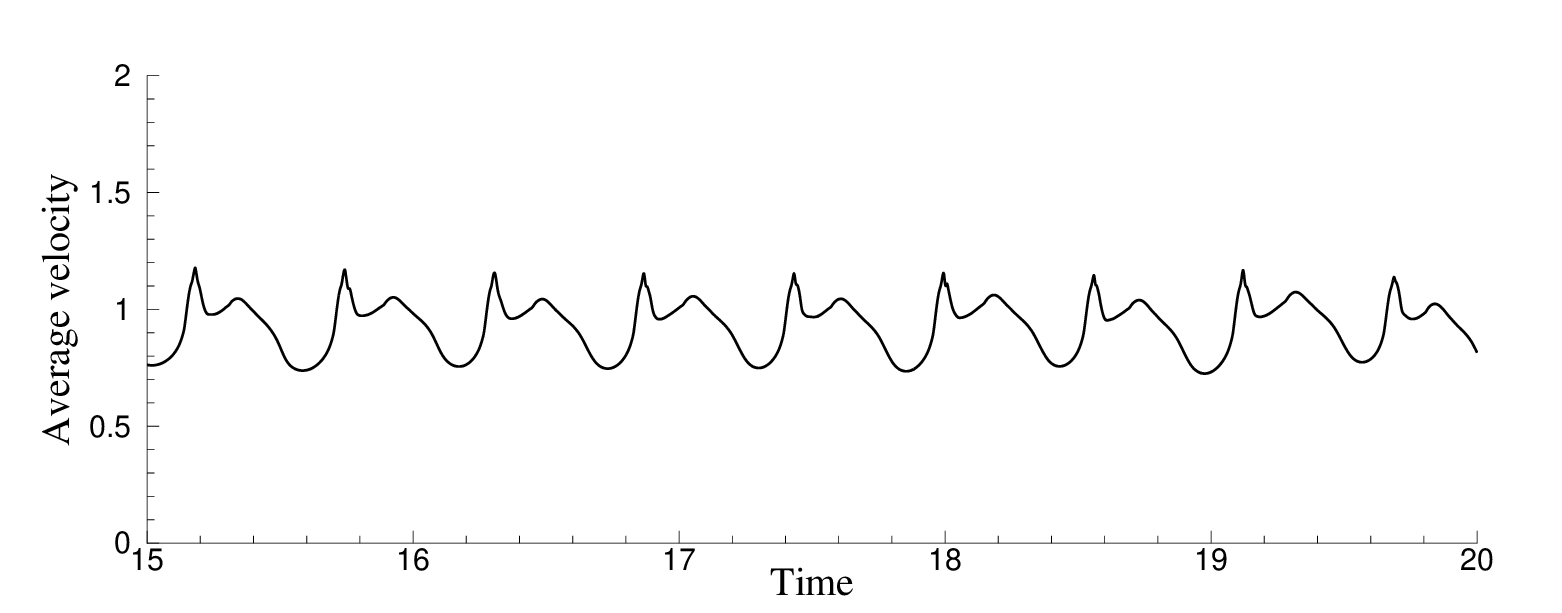}(a)
\includegraphics[width=3.5in]{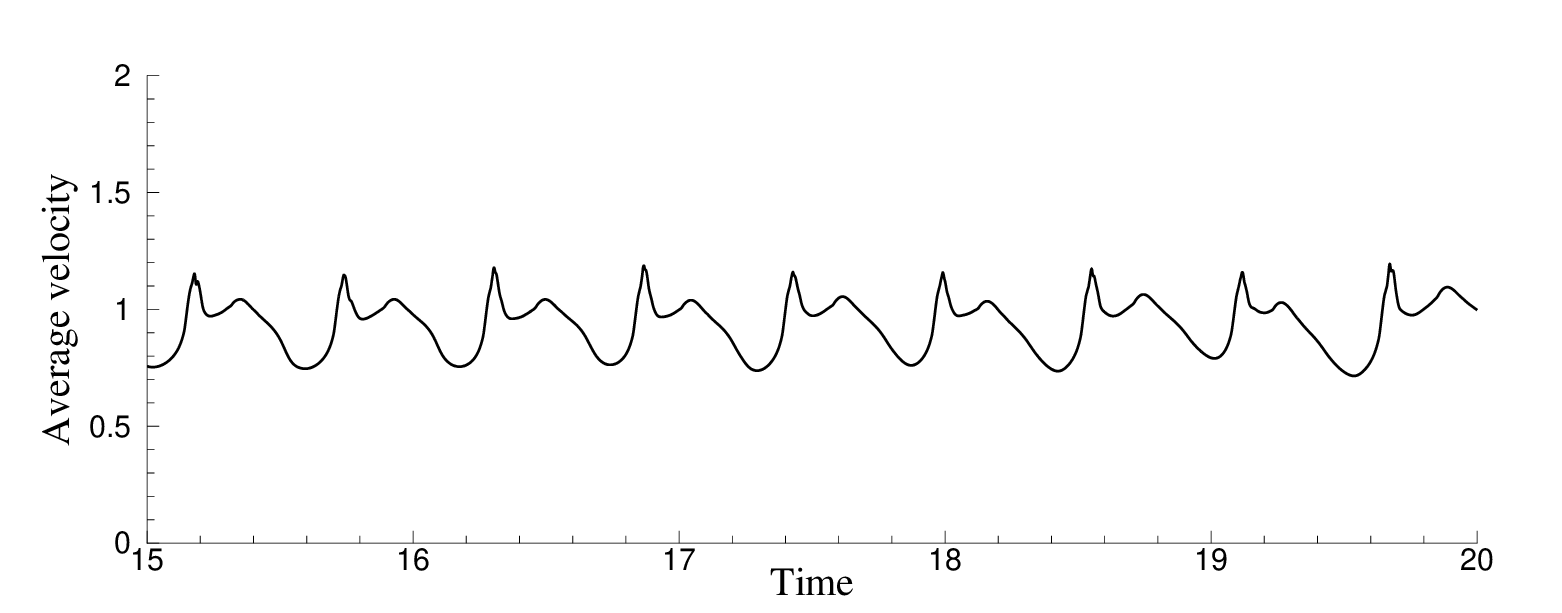}(b)
\includegraphics[width=3.5in]{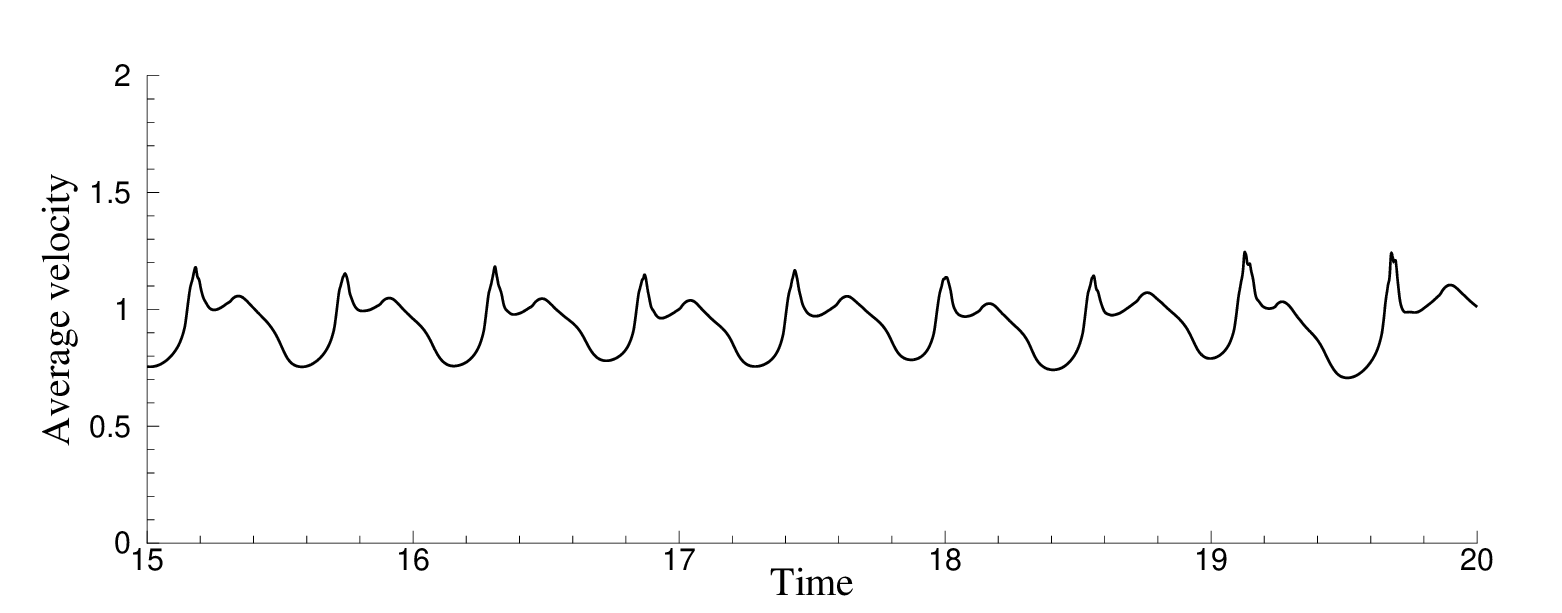}(c)
\includegraphics[width=3.5in]{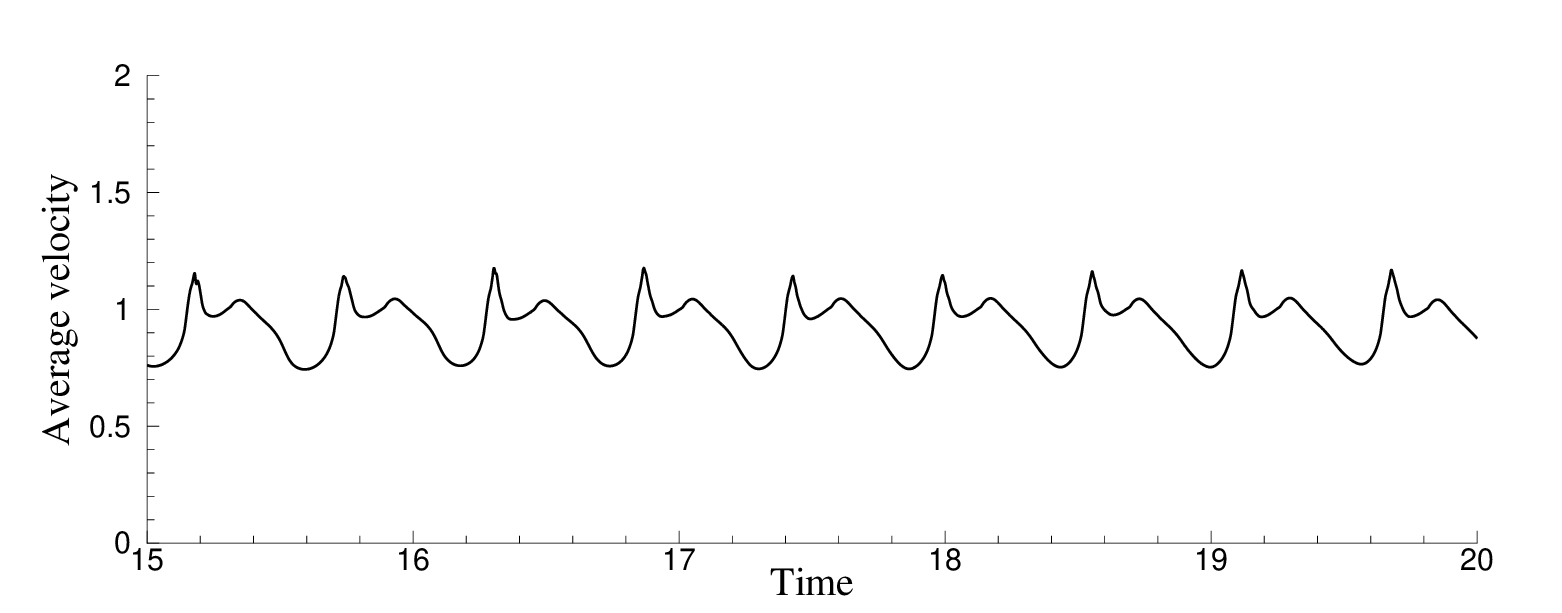}(d)
\caption{
Time histories of average vertical-velocity magnitude
obtained using different open boundary conditions:
(a) OBC \eqref{equ:obc_v_C},
(b) OBC \eqref{equ:obc_v_D}, 
(c) OBC \eqref{equ:obc_v_A},
(d) OBC \eqref{equ:obc_v_B},
}
\label{fig:airjet_vel_hist}
\end{figure}

Let us first demonstrate the long-term stability of the
computation.
We have performed long-time simulations of
this problem using different open boundary conditions.
Figure \ref{fig:airjet_vel_hist}
shows a window of the time histories of
the average vertical velocity magnitudes $V_{\text{avg}}(t)$,
\begin{equation}
V_{\text{avg}}(t) = \left(
  \frac{1}{V_{\Omega}}\int_{\Omega} |v|^2 d\Omega
\right)^{\frac{1}{2}},
\end{equation}
where $v$ is the $y$ velocity component and
$V_{\Omega}=\int_{\Omega}d\Omega$ is
the volume of the domain.
Results in Figures \ref{fig:airjet_vel_hist}(a)--(d)
are obtained using the open boundary conditions
\eqref{equ:obc_v_C}--\eqref{equ:obc_v_B},
respectively.
One can make two observations.
First, the average velocity magnitude $V_{\text{avg}}$
fluctuates over time about some constant mean
level and its time history signal exhibits
a quasi-periodic nature.
This indicates that the flow is at a statistically 
stationary state, and that 
the computations using our algorithm and the several outflow
boundary conditions are stable over a long time.
Second, the time-history curves
obtained with different open boundary
conditions \eqref{equ:obc_v_C}--\eqref{equ:obc_v_B}
are qualitatively similar,
indicating that these two-phase boundary conditions
lead to similar results about the flow.


\begin{figure}
\centerline{
\includegraphics[width=1.4in]{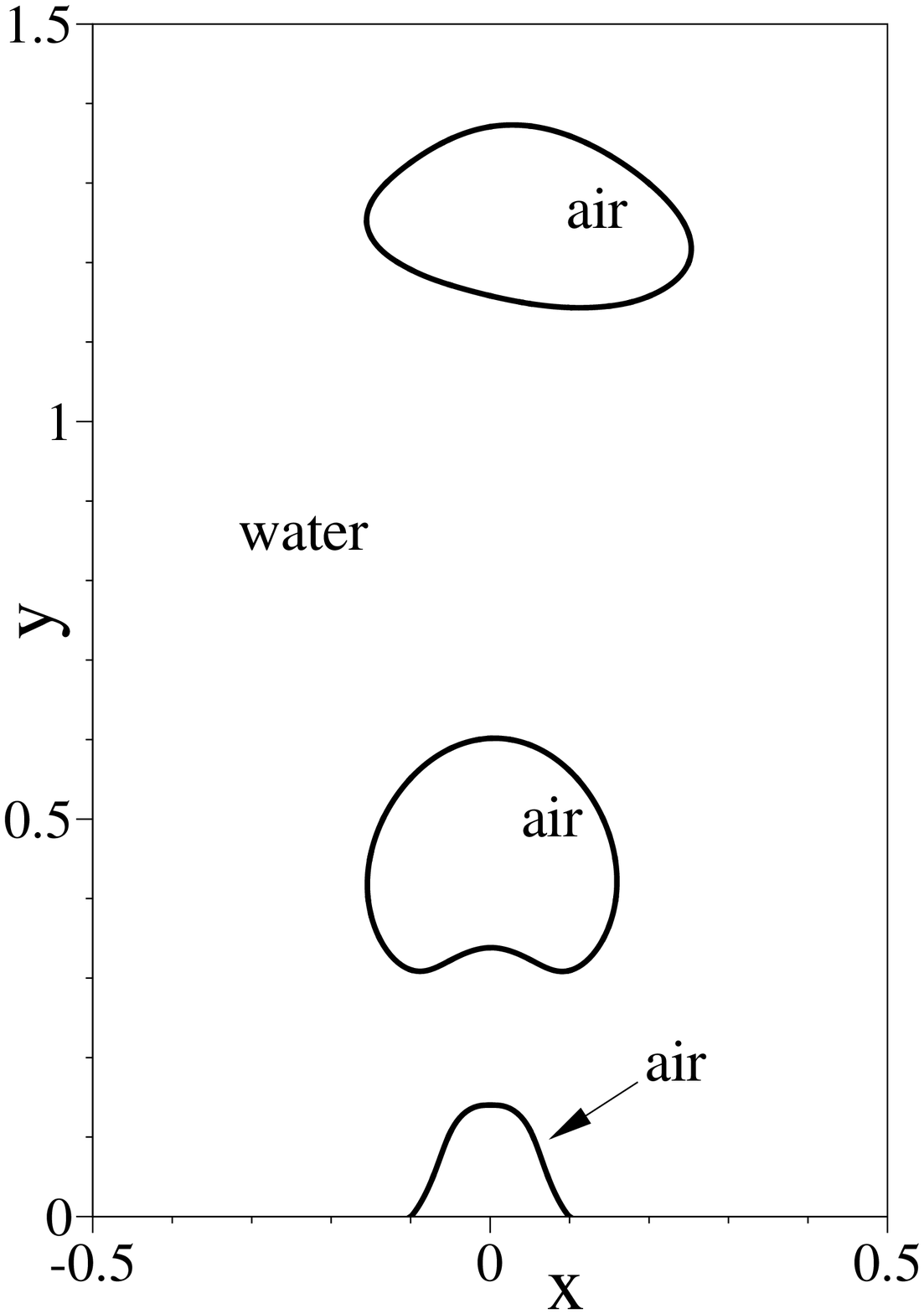}(a)
\includegraphics[width=1.4in]{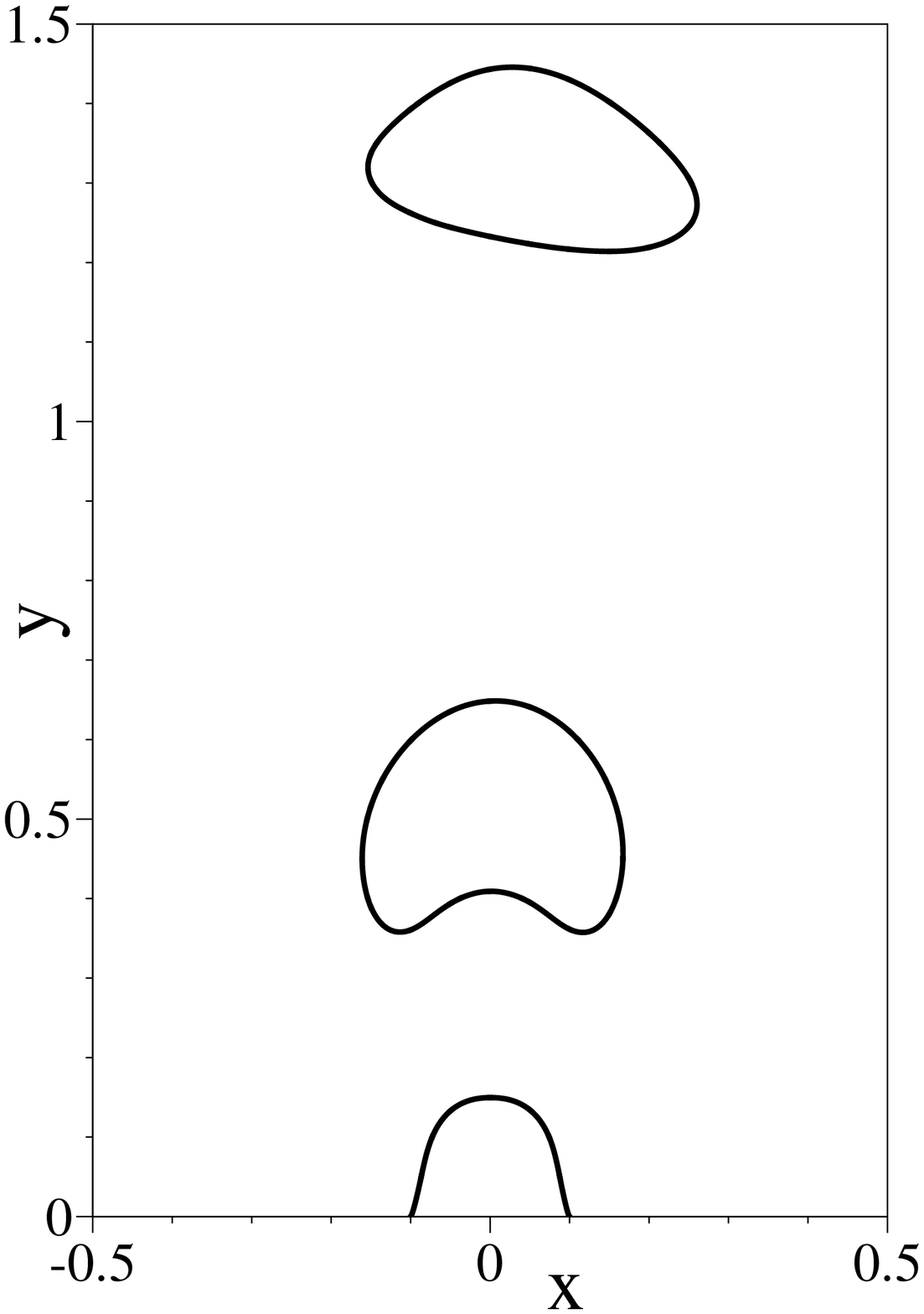}(b)
\includegraphics[width=1.4in]{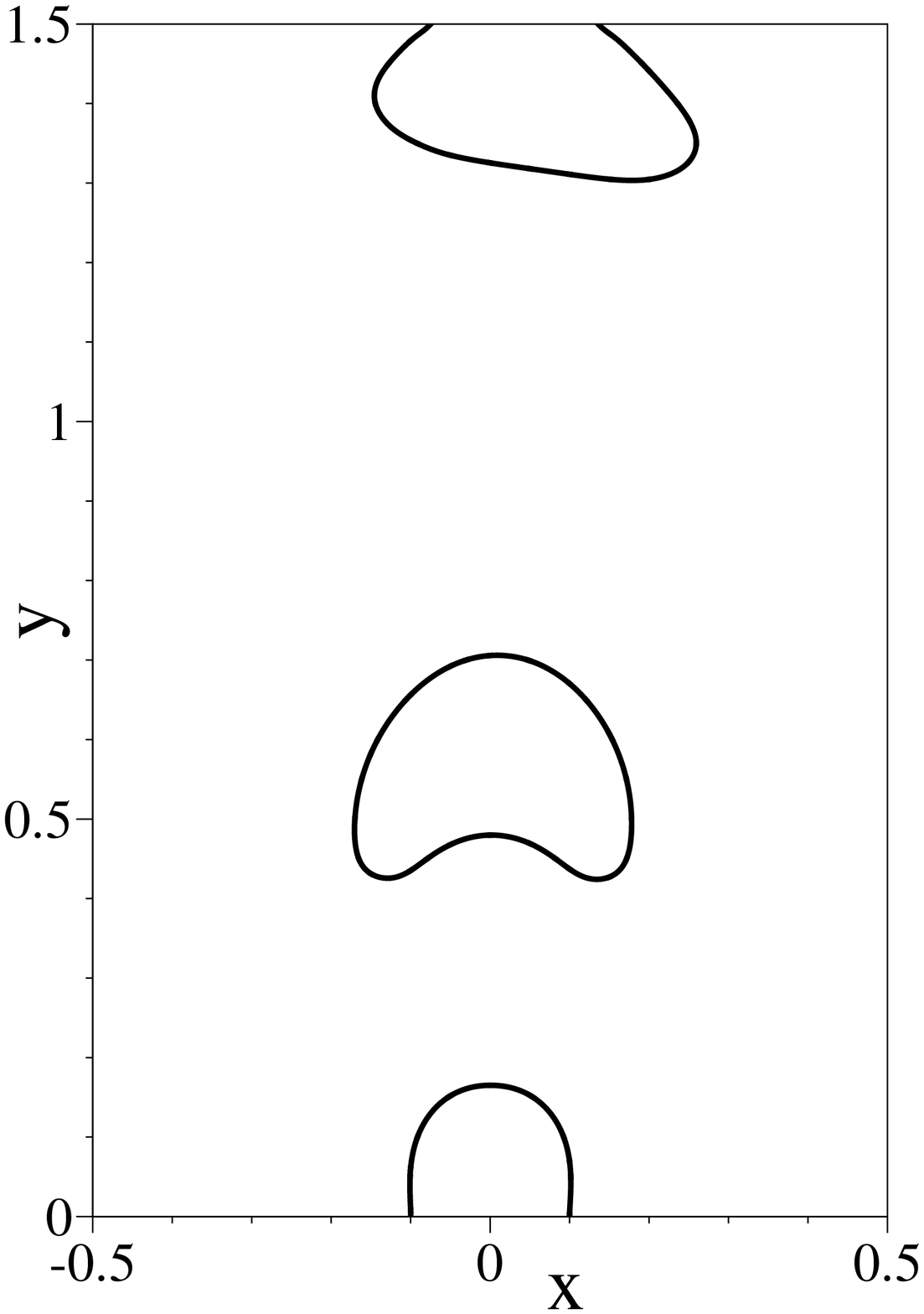}(c)
\includegraphics[width=1.4in]{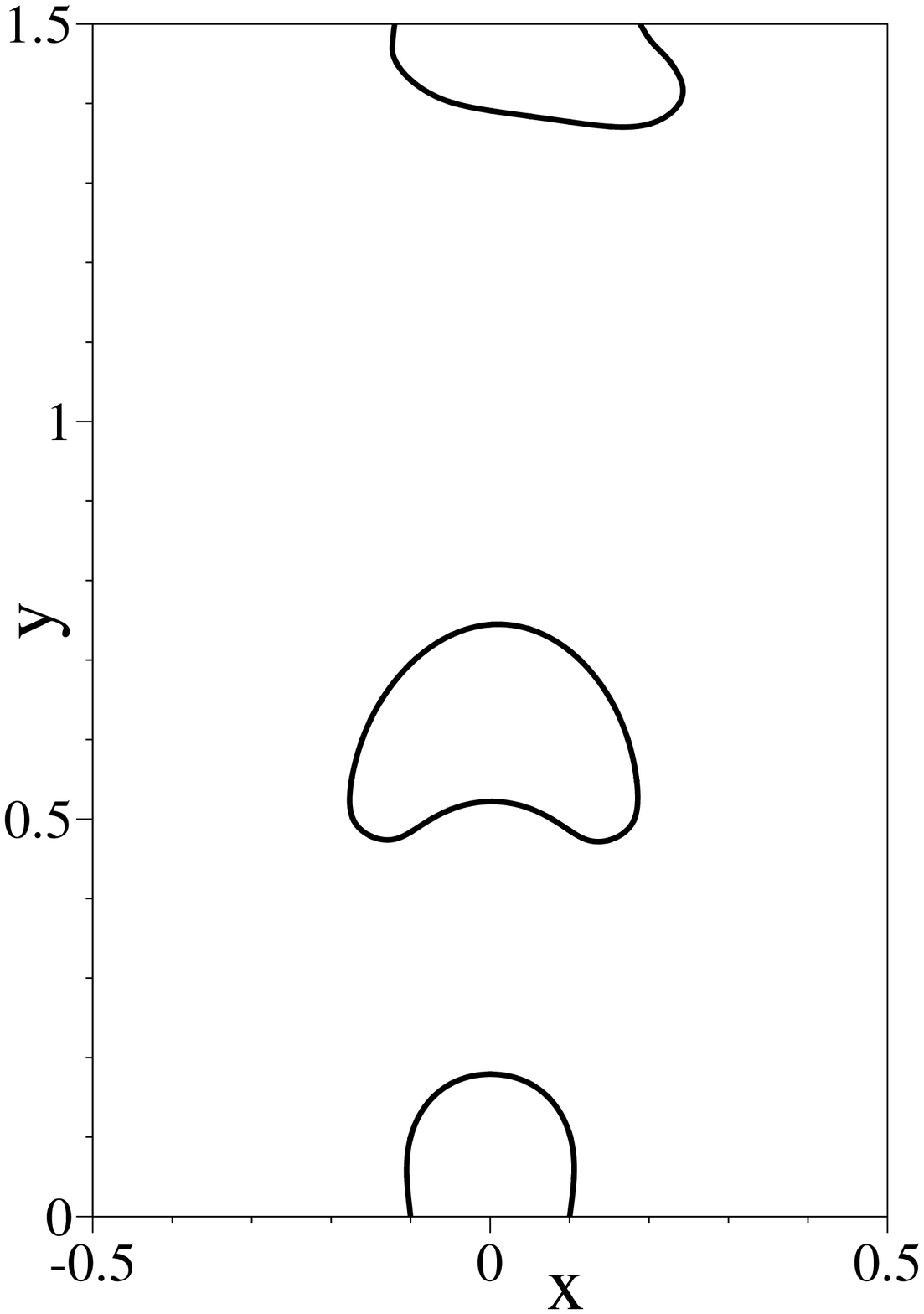}(d)
}
\centerline{
\includegraphics[width=1.4in]{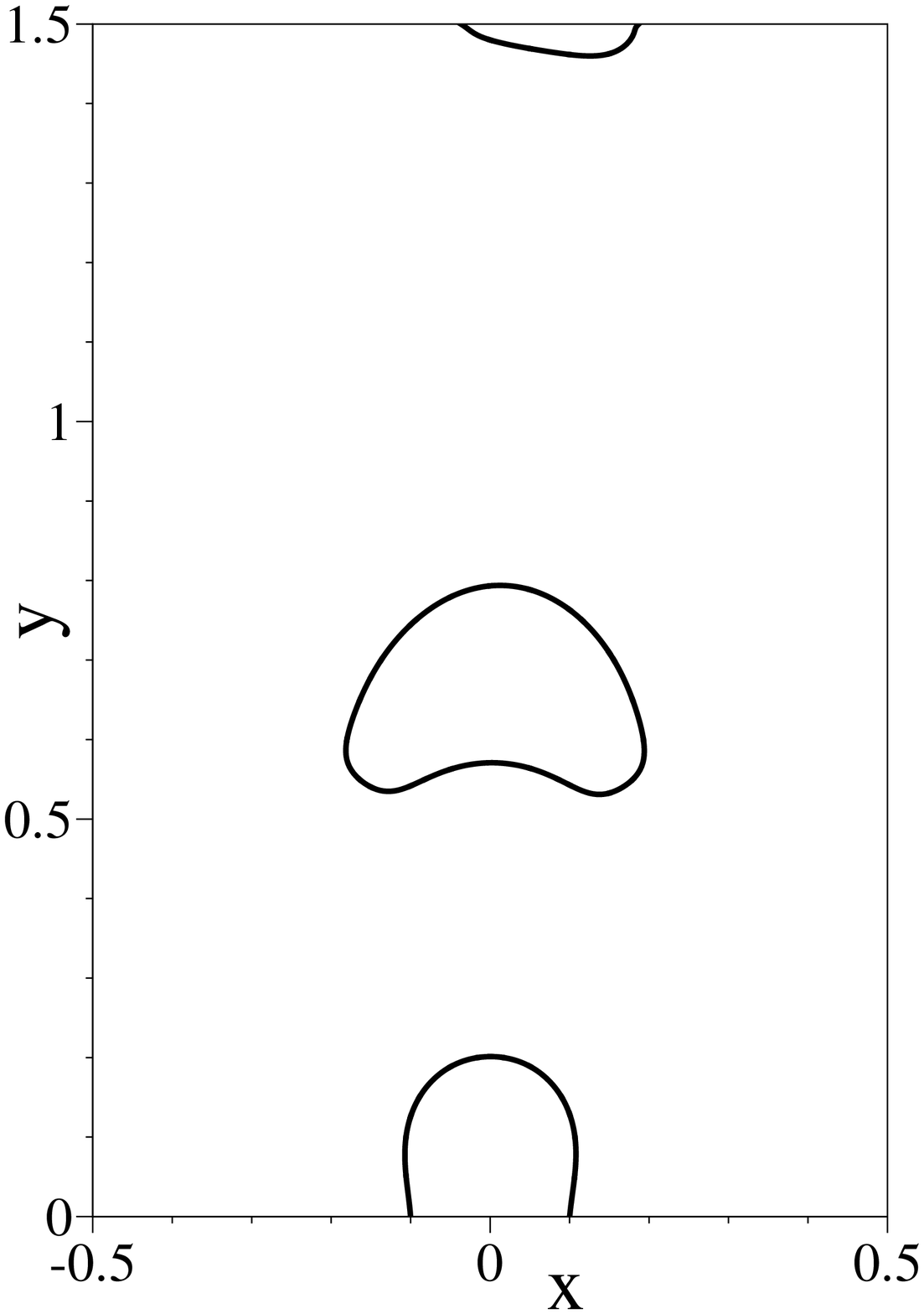}(e)
\includegraphics[width=1.4in]{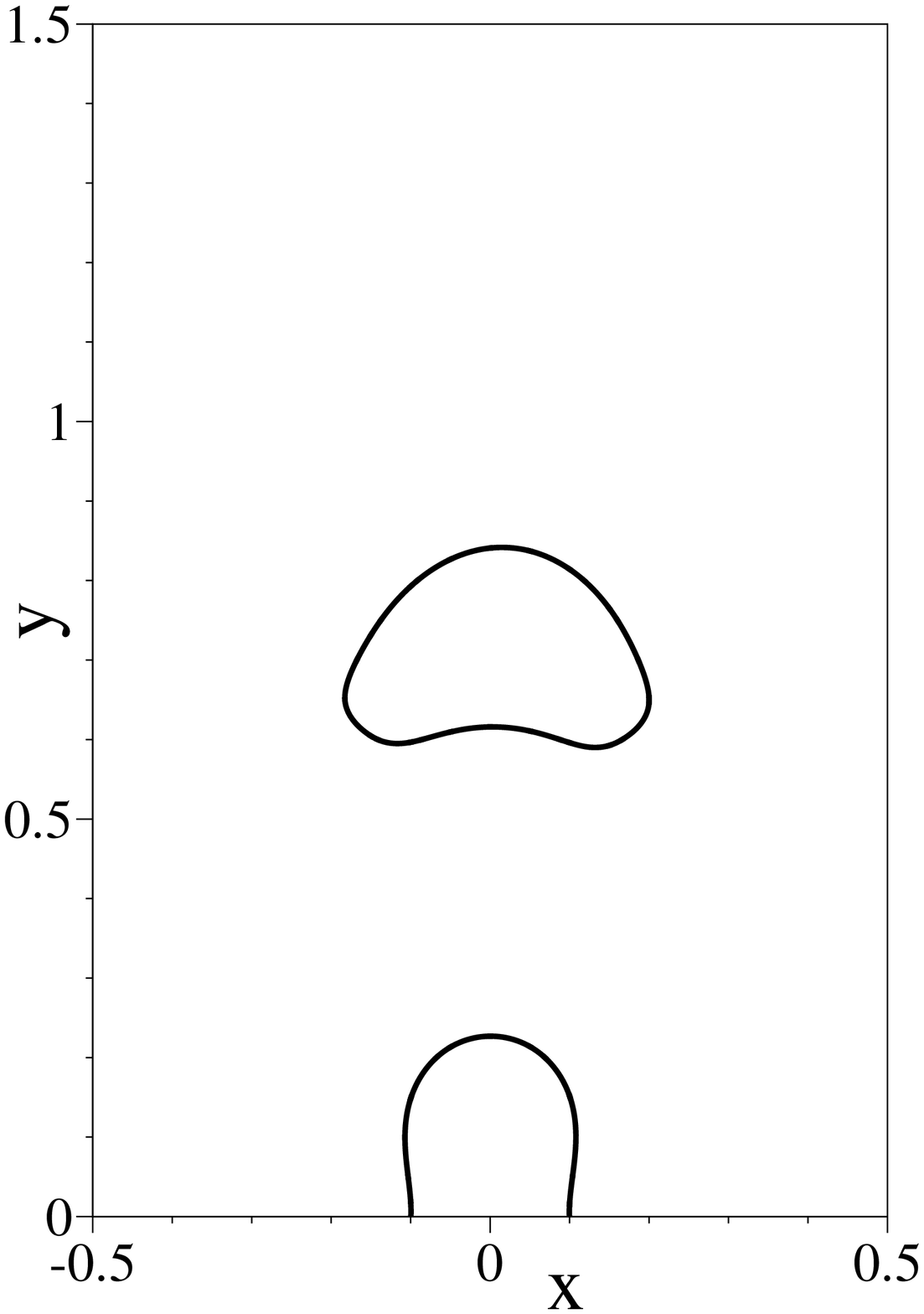}(f)
\includegraphics[width=1.4in]{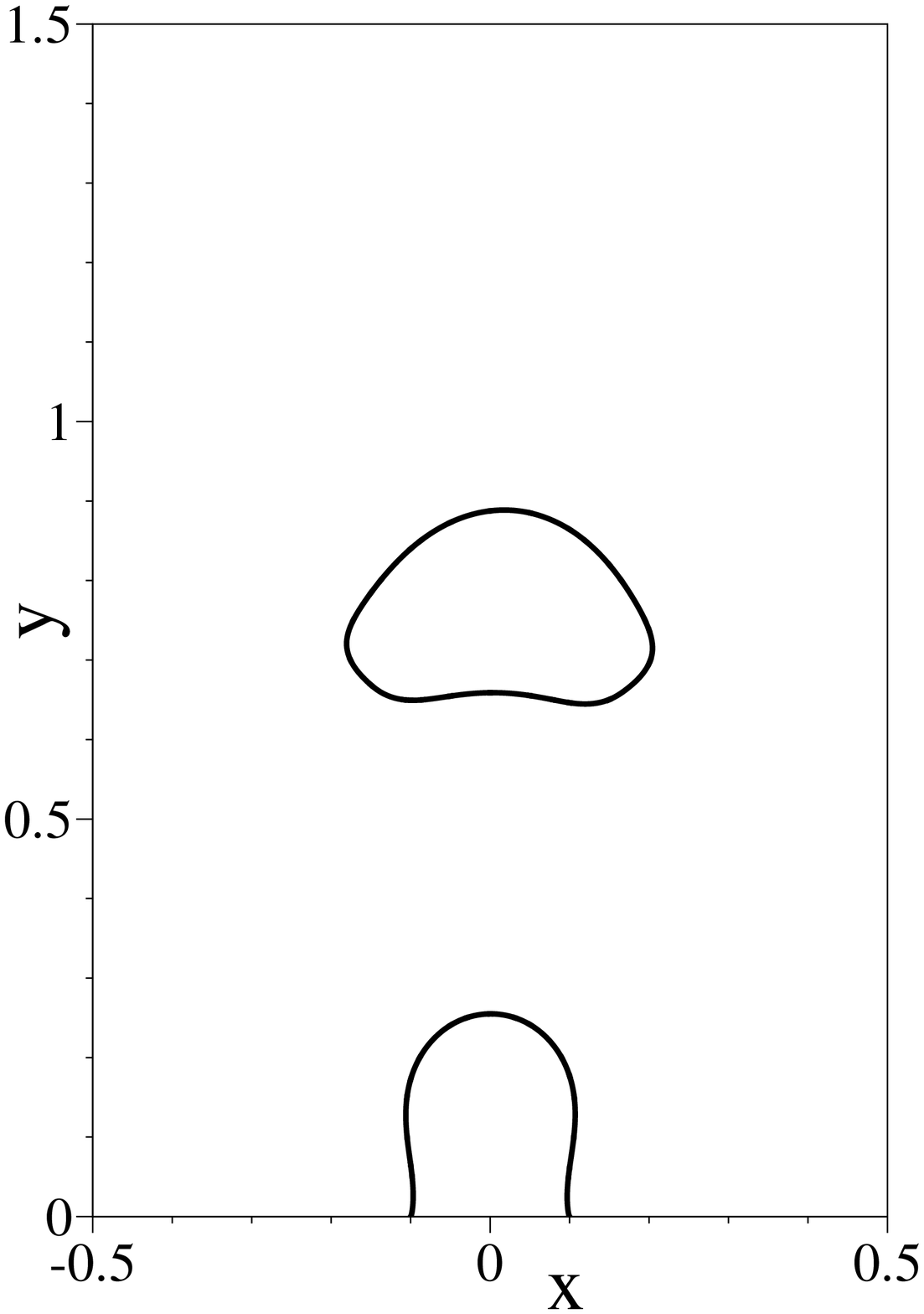}(g)
\includegraphics[width=1.4in]{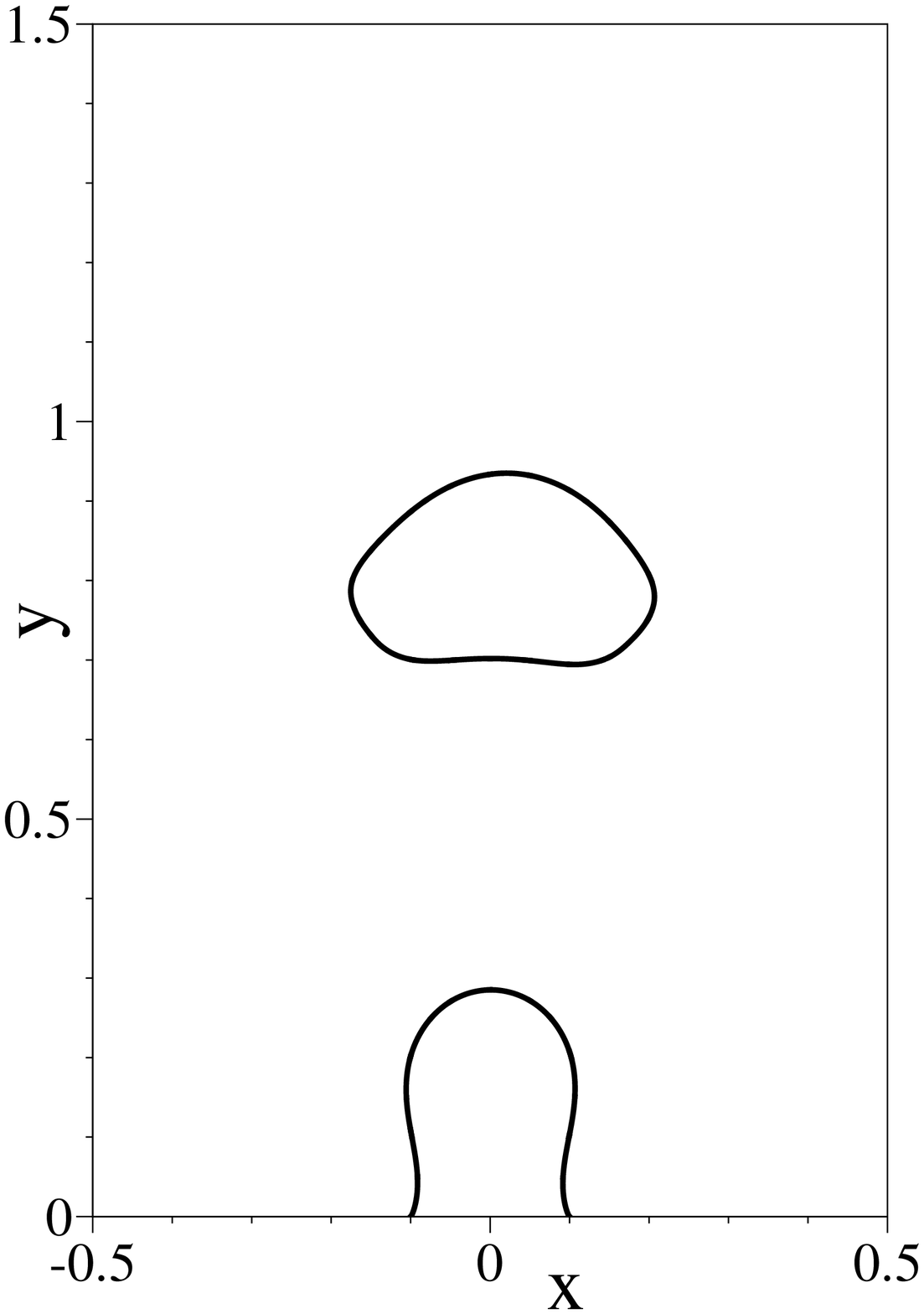}(h)
}
\caption{
Air jet in water: temporal sequence of snapshots of 
the air-water interface at time instants
(a) $t=16.9397$,
(b) $t=16.9772$,
(c) $t=17.0222$,
(d) $t=17.0522$,
(e) $t=17.0897$,
(f) $t=17.1272$,
(g) $t=17.1647$,
(h) $t=17.2022$.
Results are obtained using the boundary
condition \eqref{equ:obc_v_D} on the open
boundaries.
}
\label{fig:airjet_dynamics_1}
\end{figure}

The dynamics of this air-water flow is
illustrated by Figure \ref{fig:airjet_dynamics_1},
in which we show a temporal sequence of
snapshots of the air-water interface
in a time-window between
$t=16.9397$ and $t=17.2022$.
The fluid interface is visualized using
the contour level $\phi(\mathbf{x},t)=0$
in the plots.
These results are obtained with 
the open boundary condition \eqref{equ:obc_v_D},
corresponding to the time history in
Figure \ref{fig:airjet_vel_hist}(b).
These plots demonstrate the process of
free air bubbles generated at the wall rising
through water and crossing 
the upper domain boundary to migrate out of
the domain.
Figures \ref{fig:airjet_dynamics_1}(a)--(e)
show the leading air bubble passing
through the upper open boundary
of the domain. 
They demonstrate that the boundary
condition and the numerical algorithm we
developed in Section \ref{sec:method}
can effectively
allow the fluid interface to pass through
the open/outflow boundary in
a smooth fashion.
Simultaneously, one can observe that 
the trailing free bubble rises through 
the water, and that a new air bubble is forming
at the bottom wall
(Figures \ref{fig:airjet_dynamics_1}(b)--(h)).
Subsequently, the air bubble at the wall breaks
free, and the above process will repeat itself.

\begin{figure}
\centerline{
\includegraphics[width=1.4in]{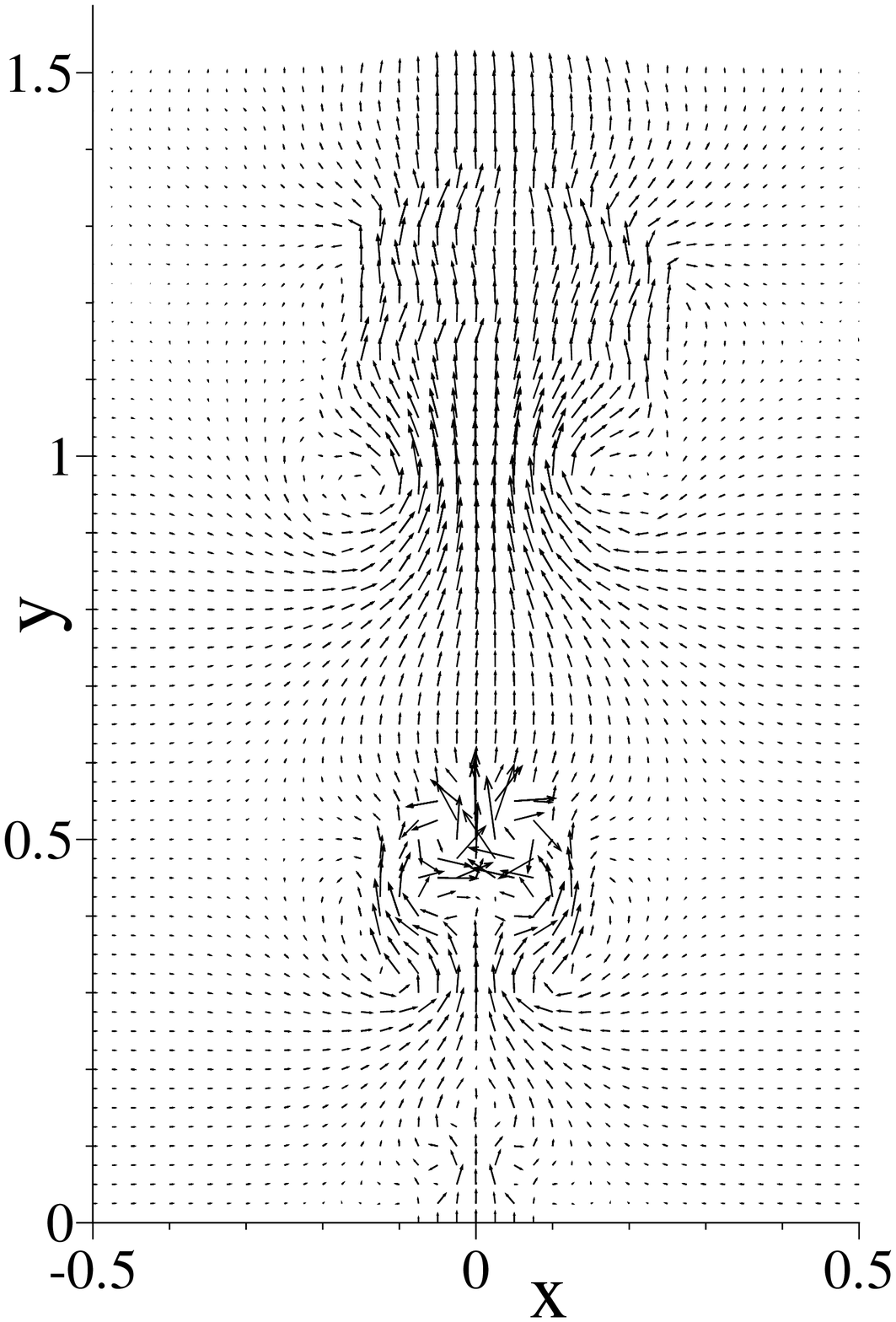}(a)
\includegraphics[width=1.4in]{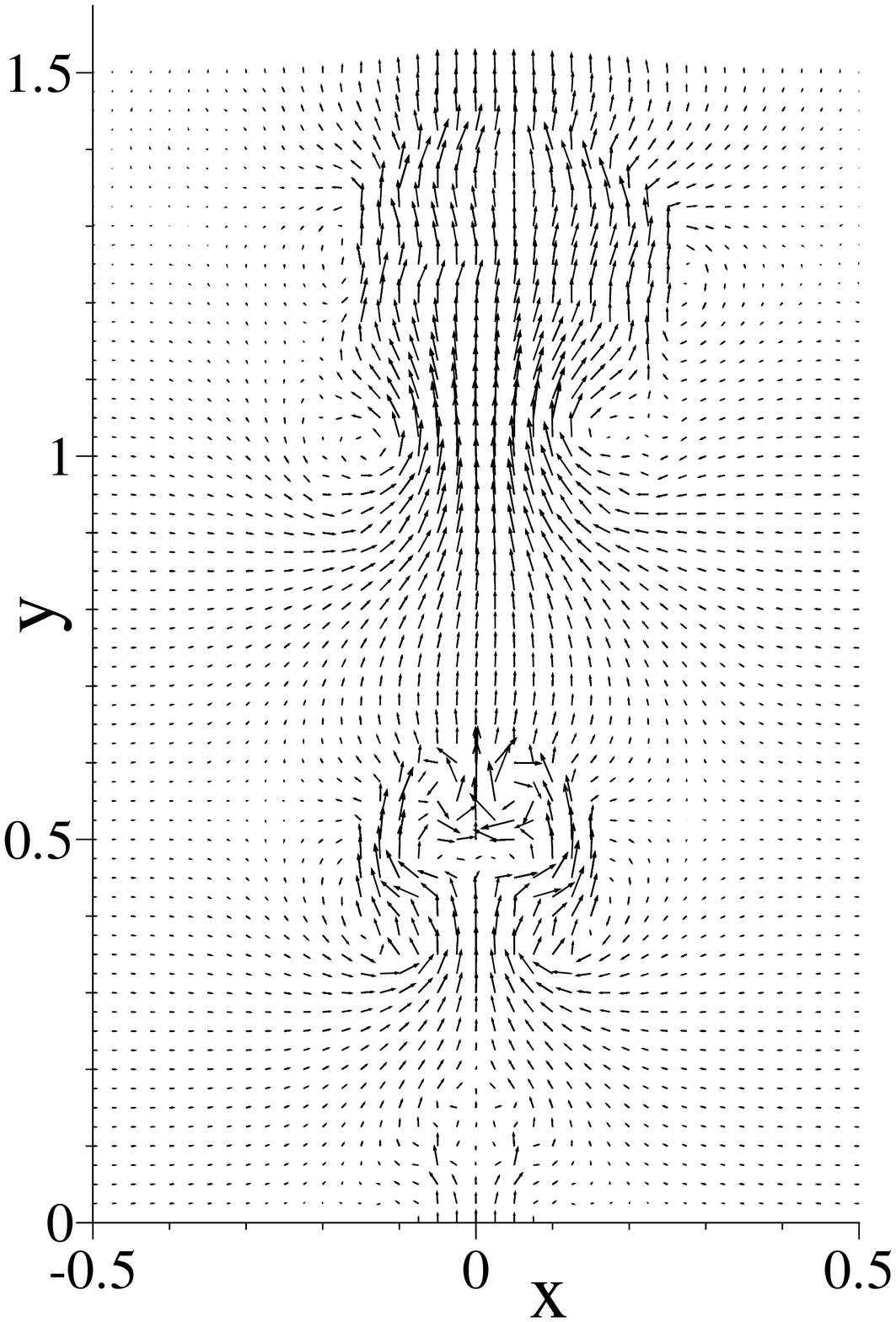}(b)
\includegraphics[width=1.4in]{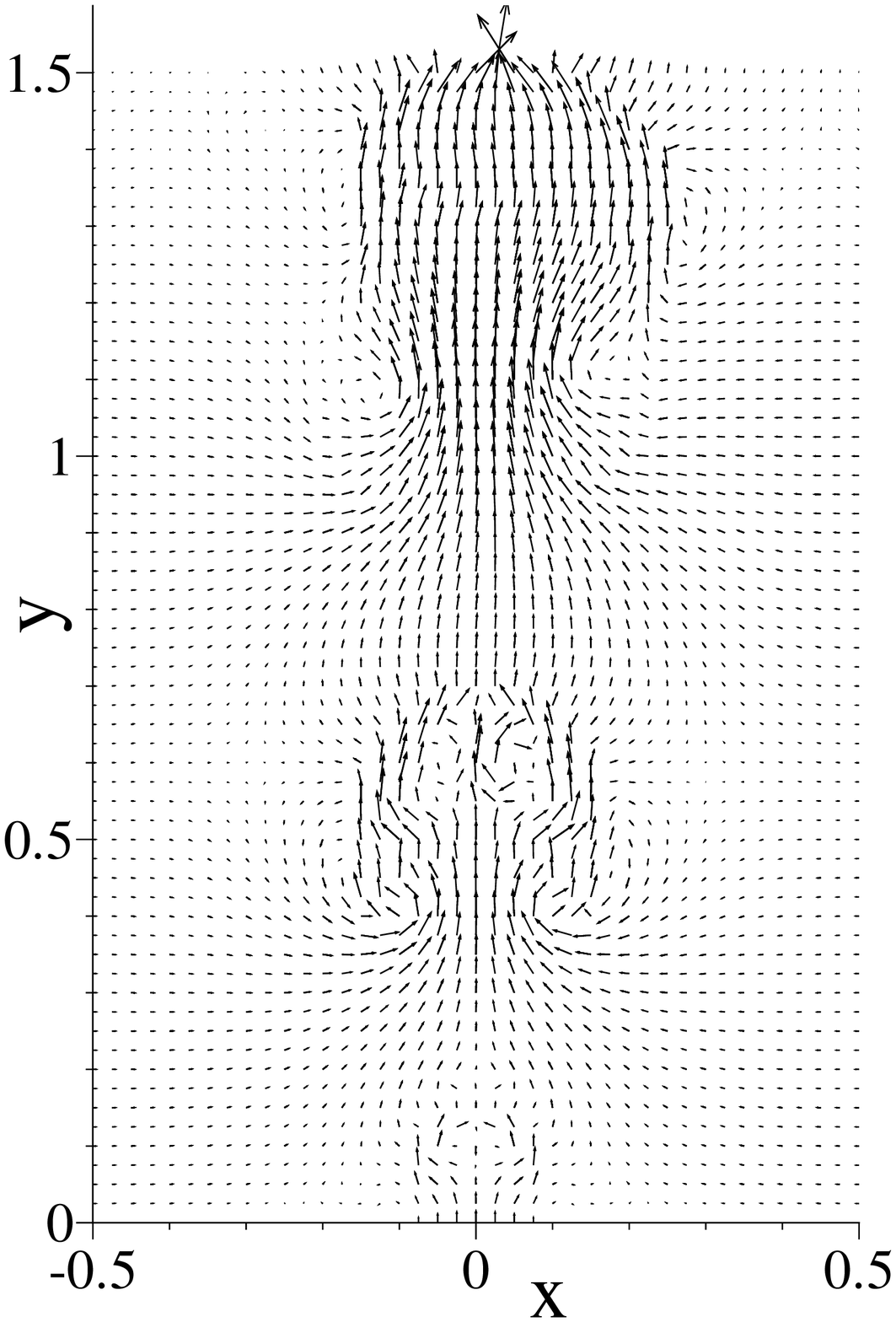}(c)
\includegraphics[width=1.4in]{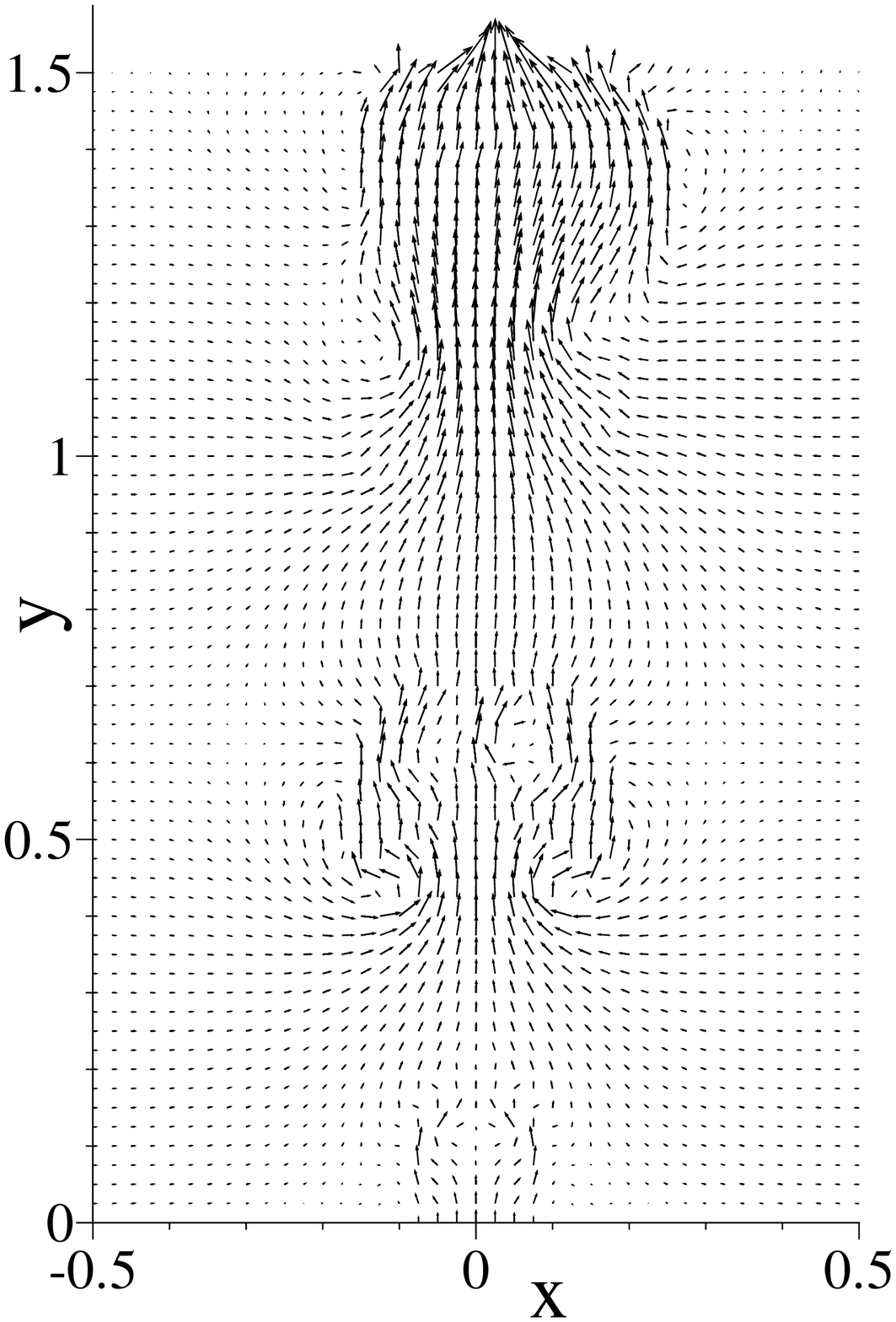}(d)
}
\centerline{
\includegraphics[width=1.4in]{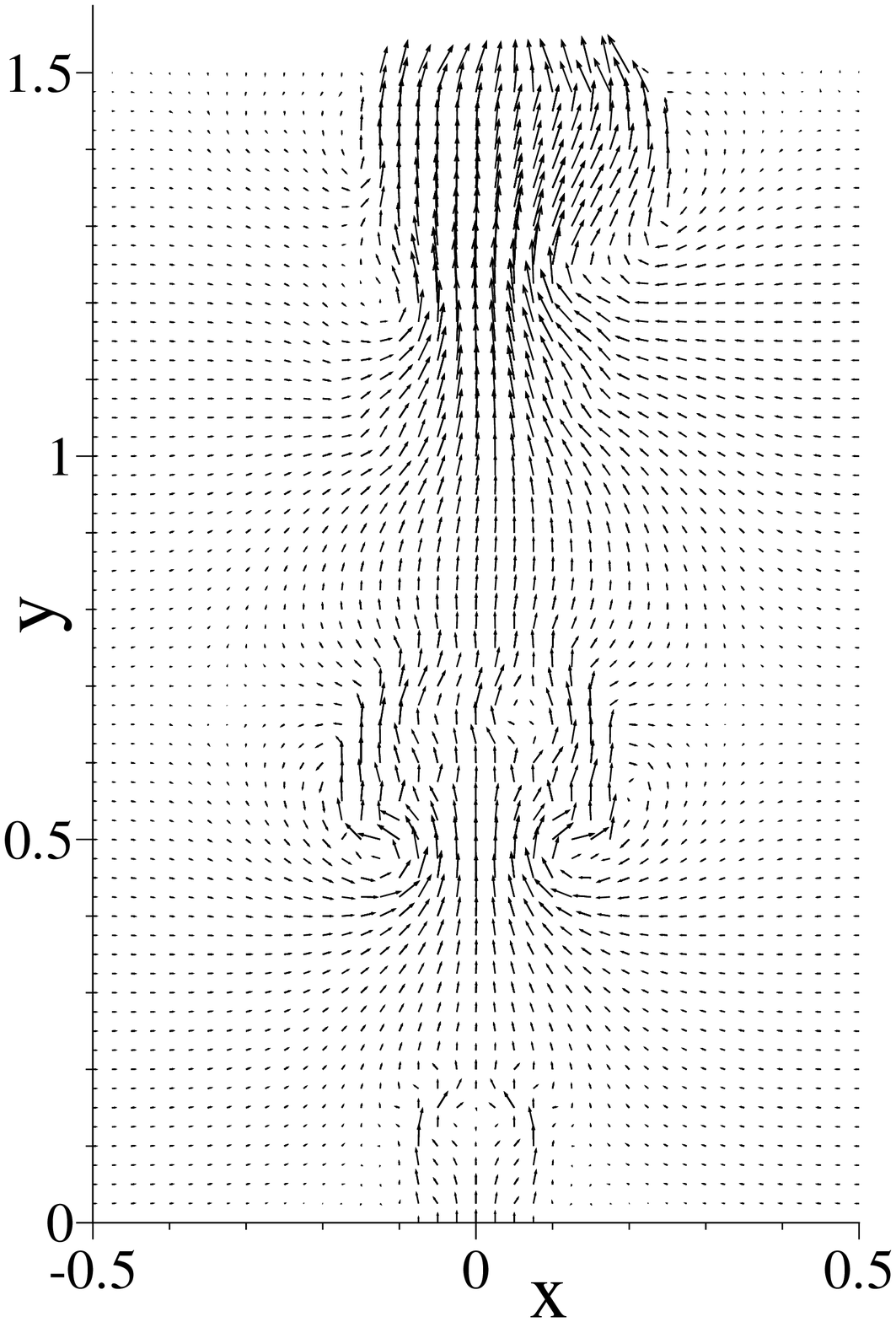}(e)
\includegraphics[width=1.4in]{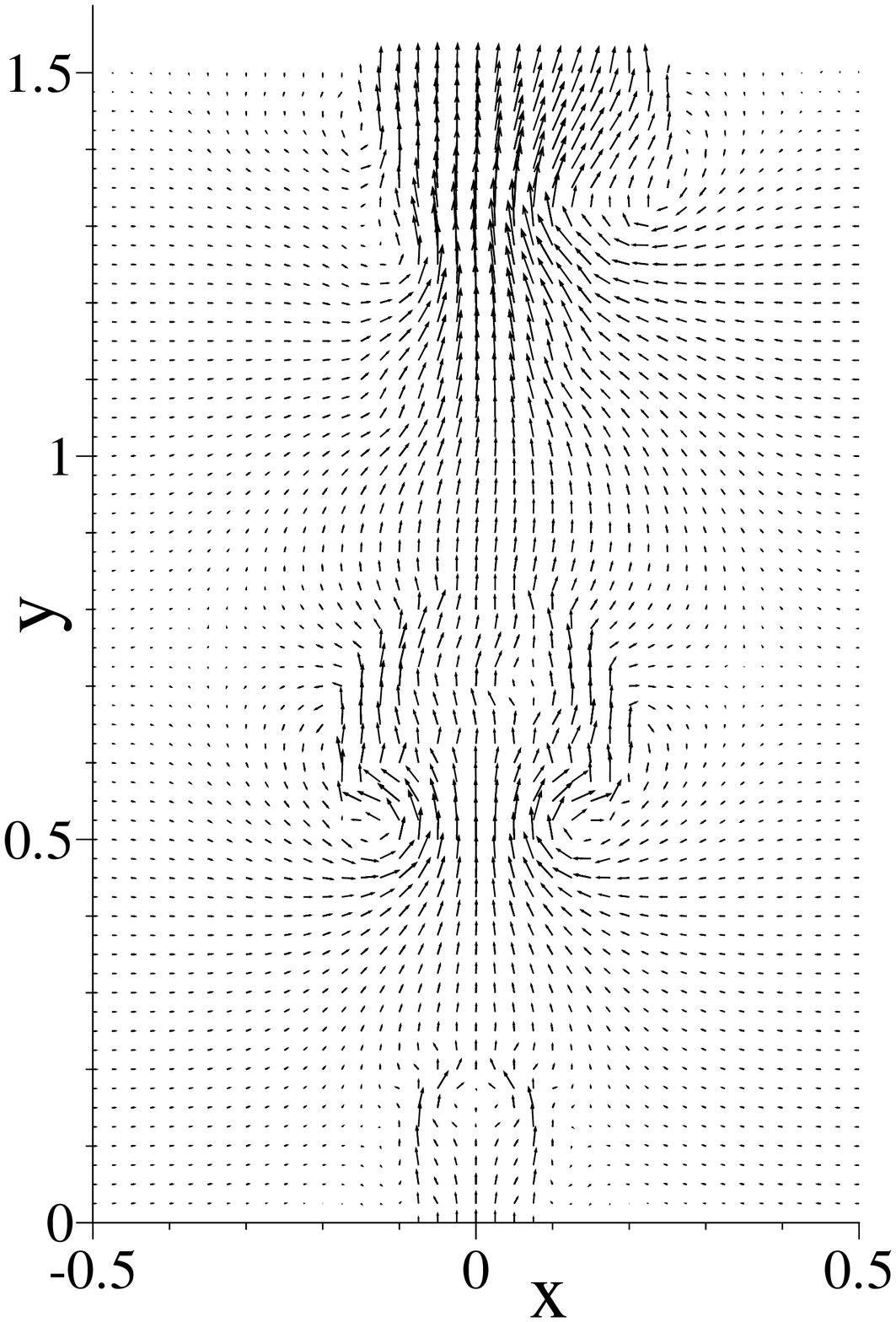}(f)
\includegraphics[width=1.4in]{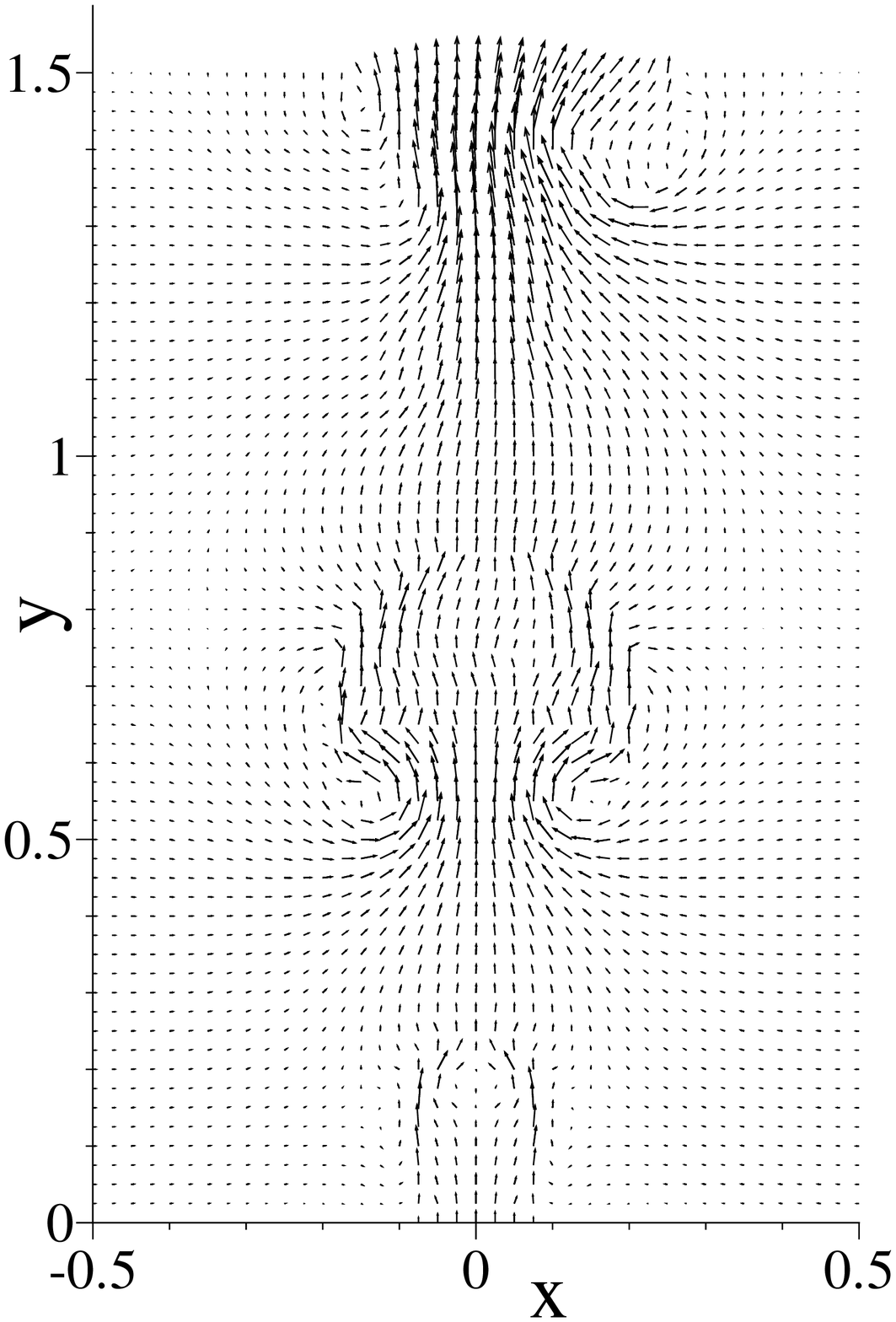}(g)
\includegraphics[width=1.4in]{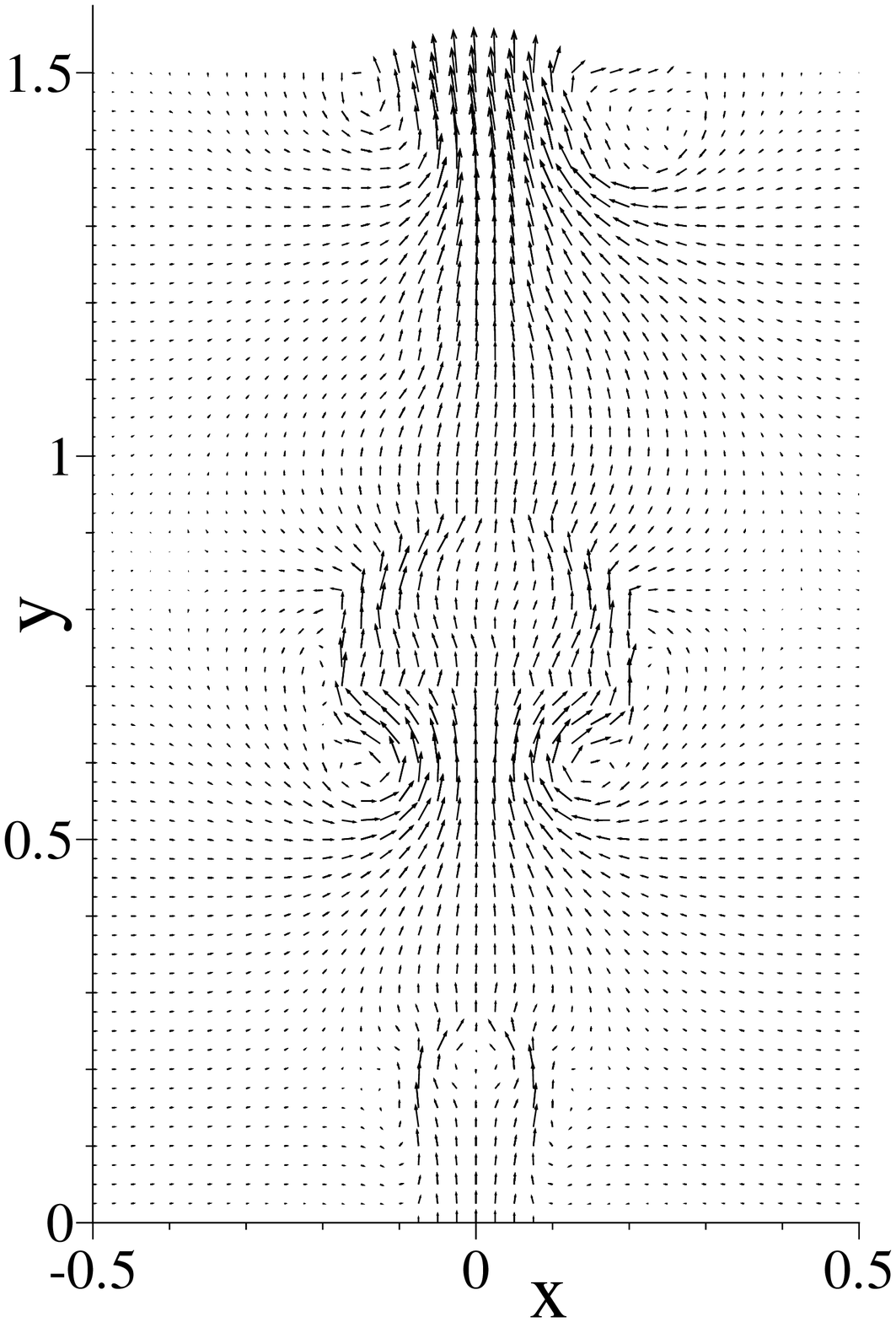}(h)
}
\caption{
Air jet in water: temporal sequence of snapshots of 
the velocity field at time instants
(a) $t=16.9397$,
(b) $t=16.9772$,
(c) $t=17.0222$,
(d) $t=17.0522$,
(e) $t=17.0897$,
(f) $t=17.1272$,
(g) $t=17.1647$,
(h) $t=17.2022$.
Velocity vectors are plotted on every ninth quadrature
points in each direction within each element.
Results are obtained using the boundary
condition \eqref{equ:obc_v_D} on the open
boundaries.
}
\label{fig:airjet_vel_1}
\end{figure}

We further illustrate
the flow dynamics using
instantaneous velocity distributions.
Figure \ref{fig:airjet_vel_1} is a temporal 
sequence of snapshots of the velocity fields
at identical time instants as those
of the interfacial plots of Figure \ref{fig:airjet_dynamics_1}.
One can observe that a significant flow field 
is induced in the regions occupied by
the air bubbles, and that a particularly
strong velocity field exists 
inside the free air bubble as it initially breaks free
 from the wall; see the region of
the trailing free bubble in Figures \ref{fig:airjet_vel_1}(a)--(b).
On the other hand, the velocity field
in the water region is in general quite weak.
As the air bubble rises through the water,
a pair of vortices forms in the water
region trailing the air bubble;
see the region behind the second air bubble
in Figures \ref{fig:airjet_vel_1}(e)--(h).
These vortices can induce a backflow 
on portions of the outflow/open boundary after the air bubble
passes through (Figure \ref{fig:airjet_vel_1}(h)).


\begin{figure}
\centerline{
\includegraphics[width=4.5in]{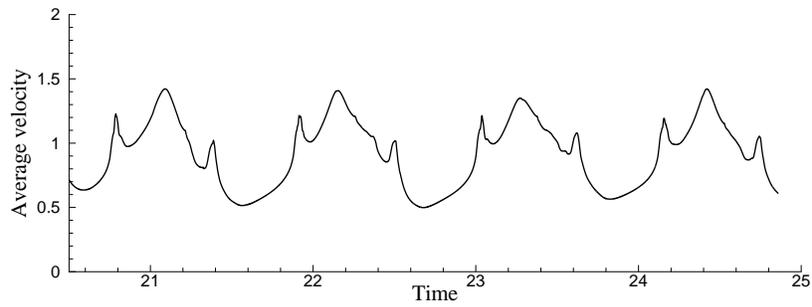}
}
\caption{
Another window of time history of the average vertical-velocity
magnitude, suggesting a somewhat different flow state. 
Result is obtained using the
open boundary condition \eqref{equ:obc_v_D}.
}
\label{fig:airjet_vel_hist_2}
\end{figure}

The above results illustrate one state of the flow.
We observe that this air-water flow can exhibit
 another state, in which the flow 
characteristics are somewhat different than those seen  above.
In Figure \ref{fig:airjet_vel_hist_2}
we show another window in the time history 
of the average magnitude of the vertical 
velocity, obtained also with the open
boundary condition \eqref{equ:obc_v_D}.
The flow evidently is at a statistically
stationary state.
Contrasting this figure  with 
Figure \ref{fig:airjet_vel_hist}(b), which
is computed using the same boundary conditions,
we can observe that
the velocity-history curves have qualitatively 
different characteristics in these figures.


\begin{figure}
\centerline{
\includegraphics[width=1.2in]{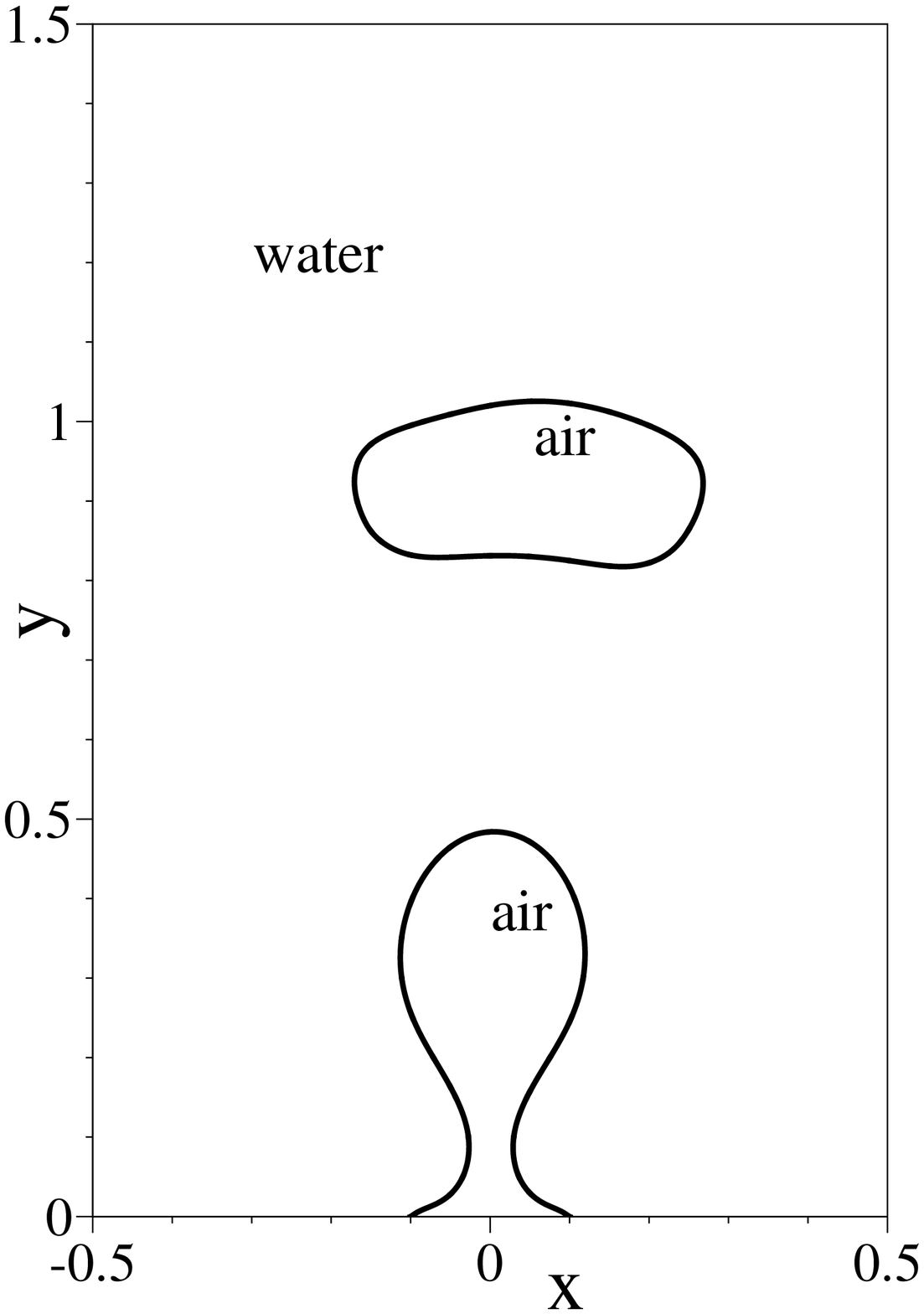}(a)
\includegraphics[width=1.2in]{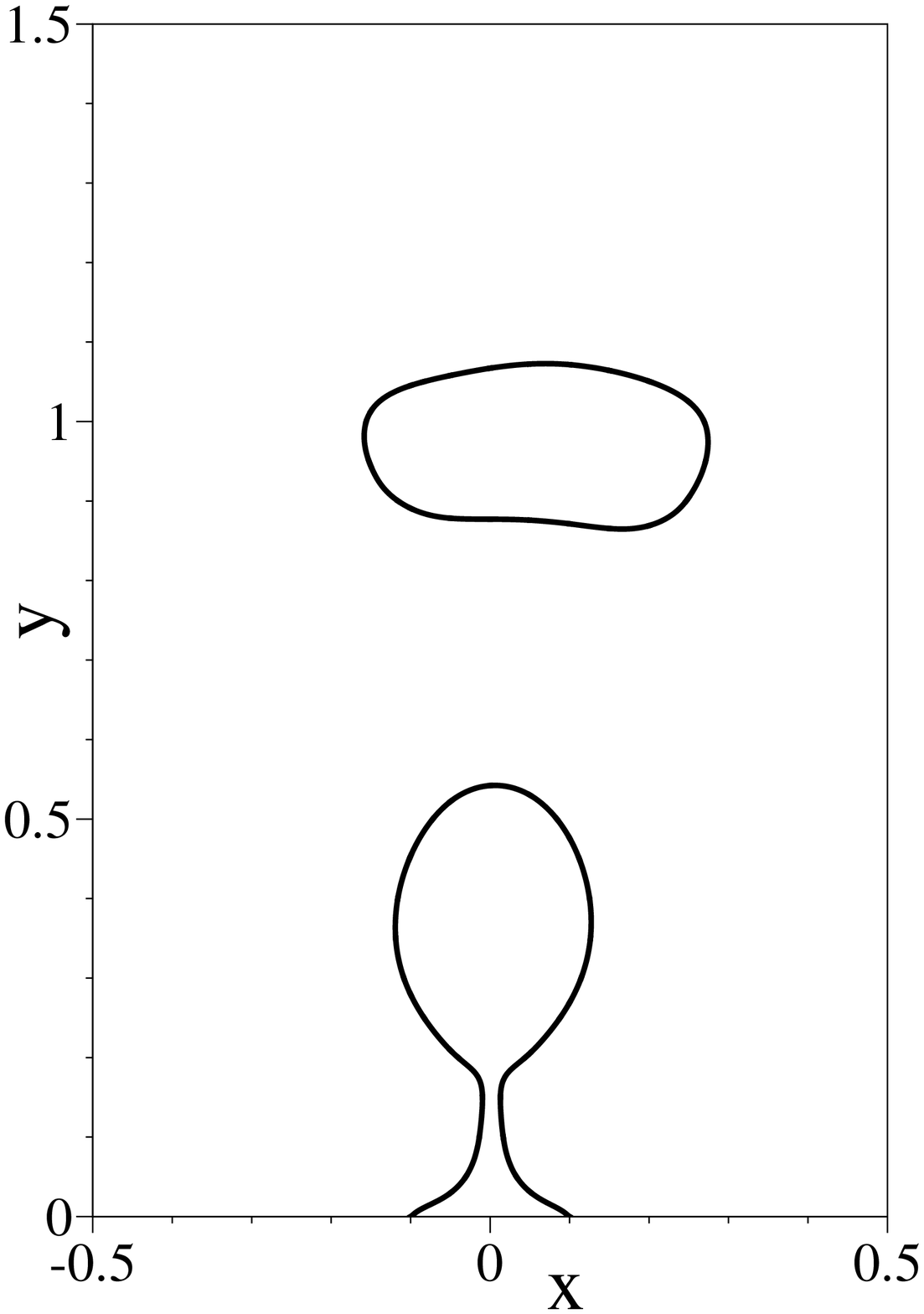}(b)
\includegraphics[width=1.2in]{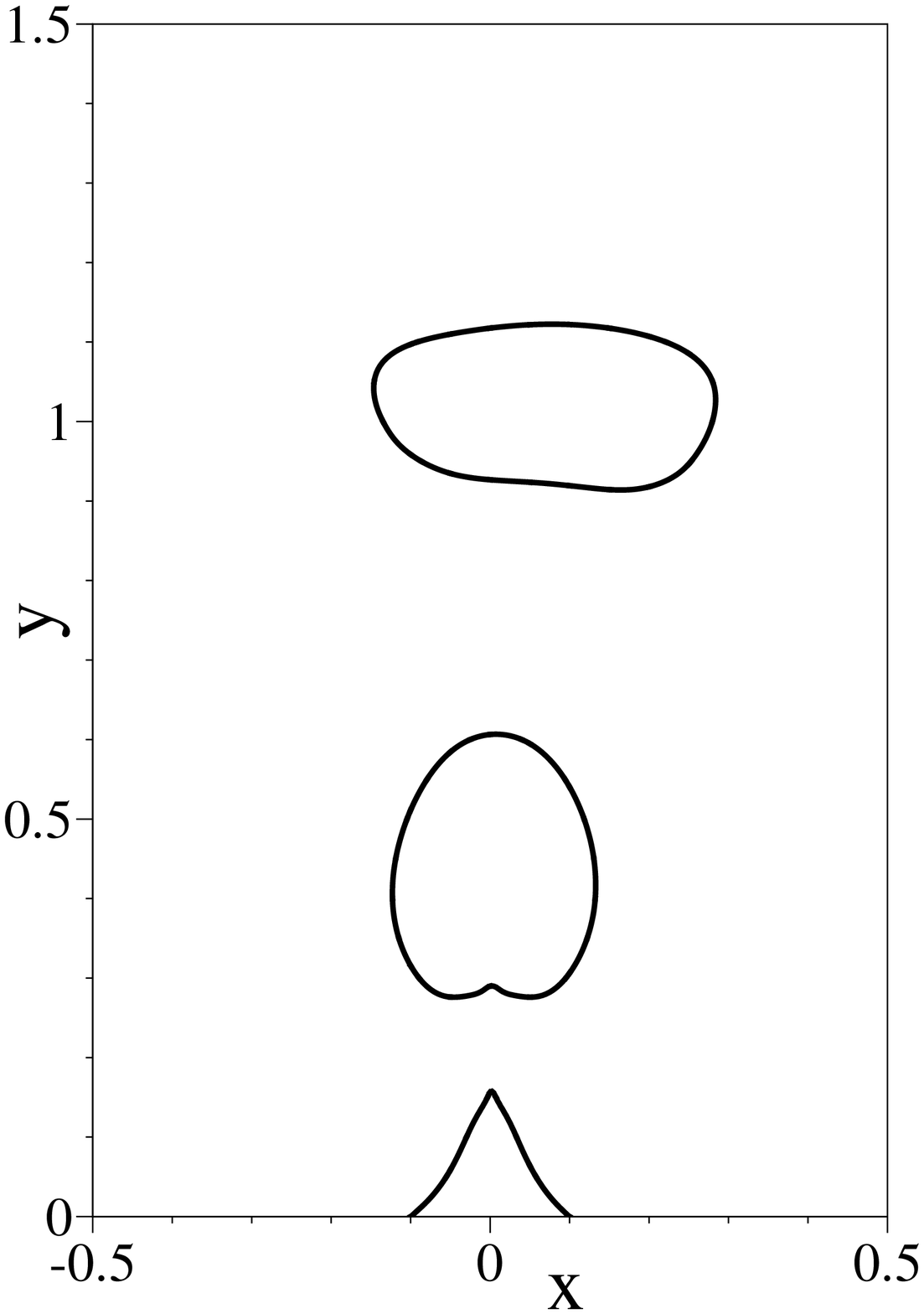}(c)
\includegraphics[width=1.2in]{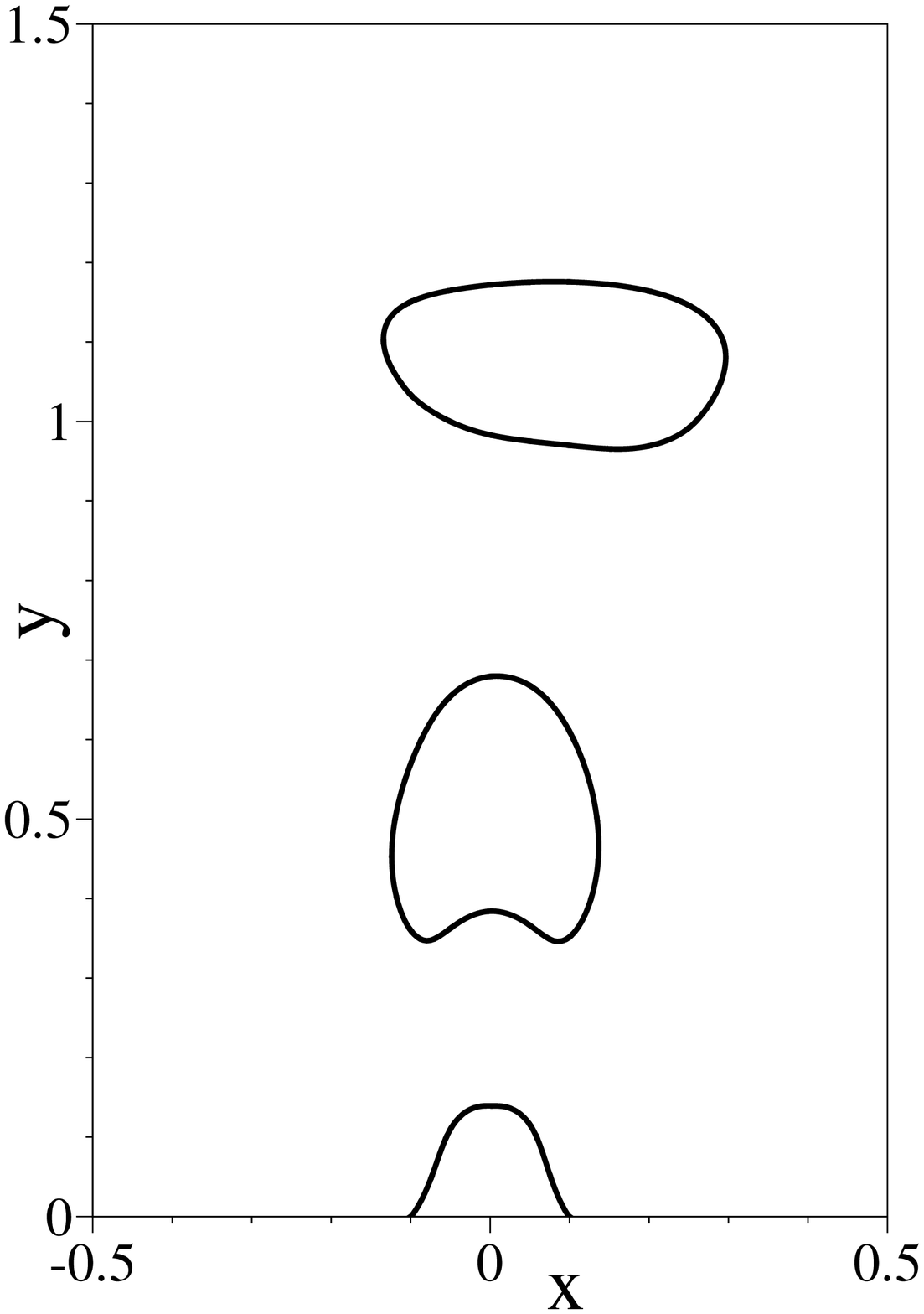}(d)
\includegraphics[width=1.2in]{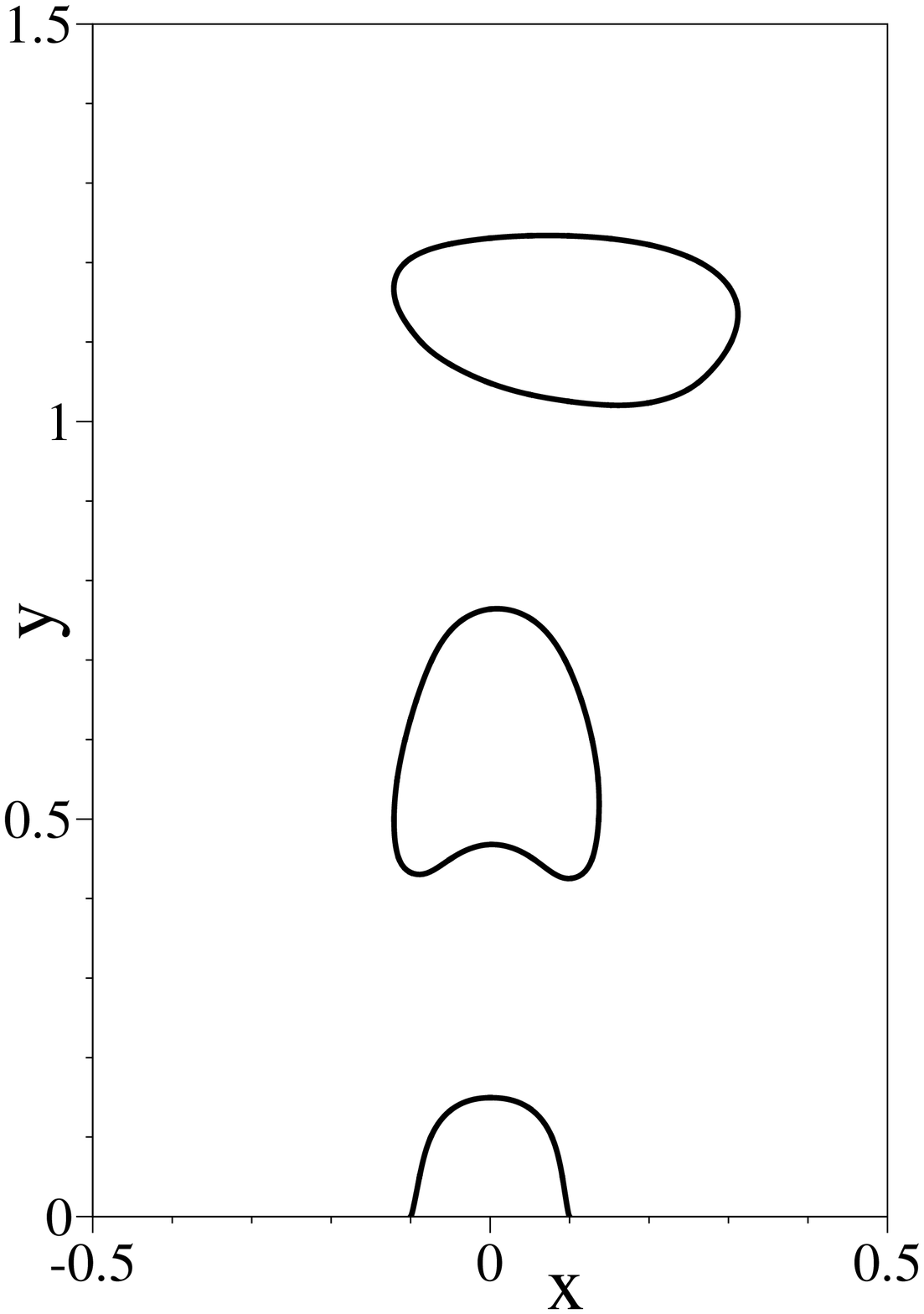}(e)
}
\centerline{
\includegraphics[width=1.2in]{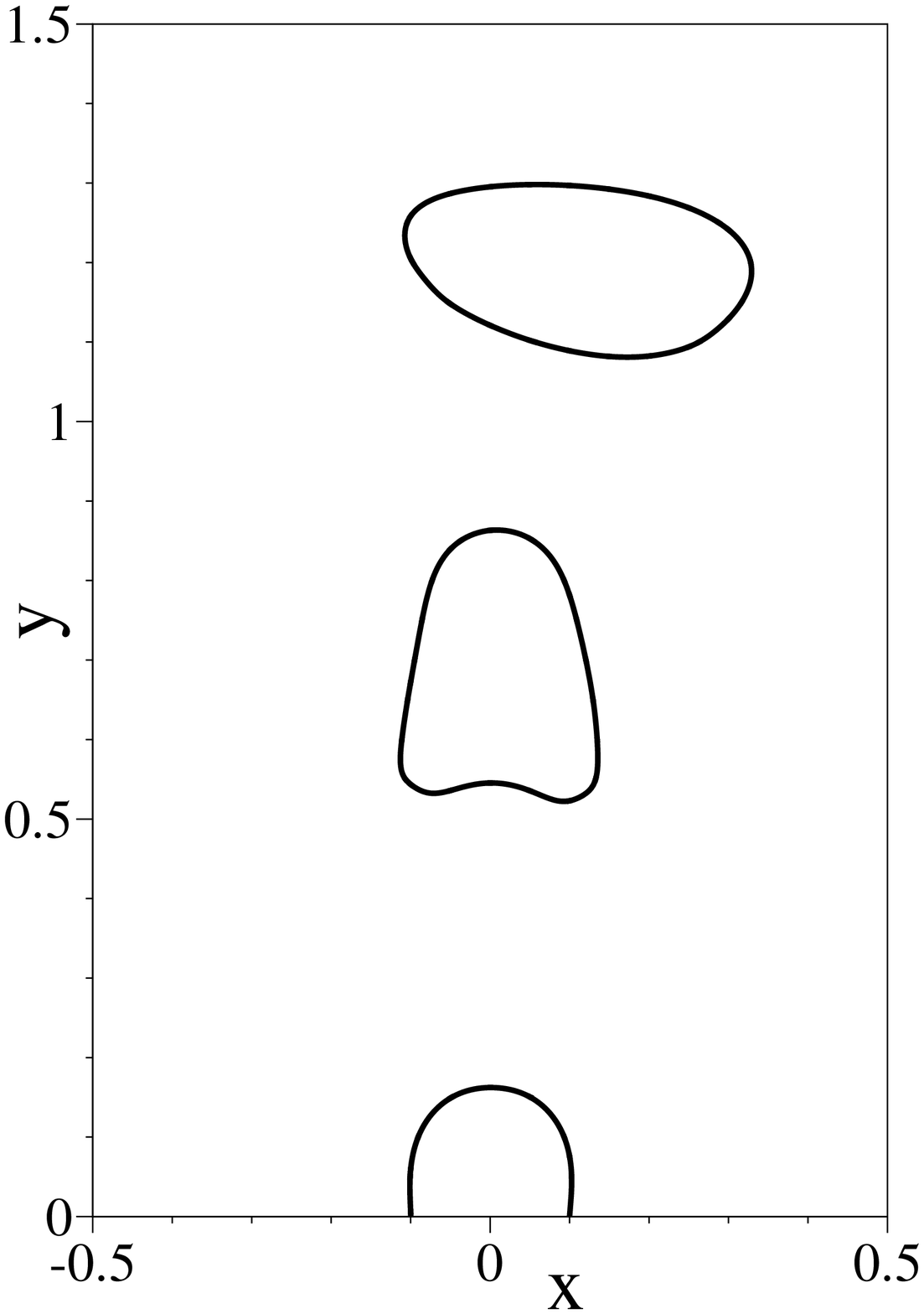}(f)
\includegraphics[width=1.2in]{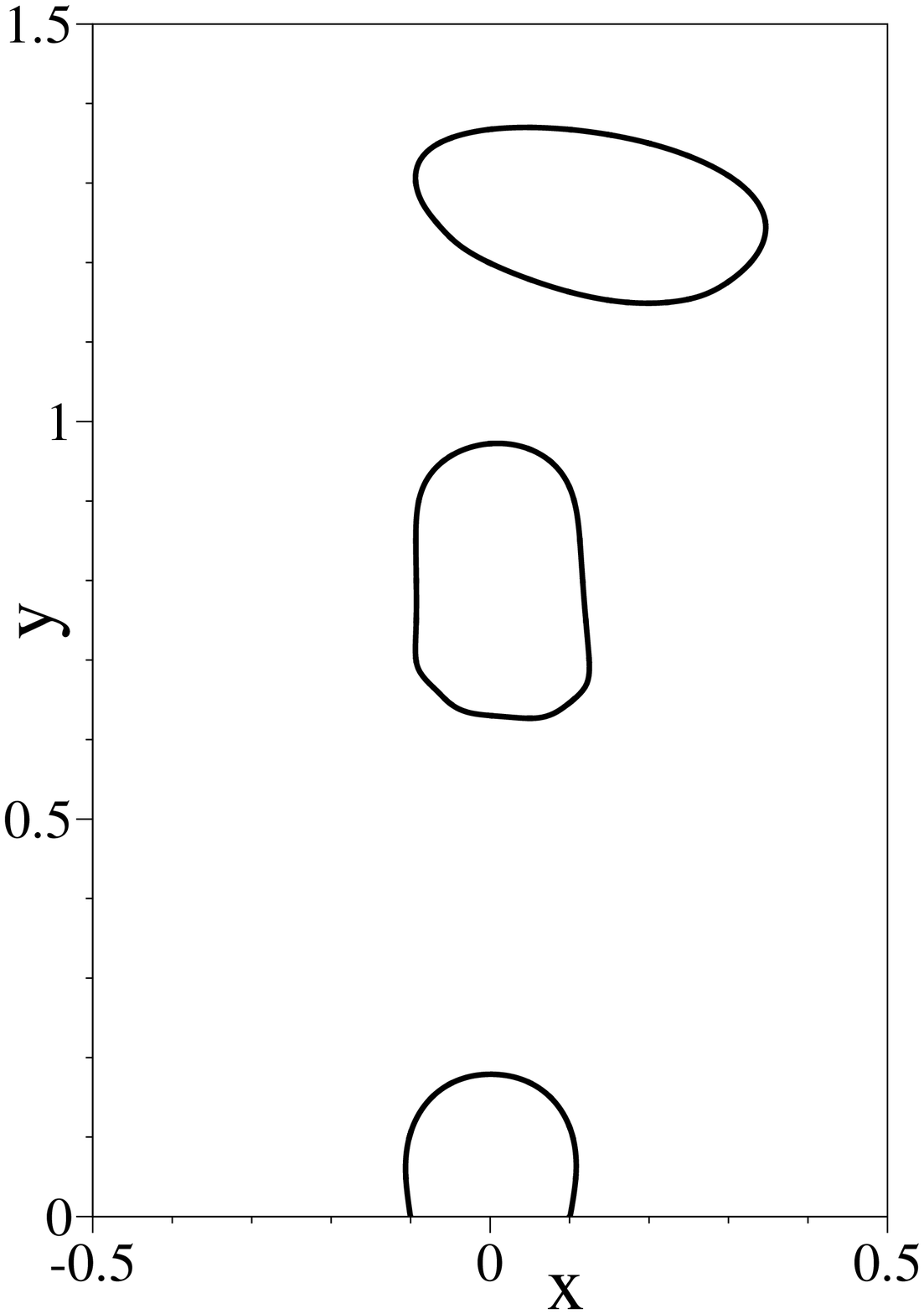}(g)
\includegraphics[width=1.2in]{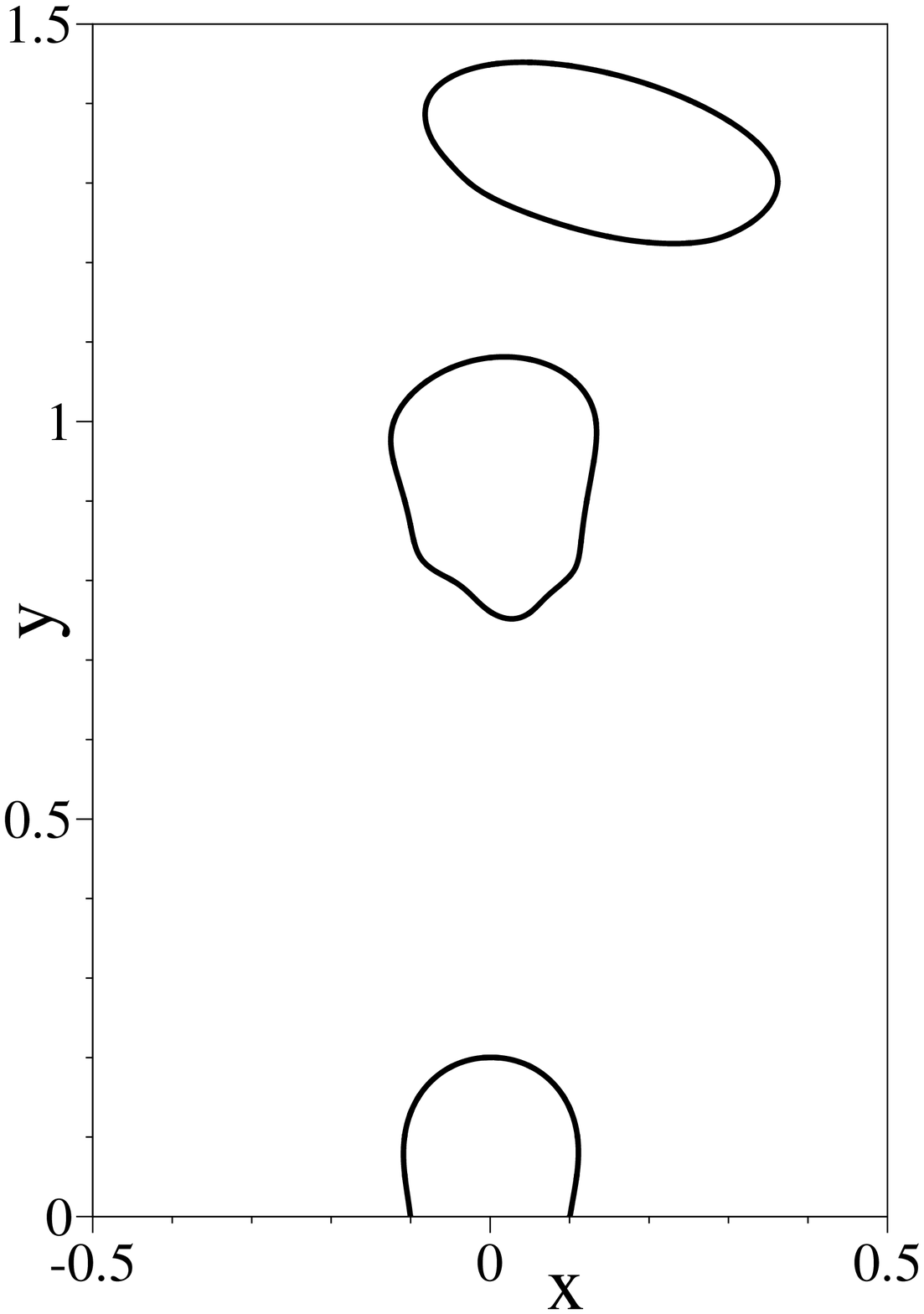}(h)
\includegraphics[width=1.2in]{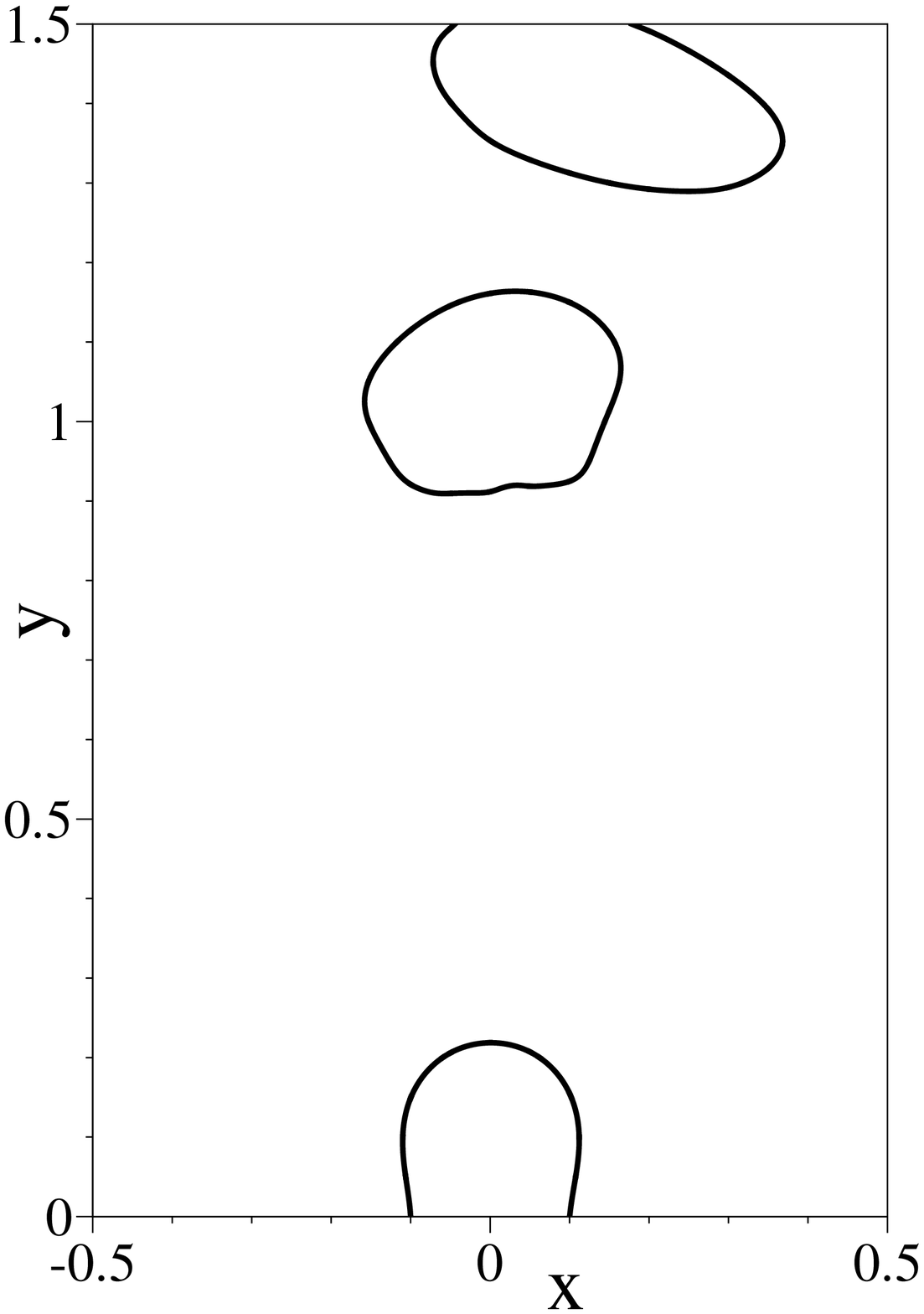}(i)
\includegraphics[width=1.2in]{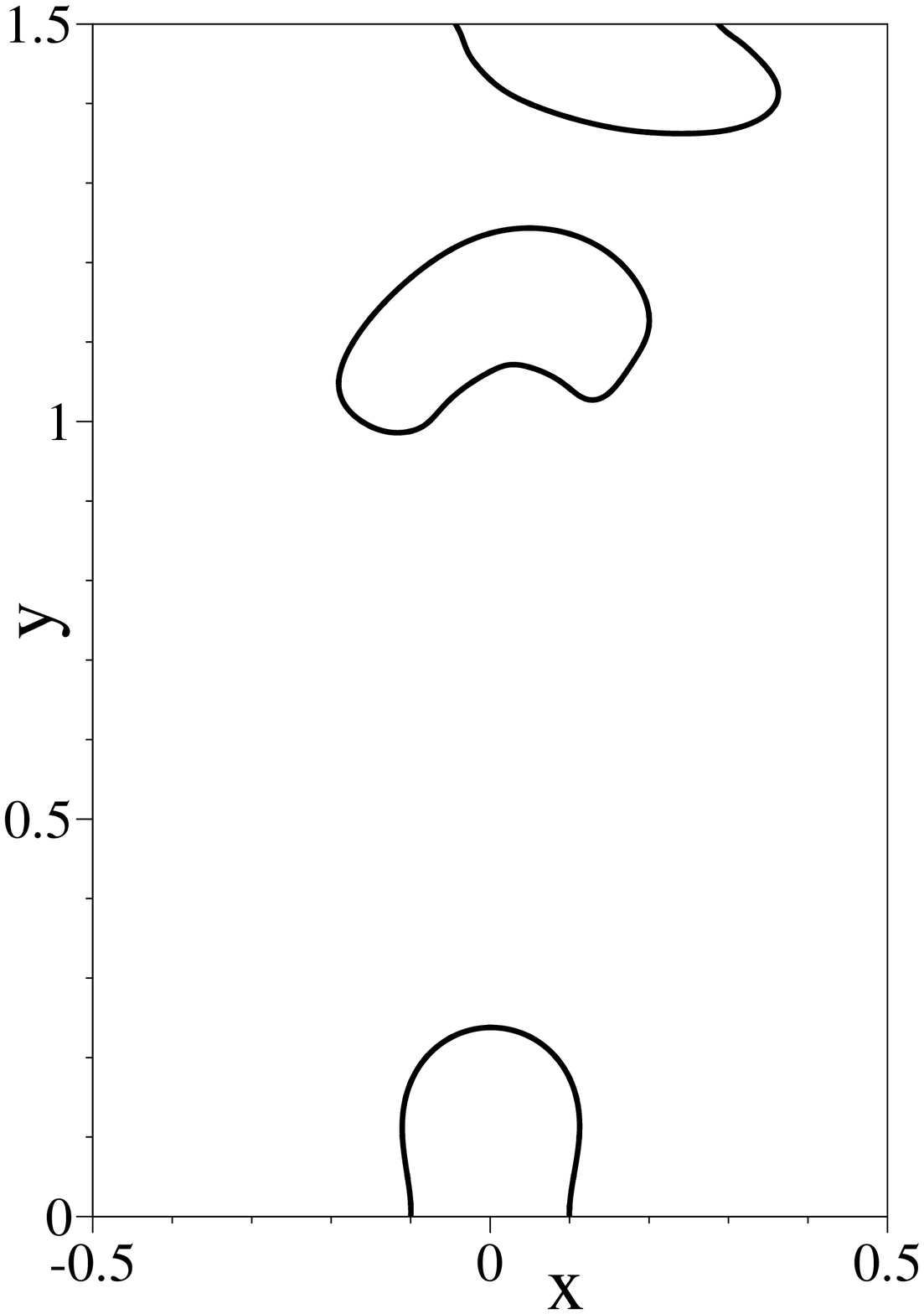}(j)
}
\centerline{
\includegraphics[width=1.2in]{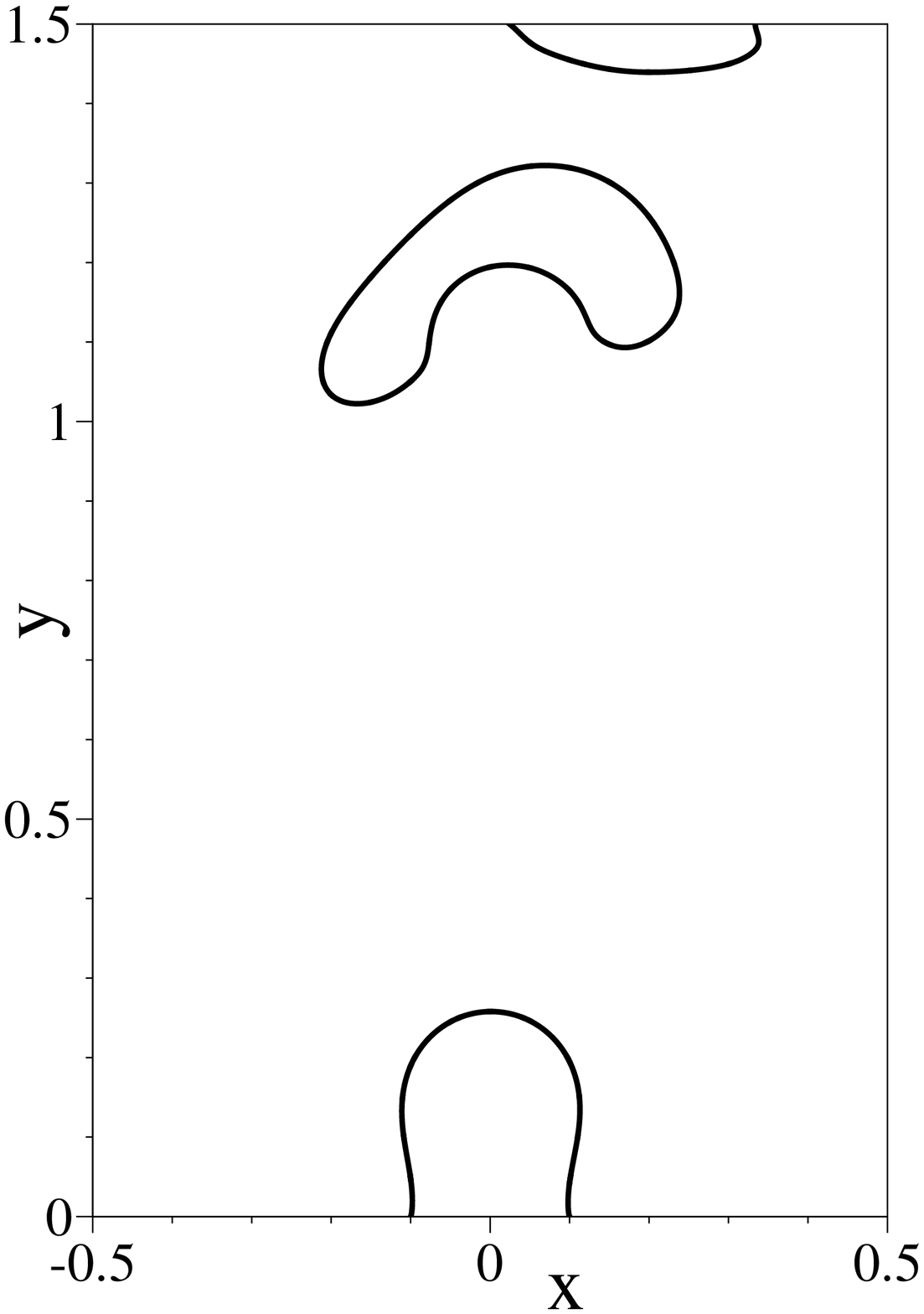}(k)
\includegraphics[width=1.2in]{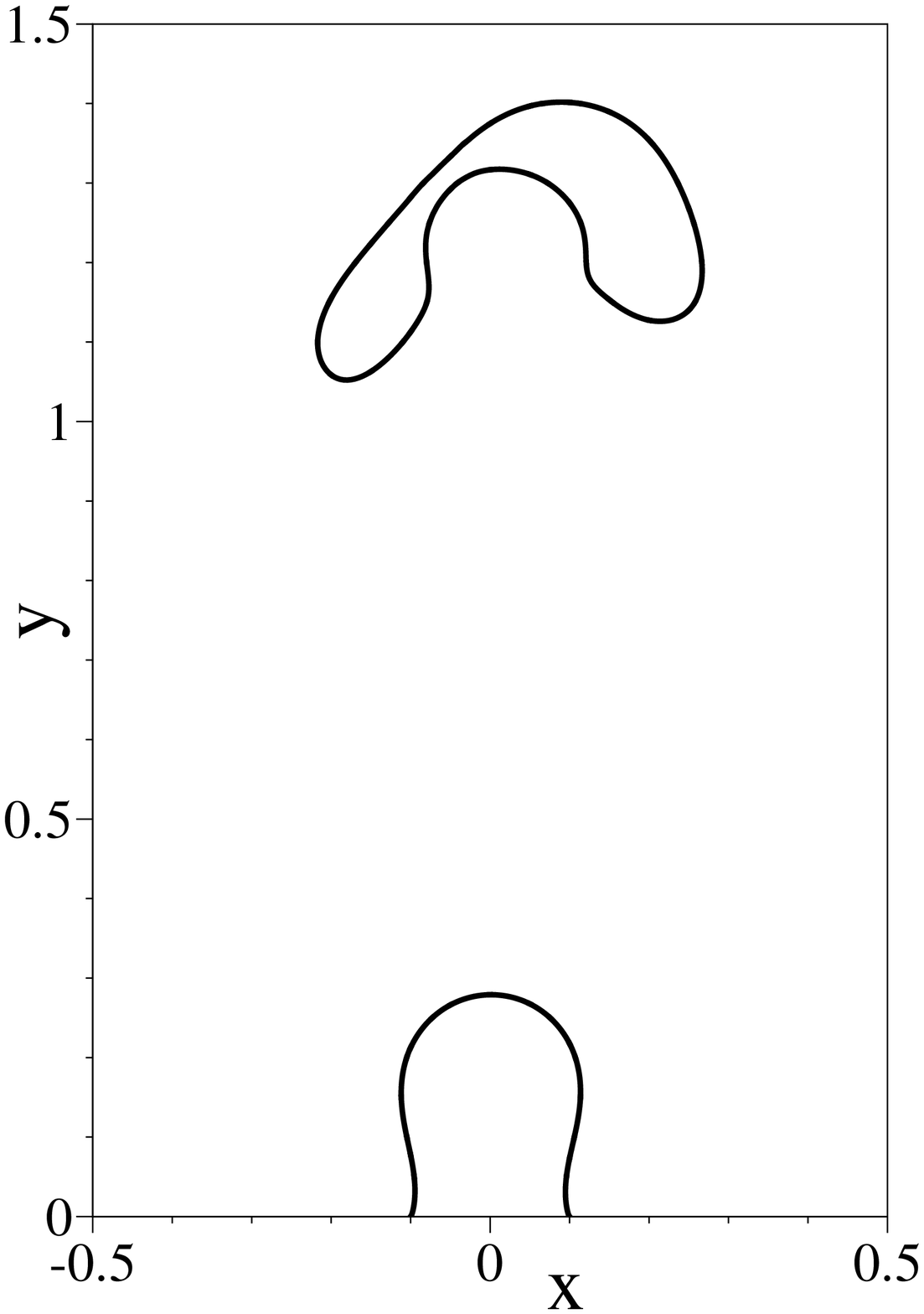}(l)
\includegraphics[width=1.2in]{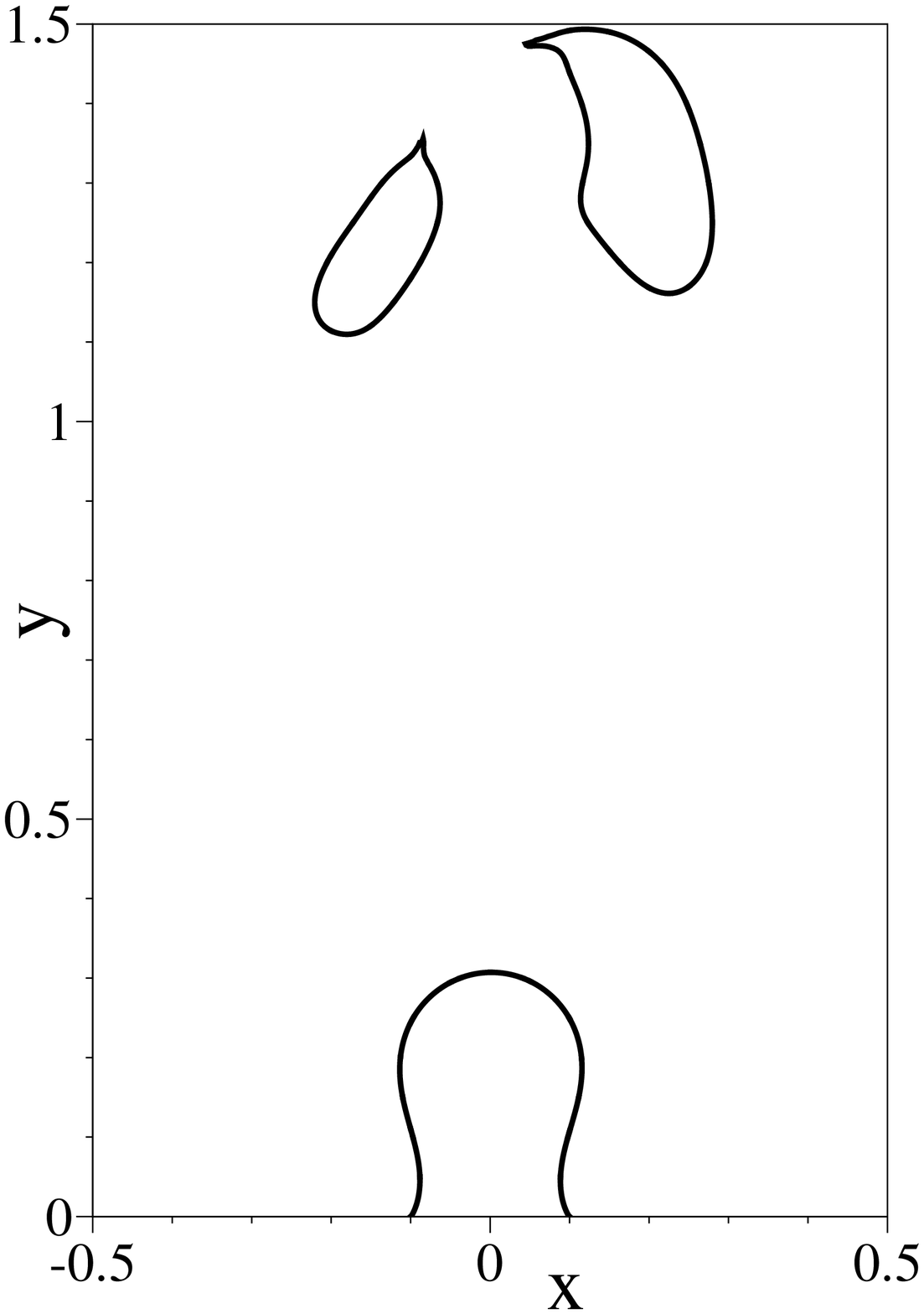}(m)
\includegraphics[width=1.2in]{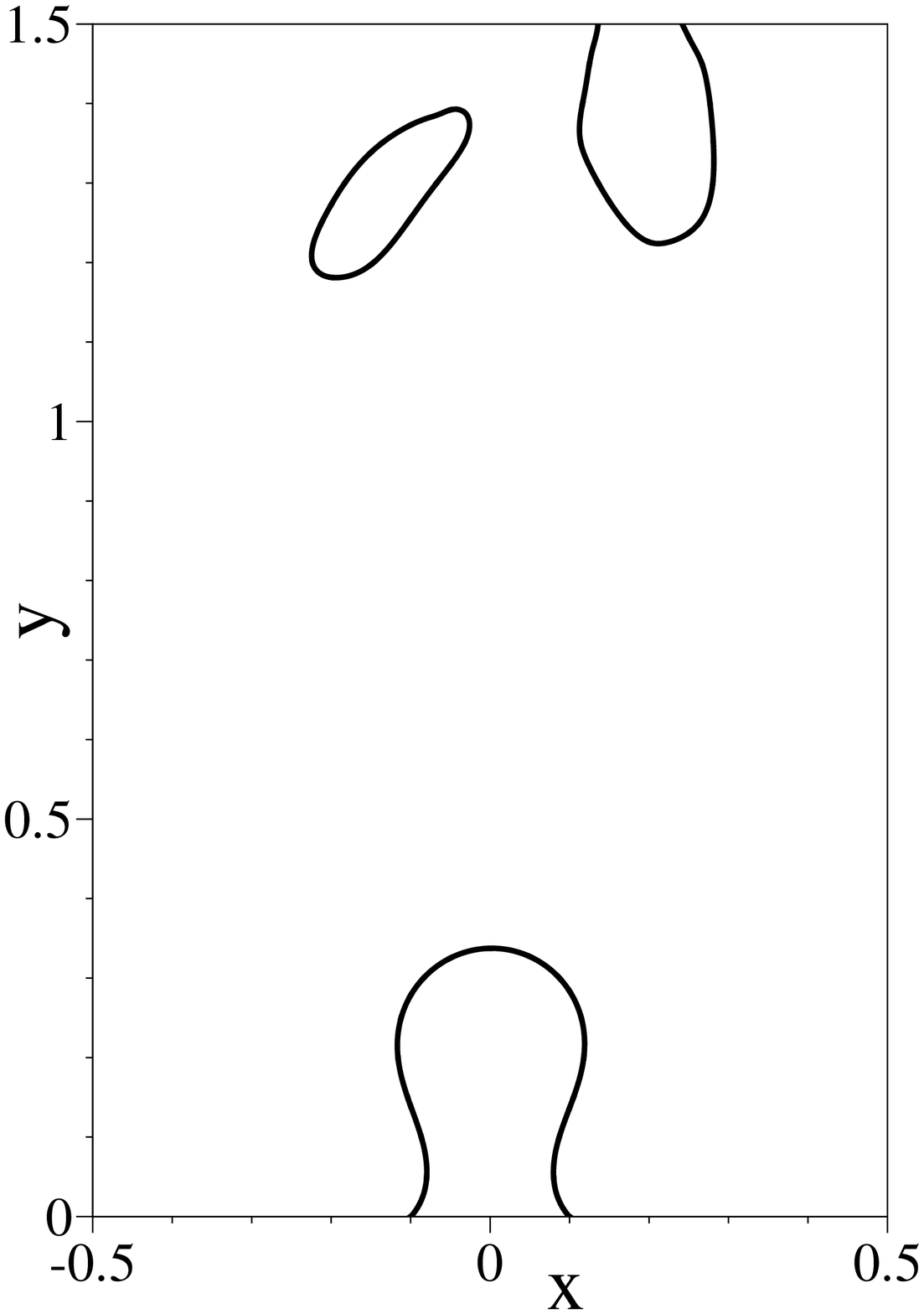}(n)
\includegraphics[width=1.2in]{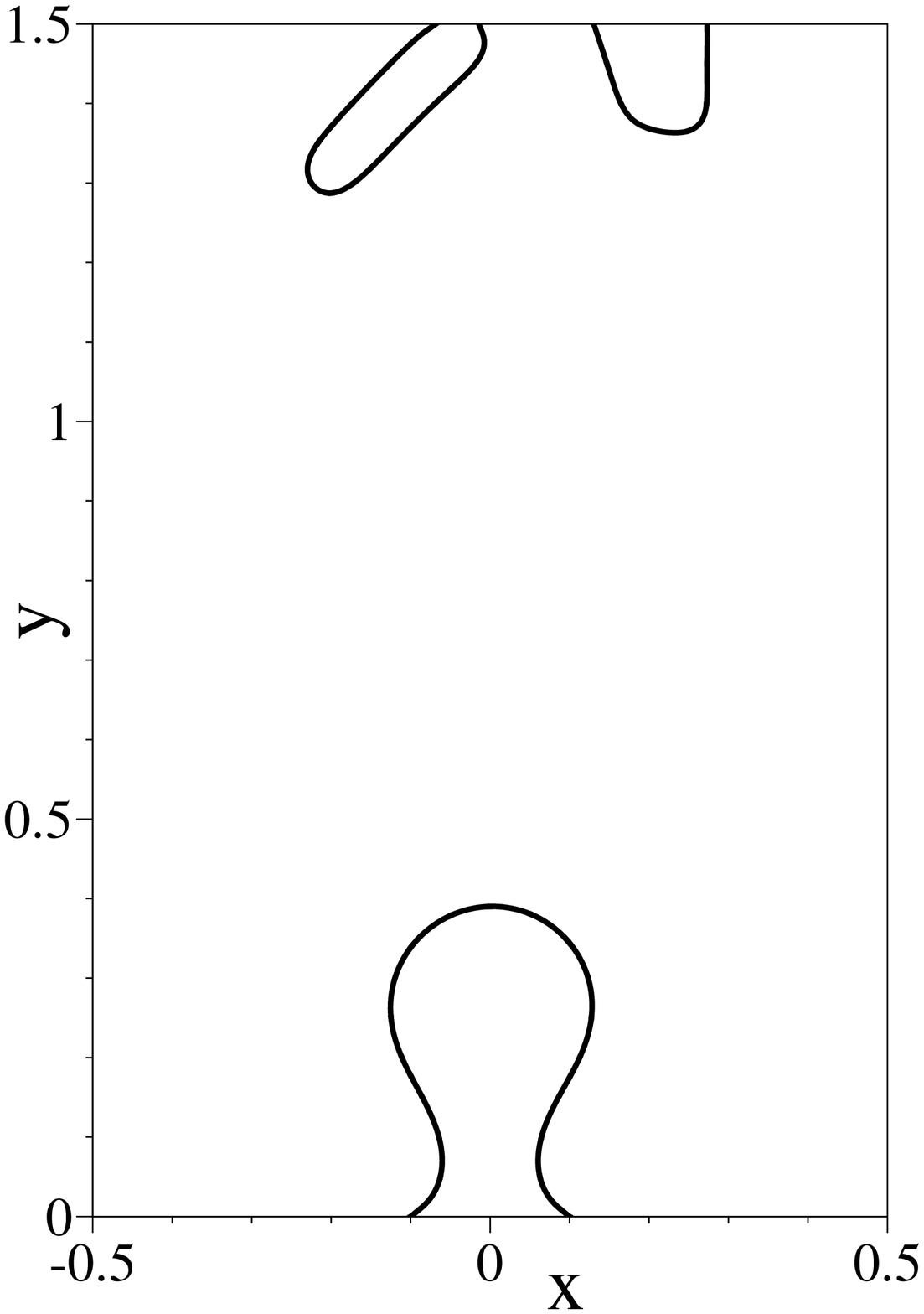}(o)
}
\caption{
Air jet in water: Temporal sequence of snapshots
of the air-water interface at time instants:
(a) $t=22.9997$,
(b) $t=23.0372$,
(c) $t=23.0747$,
(d) $t=23.1122$,
(e) $t=23.1497$,
(f) $t=23.1872$,
(g) $t=23.2247$,
(h) $t=23.2622$,
(i) $t=23,2922$,
(j) $t=23.3222$,
(k) $t=23.3522$,
(l) $t=23.3822$,
(m) $t=23.4197$,
(n) $t=23.4572$,
(o) $t=23.5172$.
Results are obtained using the boundary condition
\eqref{equ:obc_v_D} on the open boundaries.
}
\label{fig:airjet_phase_snap_2}
\end{figure}

This different flow state is further illustrated by
the temporal sequence of snapshots of
the air-water interface shown in
Figure \ref{fig:airjet_phase_snap_2},
which covers a time window
between $t\approx 23$ and $t\approx 23.5$
in the history plot of Figure \ref{fig:airjet_vel_hist_2}.
These results correspond to the 
open boundary condition \eqref{equ:obc_v_D}.
The plots clearly show the breakaway 
the air bubble from the wall
(Figures \ref{fig:airjet_phase_snap_2}(a)--(c))
and the bubble motion across the domain
and the upper open boundary
(Figures \ref{fig:airjet_phase_snap_2}(d)--(k)).
%
The crucial difference, when compared
with Figure \ref{fig:airjet_dynamics_1},
lies in the following.
When multiple free bubbles are present in
the domain, the interaction between
the leading-bubble wake and the trailing
bubble appears to have caused the trailing bubble
to accelerate and nearly catch up with
the leading one; see
Figures \ref{fig:airjet_phase_snap_2}(e)--(j).
This has also induced significant deformations
in the trailing bubble (Figures \ref{fig:airjet_phase_snap_2}(j)--(l)),
and caused it to subsequently break up 
(Figures \ref{fig:airjet_phase_snap_2}(m)--(o)).
As the free bubbles (and their daughter bubbles)
quickly move out of the domain,
one can observe that
another bubble is forming, but
still attached to the wall (Figure \ref{fig:airjet_phase_snap_2}(o)).
Consequently, the flow domain will
be depleted of free bubbles for
a period of time
beyond the time instant
corresponding to Figure \ref{fig:airjet_phase_snap_2}(o),
until the air bubble attached to the wall breaks free.
This scenario is more similar to 
the one discussed in \cite{Dong2014obc},
but is quite different from that shown by
Figure \ref{fig:airjet_dynamics_1}.
From Figures \ref{fig:airjet_phase_snap_2}(i)--(k)
we can again observe that our method allows the
air bubble and the air-water interface
to cross the open domain
 boundary in a smooth fashion.

%

The air jet in water problem is a stringent
test to the open boundary conditions.
The presence of two-phase open boundary,
combined with the large density ratio
between air and water, makes this problem
extremely challenging to simulate.
The results of this section show
that the two-phase open boundary conditions
and the numerical algorithm developed
in the current work are effective
for two-phase outflows  with 
large density and viscosity contrasts
at the outflow boundaries.
%
The $\mathbf{E}(\rho,\mathbf{n},\mathbf{u})$
term in the open boundary condition
\eqref{equ:obc_v_trans} is critical
to the stability for this problem. 
We observe that the computation
using an open boundary condition
without this term is unstable
for this problem, that is,
\begin{equation}
-p\mathbf{n} + \mu\mathbf{n}\cdot\mathbf{D}(\mathbf{u})
-\left[
  \frac{\lambda}{2}\nabla\phi\cdot\nabla\phi
  + F(\phi)
\right]\mathbf{n}
= 0, \quad \text{on} \ \partial\Omega_o,
\end{equation}
due to
the backflows induced by the
vortices at the outflow
boundary.
It is observed that increasing
$\nu_m$ in the algorithm tends to improve
the stability, and that a
larger $\mu_0$ in \eqref{equ:velocity_3}
for the numerical treatment of
the open boundary condition
also improves the stability
for the current pressure-correction based
 scheme. 
This observation concerning $\mu_0$
seems different from the trend 
observed in \cite{Dong2014obc},
which is for a velocity-correction based
algorithm.


\section{Concluding Remarks}
\label{sec:summary}


We have presented 
several new open boundary conditions 
for two-phase outflows, and a 
rotational pressure-correction based
algorithm for solving the two-phase momentum equations
in conjunction with the proposed 
open boundary conditions.
These techniques are then combined with a solver
for the phase-field equation 
to form an efficient and effective method
for simulating incompressible two-phase flows
involving open/outflow boundaries.

The two-phase open boundary conditions
presented here are inspired by the 
two-phase energy balance 
discussed in the previous work \cite{Dong2014obc}.
The current work has provided a generalization
and several new forms for
the open boundary condition
beyond those developed in \cite{Dong2014obc}
for the momentum equations.


The algorithm presented herein
for the two-phase momentum equations
is based on a rotational
pressure correction-type strategy 
for de-coupling the velocity/pressure
computations.
More importantly, the current algorithm
results in velocity and the pressure linear 
algebraic systems with {\em constant} and
{\em time-independent} coefficient matrices
after discretization, despite the 
variable nature of the mixture density and
mixture viscosity.
Therefore, these coefficient matrices can be
pre-computed during pre-processing. 
In a previous work \cite{DongS2012} we have developed
a {\em velocity correction-based} algorithm
for the variable-density Navier-Stokes equations
that possesses similar properties (leading to
constant coefficient matrices for pressure/velocity
linear systems); see also subsequent applications and 
further developments 
based on that algorithm in \cite{Dong2012,Dong2014obc,Dong2014nphase}.
The algorithm developed herein  
 in a sense can be considered as the {\em pressure-correction
counterpart} to the scheme of \cite{DongS2012}.
The implementation of the algorithm 
presented herein is suitable for $C^0$ 
spectral elements, and with no change it
also applies to conventional finite elements.
It should be noted that the rotational
pressure correction
formulation embodied in the current algorithm
has a difference to the usual pressure correction
formulations (see e.g. \cite{GuermondMS2006}),
in that apart from the velocity/pressure
we have introduced a discrete equation
and the corresponding boundary conditions
for another field variable $\xi^{n+1}$
in the algorithmic formulation.


The numerical treatments for the 
 open boundary conditions proposed
herein involve imposing a discrete  Neumann type 
condition on the outflow boundary at
the velocity substep, and two discrete Dirichlet type
conditions on the outflow boundary
at the substeps for $\xi^{n+1}$ and pressure 
respectively.
The discrete velocity-Neumann  and 
the pressure-Dirichlet conditions on
the outflow boundary
stem largely from the 
continuous open boundary condition.
But they contain modifications and additional terms
that are essential to
the stability of the algorithm.


To demonstrate the physical accuracy of
the method developed herein, we have considered
the capillary wave problem and compared quantitatively
the numerical solution 
with the two-phase exact physical solution
by \cite{Prosperetti1981} for
a range of density ratios (up to $1000$).
The comparisons show that our method has produced
physically accurate results.
We have also considered the bounce of 
a water droplet on a superhydrophobic 
surface, and compared the restitution coefficients
from the simulations and 
the experimental measurement
of \cite{RichardQ2000}.
The simulation results are in good agreement with
the experimental data.


We have further simulated the air jet in water problem
to test the effectiveness of the open boundary
conditions and the numerical algorithm 
 for two-phase problems
involving outflow or open boundaries.
This problem involves large density ratio, large
viscosity ratio, and backflows or vortices
at the two-phase open boundary.
The results demonstrate the long-time stability
of the method presented herein. 
It is also shown that
our  boundary conditions allow 
the fluid interface to pass through
the open domain boundary in a smooth
and seamless fashion.


\section*{Acknowledgement}
The support from NSF (DMS-1318820) and 
ONR (N000141110028) is gratefully acknowledged.

\section*{Appendix: Algorithm for the Phase-Field Equation}

This Appendix provides a summary of the algorithm
we developed in \cite{DongS2012} for solving the phase field
equation \eqref{equ:CH}. We refer the reader to \cite{DongS2012}
for more detailed discussions.
The notation in this appendix follows that
of Section \ref{sec:method}.

Consider the system consisting of 
the phase field equation \eqref{equ:CH},
 the inflow boundary conditions
\eqref{equ:ibc_phi_1}--\eqref{equ:ibc_phi_2},
the wall boundary conditions
\eqref{equ:wbc_phi_1}--\eqref{equ:wbc_phi_2},
and the outflow/open boundary conditions
\eqref{equ:obc_phi_1}--\eqref{equ:obc_phi_2}.
Given $(\tilde{\mathbf{u}}^n,\phi^n)$,
where $\tilde{\mathbf{u}}^n$ is the approximation 
velocity from the algorithm discussed 
in Section \ref{sec:alg_momentum},
we discretize this system
as follows:
\begin{subequations}
\begin{equation}
\frac{\gamma_0\phi^{n+1}-\hat{\phi}}{\Delta t}
+ \tilde{\mathbf{u}}^{*,n+1}\cdot\nabla\phi^{*,n+1}
= 
-\lambda\gamma_1\nabla^2\left[
 \nabla^2\phi^{n+1}
 - \frac{S}{\eta^2}(\phi^{n+1}-\phi^{*,n+1})
 - h(\phi^{*,n+1})
\right]
+ g^{n+1}
\label{equ:phi_1}
\end{equation}
\begin{equation}
\phi^{n+1} = \phi_b^{n+1},
\qquad \text{on} \ \partial\Omega_i
\label{equ:phi_2}
\end{equation}
\begin{equation}
\nabla^2\phi^{n+1} - h(\phi^{n+1}) = g_b^{n+1},
\qquad \text{on} \ \partial\Omega_i
\label{equ:phi_3}
\end{equation}
\begin{equation}
\mathbf{n}\cdot\nabla\left[
  \nabla^2\phi^{n+1} 
  - \frac{S}{\eta^2}(\phi^{n+1}-\phi^{*,n+1})-h(\phi^{*,n+1}) 
\right]
= g_{c1}^{n+1},
\quad \text{on} \ \partial\Omega_w
\label{equ:phi_4}
\end{equation}
\begin{equation}
\mathbf{n}\cdot\nabla\phi^{n+1}
=
\frac{3\sigma}{4\lambda}\cos\theta_s
\left[1-\left(\phi^{*,n+1}\right)^2\right]
+ g_{c2}^{n+1},
\quad \text{on} \ \partial\Omega_w.
\label{equ:phi_5}
\end{equation}
\begin{equation}
\mathbf{n}\cdot\nabla\left[
  \nabla^2\phi^{n+1} 
  - \frac{S}{\eta^2}(\phi^{n+1}-\phi^{*,n+1})-h(\phi^{*,n+1}) 
\right]
= g_{a1}^{n+1},
\quad \text{on} \ \partial\Omega_o
\label{equ:phi_6}
\end{equation}
\begin{equation}
\mathbf{n}\cdot\nabla\phi^{n+1}
= -D_0 \left.\frac{\partial\phi}{\partial t} \right|^{*,n+1}
+ g_{a2}^{n+1},
\quad \text{on} \ \partial\Omega_o
\label{equ:phi_7}
\end{equation}
\begin{equation}
\mathbf{n}\cdot\nabla\phi^{n+1}
= -D_0 \frac{\gamma_0\phi^{n+1}-\hat{\phi}}{\Delta t}
+ g_{a2}^{n+1},
\quad \text{on} \ \partial\Omega_o.
\label{equ:phi_8}
\end{equation}
\end{subequations}
%
In the above equations, $\hat{\phi}$
is defined in \eqref{equ:var_hat_def},
$\tilde{\mathbf{u}}^{*,n+1}$ and $\phi^{*,n+1}$ are
 defined in \eqref{equ:var_star_def},
and $S$ is a chosen constant that must satisfy
the condition
\begin{equation}
S \geqslant \eta^2\sqrt{\frac{4\gamma_0}{\lambda\gamma_1\Delta t}}.
\label{equ:S_condition}
\end{equation}
$\left.\frac{\partial\phi}{\partial t}  \right|^{*,n+1}$
is an explicit approximation of
$\frac{\partial\phi}{\partial t}$ at
time step $(n+1)$, given by
\begin{equation}
\left.\frac{\partial\phi}{\partial t}  \right|^{*,n+1}
= \left\{
\begin{array}{ll}
\frac{1}{\Delta t}(\phi^n - \phi^{n-1}), & \text{if} \ J=1 \\
\frac{1}{\Delta t}\left(
  \frac{5}{2}\phi^n - 4\phi^{n-1} + \frac{3}{2}\phi^{n-2}
\right), & \text{if} \ J=2
\end{array}
\right.
\label{equ:phi_deriv_expr}
\end{equation}
where $J$ is the order of temporal accuracy
defined in Section \ref{sec:alg_momentum}.
Note that equations \eqref{equ:phi_7} and \eqref{equ:phi_8}
are two different discretizations of
the boundary condition \eqref{equ:obc_phi_2},
and they will be used in different stages of 
the implementation as discussed below.

Rewrite \eqref{equ:phi_1} into
\begin{equation}
\nabla^2\left[
  \nabla^2\phi^{n+1} - \frac{S}{\eta^2}\phi^{n+1}
\right]
+ \frac{\gamma_0}{\lambda\gamma_1\Delta t}\phi^{n+1}
= Q = Q_1 + \nabla^2 Q_2,
\label{equ:phi_1_trans_1}
\end{equation}
where
\begin{equation}
Q_1 = \frac{1}{\lambda\gamma_1}\left[
  g^{n+1} - \tilde{\mathbf{u}}^{*,n+1}\cdot\nabla\phi^{*,n+1}
  + \frac{\hat{\phi}}{\Delta t}
\right],
\quad
Q_2 = -\frac{S}{\eta^2}\phi^{*,n+1}
+ h(\phi^{*,n+1}).
\end{equation}
Equation \eqref{equ:phi_1_trans_1}
can be reformulated into an equivalent form
(see \cite{YueFLS2004,DongS2012})
\begin{subequations}
\begin{equation}
\nabla^2\psi^{n+1} 
- \left(\alpha + \frac{S}{\eta^2}  \right)\psi^{n+1} = Q,
\label{equ:CH_psi}
\end{equation}
\begin{equation}
\nabla^2\phi^{n+1} + \alpha\phi^{n+1} = \psi^{n+1},
\label{equ:CH_phi}
\end{equation}
\end{subequations}
where $\psi^{n+1}$ is an auxiliary
phase field function, and the constant
$\alpha$ is given by
\begin{equation}
\alpha = -\frac{S}{2\eta^2}\left[
 1 - \sqrt{1 - \frac{4\gamma_0}{\lambda\gamma_1\Delta t}\left(\frac{\eta^2}{S} \right)^2 }
\right].
\end{equation}
Note that $\alpha<0$
 and $\alpha+\frac{S}{\eta^2}>0$
under the condition \eqref{equ:S_condition}
for the chosen constant $S$.

In light of equation \eqref{equ:CH_phi}, we can 
transform \eqref{equ:phi_3} into
\begin{equation}
\psi^{n+1} = \alpha\phi_b^{n+1} + h(\phi_b^{n+1})
 - g_b^{n+1},
\quad \text{on} \ \partial\Omega_i,
\label{equ:ibc_phi_reform_1}
\end{equation}
where we have used \eqref{equ:phi_2}.
Similarly, equation \eqref{equ:phi_4} is transformed
into
\begin{equation}
\mathbf{n}\cdot\nabla\psi^{n+1} = \mathbf{n}\cdot\nabla Q_2
+ \left( 
  \alpha + \frac{S}{\eta^2}
\right)\left\{
  \frac{3\sigma}{4\lambda}\cos\theta_s
\left[1-\left(\phi^{*,n+1}\right)^2\right]
+ g_{c2}^{n+1}
\right\}
+ g_{c1}^{n+1},
\quad \text{on} \ \partial\Omega_w,
\label{equ:wbc_phi_1_reform_1}
\end{equation}
where we have used equations \eqref{equ:phi_5}
and \eqref{equ:CH_phi}.
Equation \eqref{equ:phi_6} is transformed
into 
\begin{equation}
\mathbf{n}\cdot\nabla\psi^{n+1}
= \mathbf{n}\cdot\nabla Q_2
+ \left(\alpha + \frac{S}{\eta^2}  \right)
  \mathbf{n}\cdot\nabla\phi^{n+1}
+ g_{a1}^{n+1},
\quad \text{on} \ \partial\Omega_o,
\label{equ:obc_phi_1_reform_1}
\end{equation}
where we have used \eqref{equ:CH_phi}.

We next derive the weak forms for
the equations \eqref{equ:CH_psi} and
\eqref{equ:CH_phi} in order
to facilitate the implementation
with $C^0$ spectral elements.
Let
\begin{equation}
H^1_{\phi 0}(\Omega) = \left\{ \
v\in H^1(\Omega) \ :\ v|_{\partial\Omega_i} = 0
\ \right\},
\end{equation}
and $\varpi\in H^1_{\phi 0}(\Omega)$
denote the test function.
Taking the $L^2$ inner product
between $\varpi$ and equation \eqref{equ:CH_psi}
and integrating by part,
we get the weak form about $\psi^{n+1}$,
\begin{equation}
\begin{split}
\int_{\Omega}\nabla\psi^{n+1}\cdot\nabla\varpi
&+ \left(\alpha + \frac{S}{\eta^2} \right)\int_{\Omega}\psi^{n+1}\varpi
=
-\int_{\Omega} Q_1\varpi
+ \int_{\Omega}\nabla Q_2\cdot\nabla\varpi \\
&
+ \left(
  \alpha + \frac{S}{\eta^2}
\right)\int_{\partial\Omega_w}\left\{
  \frac{3\sigma}{4\lambda}\cos\theta_s\left[
    1-(\phi^{*,n+1})^2
  \right] + g_{c2}^{n+1}
\right\} \varpi
+\int_{\partial\Omega_w} g_{c1}^{n+1}\varpi \\
&
+ \left(
  \alpha + \frac{S}{\eta^2}
\right) \int_{\partial\Omega_o}\left(
  -D_0\left.\frac{\partial\phi}{\partial t} \right|^{*,n+1}
  + g_{a2}^{n+1}
\right) \varpi
+ \int_{\partial\Omega_o} g_{a1}^{n+1}\varpi,
\quad \forall \varpi \in H^1_{\phi 0}(\Omega),
\end{split}
\label{equ:psi_weakform}
\end{equation}
where we have used \eqref{equ:wbc_phi_1_reform_1},
\eqref{equ:obc_phi_1_reform_1}, 
and \eqref{equ:phi_7}. Note that
$
\left.\frac{\partial\phi}{\partial t} \right|^{*,n+1}
$
is given by \eqref{equ:phi_deriv_expr}.

Again let $\varpi \in H_{\phi 0}^1(\Omega)$ denote
the test function.
Taking the $L^2$ inner product between $\varpi$
and equation \eqref{equ:CH_phi} and integrating
by part, we obtain the weak form about
$\phi^{n+1}$,
\begin{equation}
\begin{split}
\int_{\Omega} \nabla\phi^{n+1}\cdot\nabla\varpi
&
-\alpha \int_{\Omega}\phi^{n+1} \varpi
+ \frac{\gamma_0 D_0}{\Delta t} \int_{\partial\Omega_o}
    \phi^{n+1}\varpi
= -\int_{\Omega} \psi^{n+1}\varpi 
+ \int_{\partial\Omega_o}\left(
  \frac{D_0}{\Delta t}\hat{\phi} + g_{a2}^{n+1}
\right) \varpi \\
&
+ \int_{\partial\Omega_w}\left\{
  \frac{3\sigma}{4\lambda}\cos\theta_s\left[
    1-(\phi^{*,n+1})^2
  \right] + g_{c2}^{n+1}
\right\},
\quad \forall \varpi \in H^1_{\phi 0}(\Omega),
\end{split}
\label{equ:phi_weakform}
\end{equation}
where we have used \eqref{equ:phi_5}
and \eqref{equ:phi_8}.

Equations \eqref{equ:psi_weakform} and 
\eqref{equ:phi_weakform} are in weak forms,
and all the terms involved therein can be
computed directly using $C^0$ elements.
These equations can be discretized in space
using $C^0$ spectral elements in the 
standard fashion.
Note that these two equations are
de-coupled.

In summary, given $(\tilde{\mathbf{u}}^n, \phi^{n})$,
our final algorithm for the phase field
equation consists of the following procedure.
We refer to this procedure as
{\bf AdvancePhase}. It produces 
$(\psi^{n+1},\phi^{n+1},\nabla^2\phi^{n+1})$ in a
de-coupled fashion as follows \\
\noindent\underline{{\bf AdvancePhase} procedure:}
\begin{itemize}

\item
Solve equation \eqref{equ:psi_weakform},
together with the Dirichlet condition
\eqref{equ:ibc_phi_reform_1} on $\partial\Omega_i$,
for $\psi^{n+1}$;

\item
Solve equation \eqref{equ:phi_weakform},
together with the Dirichlet condition \eqref{equ:phi_2}
on $\partial\Omega_i$, for $\phi^{n+1}$;

\item
Compute $\nabla^2\phi^{n+1}$ according to equation
\eqref{equ:CH_phi} as follows:
\begin{equation}
\nabla^2\phi^{n+1} = \psi^{n+1} - \alpha\phi^{n+1}.
\label{equ:laplace_phi_expr}
\end{equation}

\end{itemize}


%
\bibliographystyle{plain}
\bibliography{pcorr,obc,mypub,nse,sem,contact_line,interface}

\end{document}